\documentclass[aps,prb,reprint,superscriptaddress,longbibliography,floatfix]{revtex4-2}
\usepackage{amsmath,amssymb,bm,booktabs,mathrsfs}
\usepackage[colorlinks=true,linkcolor=blue,citecolor=blue,urlcolor=blue]{hyperref}

\hypersetup{pdftitle={Quotient indices for symmetry-protected zero modes},pdfauthor={Zheng Zhang and Y. X. Zhao},pdfsubject={PRB-format manuscript draft}}
\newcommand{\R}{\mathbb R}
\newcommand{\C}{\mathbb C}
\newcommand{\Z}{\mathbb Z}
\newcommand{\HH}{\mathbb H}
\newcommand{\Cl}{\mathrm{Cl}}
\newcommand{\Res}{\operatorname{Res}}
\newcommand{\im}{\operatorname{im}}
\newcommand{\Ind}{\operatorname{Ind}}
\newcommand{\End}{\operatorname{End}}
\newcommand{\tr}{\operatorname{tr}}
\newcommand{\id}{\mathbf 1}
\newcommand{\Rep}{\mathcal R}
\newcommand{\Qind}{\mathcal I}
\newcommand{\eps}{\epsilon}
\newcommand{\coker}{\operatorname{coker}}
\begin{document}
\raggedbottom
\title{Quotient Index for Symmetry-Protected Zeros and the Tenfold Way in Mackey’s Construction of Super Corepresentations}
\author{Zheng Zhang}
\affiliation{Department of Physics, School of Science, Lanzhou University of Technology, Lanzhou 730050, China}

\affiliation{Department of Physics and HK Institute of Quantum Science \& Technology, The University of Hong Kong, Pokfulam Road, Hong Kong, China}

\author{Y. X. Zhao}
\email{yuxinphy@hku.hk}
\affiliation{Department of Physics and HK Institute of Quantum Science \& Technology, The University of Hong Kong, Pokfulam Road, Hong Kong, China}
\date{\today}
\begin{abstract}
Symmetries that anticommute with a Hamiltonian can protect zero modes by obstructing their removal through symmetry-compatible perturbations. We formulate this obstruction as a quotient of the additive group of symmetry representations by the image of the map that forgets a compatible representation grading. To compute the quotient, we use Mackey's method to construct all irreducible super corepresentations of finite graded symmetry groups, allowing antiunitary operations and projective factor systems. Starting from the unitary Hamiltonian-preserving normal subgroup, this construction yields two complex and ten real structural types, characterized by residual Altland--Zirnbauer symmetry algebras. The residual and induced graded modules have isomorphic graded commutants, and their represented symmetry algebras share a graded Morita class that is super opposite to that of the commutant. We derive character criteria that identify these types. Exhaustive enumeration of the restriction relations gives independent sector contributions $0$, $\Z$, or $\Z_2$, with only integer factors in the purely unitary case. In the finite semisimple setting, the complete index vanishes if and only if all zero modes can be lifted by a symmetry-compatible perturbation. Applications to one-dimensional end states and interpolation cylinders of disordered boundary Dirac theories recover protected multiplet counts and connect the quotient to bulk topology and sigma-model topological terms using bulk--boundary correspondence, texture-index theorems, and fermion-phase calculations.
\end{abstract}
\maketitle

\section{Introduction}
The symmetry structure of a one-body Hamiltonian can differ from that of the underlying many-body system. Besides symmetries that commute with the Hamiltonian, a one-body description can admit unitary or antiunitary operators that anticommute with it, mapping energy $E$ to $-E$. Freed and Moore emphasized the spectral condition behind this distinction: a Hamiltonian bounded below and unbounded above cannot admit such a reversing symmetry \cite{FreedMoore2013}. For a many-body Hamiltonian satisfying these spectral conditions, reflection of arbitrarily high energies would contradict the lower bound set by the ground-state energy. A one-body Hamiltonian in the Dirac--Nambu description need not satisfy these conditions and can have a spectrum symmetric about zero. Familiar examples of Hamiltonian-reversing operations are antiunitary particle-hole symmetry $C$ and the unitary chiral product $CT$ in the Altland--Zirnbauer (AZ) classification, as well as sublattice symmetry in bipartite hopping models \cite{AltlandZirnbauer1997,Ryu2010}. Anticommuting operators also occur naturally for chirally graded elliptic Dirac operators, with chirality supplying the grading. The associated zero-mode index connects the chirality imbalance to background-field topology and underlies the integrated axial anomaly \cite{AtiyahSinger1968,Fujikawa1979}.

Such symmetries make zero energy special: states away from zero occur in opposite-energy pairs, whereas zero modes can carry symmetry representations that obstruct their removal. Examples include end states of one-dimensional topological insulators and superconductors, particularly in the chiral AZ classes \cite{Schnyder2008,Ryu2010,Kitaev2009}, index-protected kernels of Dirac operators \cite{ABS1964,AtiyahSinger1969}, and superconducting gap nodes at high-symmetry momenta, where crystalline representations constrain the allowed couplings \cite{Sumita2019,YangLiu2024}. Anticommutation alone does not protect every zero mode; protection depends on the symmetry content of the entire zero-mode sector. Opposite chiralities may pair and move away from zero, while an excess of one chirality remains. With antiunitary or additional crystalline symmetries, the elementary removable content can be different. The common question is therefore which combinations of zero-mode symmetry multiplets admit an invertible symmetry-compatible perturbation.

Clifford-module index theory and symmetry-based classifications of superconducting nodes already connect representation data to topological obstructions \cite{ABS1964,AtiyahSinger1969,Sumita2019,YangLiu2024}. The problem we address is to formulate this removal question directly and uniformly for a general projective symmetry group, including unitary and antiunitary operations that either preserve or reverse the Hamiltonian. Projective symmetry actions reflect the ray nature of quantum states, with phase ambiguities encoded by factor systems \cite{Wigner1939}. In crystalline systems, gauge structures can give rise to nontrivial factor systems and projective symmetry algebras \cite{ZhaoProjectivePT2021,ShaoGauge2021,ChenGauge2023}. We seek an explicit additive invariant of the zero-mode representation, together with a construction of all removable multiplets. This separates the algebraic obstruction from the additional bulk, spatial, or analytic data needed to identify it with a topological invariant in a particular application.

Our construction is a \emph{quotient index}: the additive group of zero-mode symmetry representations, or corepresentations when antiunitary operations are present, modulo the subgroup generated by removable representations. Spectral flattening identifies removability with the existence of a compatible Hermitian involution $h$, with $h^2=\id$. Indeed, an invertible perturbation can be flattened to $h$, and conversely a small multiple of $h$ lifts every mode in the sector. The eigenspaces of $h$ supply a compatible $\Z_2$ grading. The removable representations are therefore precisely the underlying ungraded representations of compatible super representations, and the denominator of the quotient is the \emph{image} of the map that forgets this grading. 

Computing the quotient thus leads naturally to the representation theory of graded symmetry groups. We use the commutativity grading $c(g)=\pm1$ to record whether a symmetry preserves or reverses the Hamiltonian, and the independent unitarity grading $\tau(g)=\pm1$ to distinguish unitary and antiunitary actions. A representation with a compatible module grading is a super representation; with antiunitary actions included, it is a super corepresentation. Adjoining the lifting involution as a generator converts these objects into projective corepresentations of an enlarged group $G^h$, whose factor system records the prescribed commutation signs. We use established induction and graded-algebra methods to construct these irreducible corepresentations, or equivalently the irreducible super corepresentations of $G$, and compute their contributions to the quotient for fixed gradings and a fixed projective factor system in the finite semisimple setting.

The relevant algebraic classifications are established results. Finite-dimensional real graded-simple algebras have ten graded Morita classes, each with a unique super division-algebra representative up to graded isomorphism: eight have a real graded center and two have a complex graded center \cite{Wall1964,DeligneMorgan1999,DeligneSpinors1999,GeikoMoore2021}. Morita equivalence identifies the graded module structure while allowing matrix multiplicities to change. The super Schur lemma states that the graded commutant of an irreducible graded module is a super division algebra, meaning that every nonzero homogeneous endomorphism is invertible \cite{DeligneMorgan1999,FreedMoore2013}. Its isomorphism class therefore assigns one of ten structural types to an irreducible super corepresentation. This extends the role of $\R$, $\C$, and $\HH$ in the ordinary threefold classification \cite{Dyson1962}.

The relation between these classifications is fixed by the graded double-centralizer theorem. If a graded-simple algebra $\mathscr A$ acts on an irreducible graded module with graded commutant $\mathscr D$, then $\mathscr A\sim_{\mathrm{grMorita}}\mathscr D^{\mathrm{sop}}$ \cite{DeligneSpinors1999}. Here $\mathrm{sop}$ denotes the super opposite: reversing the order of two homogeneous factors includes the sign determined by their parities. Thus the algebra and its commutant have super-opposite graded Morita classes. 

The ten AZ classes provide a physical realization of this algebraic tenfold way. The associated $CT$ symmetry algebras are generated by the available preserving and reversing symmetries, together with the complex structure, and are graded by their commutation with the Hamiltonian. These algebras represent the ten graded Morita classes \cite{FreedMoore2013,GeikoMoore2021}. They consequently supply AZ labels for the ten super division algebras and, through the graded commutant, for the structural types of irreducible super corepresentations. Such a label does not distinguish inequivalent irreducible representations within one type. We use the same Morita-class labeling for symmetry algebras and commutant algebras, so their labels are related by super opposition.

Our contribution is the explicit formulation of the zero-mode removal obstruction as a quotient index and its exhaustive computation organized by the unitary Hamiltonian-preserving subgroup. Using Mackey's orbit-and-little-group method \cite{Mackey1958} and its extensions to projective corepresentations with antiunitary symmetries \cite{Parthasarathy1969,ShawLever1974,Zhang2026}, we begin with irreducible projective representations $\Delta$ of the unitary Hamiltonian-preserving normal subgroup $N=\ker\tau\cap\ker c$. The quotient $G/N$ acts on these representations. For each orbit, stabilizer intertwiners remove the reference-space action and leave a projective symmetry algebra on the residual multiplicity space. The five possible little cogroups, which are subgroups of $\Z_2\times\Z_2$, and their residual factor classes yield ten types of $N$-irrep sectors, each carrying one of the AZ $CT$ algebras as its residual symmetry algebra. Tensoring the residual modules with the intertwiners and inducing to $G^h$ constructs all super irreps, including for nonsplit extensions. A particular group realizes only the types permitted by its orbits and factor system; its residual AZ labels need not coincide with the AZ label of the full physical system.

The construction also determines the graded commutants. Let $E$ be the irreducible graded residual module, $\mathscr A_\Delta$ its residual symmetry algebra, and $V$ the induced super irrep. We prove that their graded commutants are isomorphic, $\mathscr D(V)\cong_{\mathrm{gr}}\mathscr D(E)$. Applying the general algebra--commutant relation described above to both modules then gives
\[
 \mathscr B_V\sim_{\mathrm{grMorita}}\mathscr A_\Delta
 \sim_{\mathrm{grMorita}}\mathscr D(V)^{\mathrm{sop}},
\]
where $\mathscr B_V$ is the image of the original symmetry algebra on $V$. Thus the residual and represented symmetry algebras have the same graded Morita class, while their commutants have the super-opposite class. This relates their AZ labels by an involution that exchanges six of the ten labels and fixes the other four. Combining Wigner--Frobenius--Schur indicators with character overlaps \cite{Dyson1962,Sumita2019}, we formulate complete criteria for these structural types in the present projective setting with independent $\tau$ and $c$ gradings, both on the starting $N$-irrep and on the constructed super irrep; the tests on the constructed super irrep retain the orbit-size and constituent-multiplicity factors. When every symmetry is unitary, the same construction reduces to the two complex types A and AIII.

To obtain the full quotient index group, we exhaustively enumerate all super irreps and their ungraded restrictions. Equivalently, the Mackey construction enumerates all $G/N$ orbits of irreps of the unitary Hamiltonian-preserving subgroup $N$, together with the residual modules for each orbit. Each independent orbit sector contributes one component, $0$, $\Z$, or $\Z_2$; equivalent restrictions of different super irreps are counted only once. The three possibilities arise because the elementary removable content is, respectively, one already gappable multiplet, a pair of inequivalent irreducible multiplets, or two copies of one multiplet. The corresponding index is, respectively, trivial, an integer multiplicity difference, or a multiplicity parity. Taking the direct sum over all orbit sectors gives the full quotient group, a direct sum of $\Z$ and $\Z_2$ factors; the purely unitary case has only integer factors. In the finite semisimple setting, the complete index vanishes if and only if the zero-mode representation admits a compatible invertible perturbation. These statements concern unrestricted symmetry-compatible perturbations on the sector; microscopic locality requires further input.

We illustrate the construction in two applications. For end modes of one-dimensional insulators and superconductors, it yields chiral imbalances, Majorana parities, and Majorana Kramers-pair parities \cite{Kitaev2001,ZhangKaneMele2013,ZhaoWang2014}, whose relation to bulk topology is supplied by bulk--boundary correspondence \cite{Schnyder2008,Ryu2010,Kitaev2009}. For interpolation cylinders of disordered boundary Dirac theories, we determine the residual symmetry sectors and their quotient indices. A normalized texture-index theorem and a regulated fermion-phase or anomaly calculation then connect the cylinder-kernel class to sigma-model theta or Wess--Zumino terms \cite{ZhaoSigma2026}.

Section~\ref{sec:motivation} introduces the quotient through end modes of semi-infinite one-dimensional insulators in class AIII, and Sec.~\ref{sec:quotient} defines the general quotient index and establishes its stability. Sections~\ref{sec:unitary} and \ref{sec:tenfold} develop the complex and real Mackey constructions, graded commutants, and restriction indices; the complex character tests appear in Sec.~\ref{sec:unitary}, and Sec.~\ref{sec:characters} gives the real character tests. Sections~\ref{sec:examples} and \ref{sec:application} present the two applications, followed by the discussion and conclusions. Appendices~\ref{app:complexalternative} and \ref{app:realalternative} explain how alternative starting subgroups change auxiliary residual labels while preserving the physical graded commutants and quotient indices. Appendix~\ref{app:complexalternative} starts Mackey's construction from the whole group $G$, whereas Appendix~\ref{app:realalternative} starts from the unitary subgroup $G_0$, also used in the antilinear representation theory of Rumynin and Taylor \cite{RumyninTaylor2021}.

\section{Motivation: End zero modes of a one-dimensional class-AIII insulator}
We use the end zero modes of a semi-infinite one-dimensional insulator in AZ class AIII to illustrate the basic ideas behind the quotient index.
\label{sec:motivation}
\subsection{Chirality and the stability of end states}
Consider a one-dimensional noninteracting insulator with a bulk gap at zero energy and unitary chiral symmetry, placing it in AZ class AIII. We focus on the end of a semi-infinite system. The single-particle Hamiltonian satisfies
\begin{equation}
 \Gamma H=-H\Gamma,\qquad
 \Gamma^\dagger=\Gamma,\qquad \Gamma^2=\id,
 \label{eq:endchiral}
\end{equation}
where $\Gamma$ is the chiral symmetry operator. We ask which combinations of end zero modes can be shifted away from zero by a perturbation near the end that preserves chiral symmetry and keeps the bulk gapped.

For a concrete illustration, consider the semi-infinite Su--Schrieffer--Heeger (SSH) chain with real hopping amplitudes $t_1,t_2$ \cite{SuSchriefferHeeger1979,SuSchriefferHeeger1980}:
\begin{equation}
 \begin{aligned}
 H_{\mathrm{SSH}}=\sum_{n=1}^{\infty}\bigl(&t_1|A_n\rangle\langle B_n|\\
 &+t_2|B_n\rangle\langle A_{n+1}|+\mathrm{H.c.}\bigr).
 \end{aligned}
 \label{eq:sshHamiltonian}
\end{equation}
Here $A_n,B_n$ are the two sites of unit cell $n$. In the ordered basis $(|A_1\rangle,|B_1\rangle,|A_2\rangle,|B_2\rangle,\ldots)$, the Hamiltonian is the semi-infinite matrix
\begin{equation}
 H_{\mathrm{SSH}}=
 \begin{pmatrix}
 0&t_1&0&0&0&0&\cdots\\
 t_1&0&t_2&0&0&0&\cdots\\
 0&t_2&0&t_1&0&0&\cdots\\
 0&0&t_1&0&t_2&0&\cdots\\
 0&0&0&t_2&0&t_1&\cdots\\
 0&0&0&0&t_1&0&\ddots\\
 \vdots&\vdots&\vdots&\vdots&\vdots&\ddots&\ddots
 \end{pmatrix}.
 \label{eq:sshmatrix}
\end{equation}
The chain begins at $A_1$, with no $B_0$ site, and extends indefinitely toward the bulk. Since every hopping connects opposite sublattices, the operator
\begin{equation}
 \Gamma=\operatorname{diag}(1,-1,1,-1,\ldots)
 \label{eq:sshGamma}
\end{equation}
obeys Eq.~\eqref{eq:endchiral}. We use this model with only chiral symmetry imposed on the allowed perturbations.

For $|t_1|<|t_2|$, set $r=t_1/t_2$. This termination supports the normalized end zero mode
\begin{equation}
 \begin{gathered}
 |\psi_0\rangle=\sqrt{1-r^2}\sum_{n=1}^{\infty}(-r)^{n-1}|A_n\rangle,\\
 H_{\mathrm{SSH}}|\psi_0\rangle=0,\qquad
 \Gamma|\psi_0\rangle=|\psi_0\rangle.
 \end{gathered}
 \label{eq:sshzeromode}
\end{equation}
Indeed, the amplitudes satisfy $t_1\psi_{A_n}+t_2\psi_{A_{n+1}}=0$ and $\psi_{B_n}=0$. The condition $|r|<1$ ensures normalizability and exponential localization at the end. Thus the semi-infinite Hamiltonian has an end zero mode of definite chirality while its bulk remains gapped, illustrating the protection of zero-energy boundary states by chiral symmetry and bulk topology \cite{RyuHatsugai2002}.

More generally, assume that the space $Z=\ker H$ of end-localized zero modes is finite-dimensional. Since $\Gamma$ preserves $Z$, it splits this space into its two eigenspaces:
\begin{equation}
 Z=Z_+\oplus Z_-,\qquad
 \Gamma|_{Z_\pm}=\pm\id,\qquad n_\pm=\dim_\C Z_\pm.
 \label{eq:enddecomposition}
\end{equation}
The familiar chiral index is
\begin{equation}
 \nu=n_+-n_-\in\Z,
 \label{eq:endindex}
\end{equation}
which equals the integer-valued bulk topological invariant.

The index $\nu$ measures the obstruction to lifting all zero modes by symmetry-preserving perturbations. Consider zero modes $|\psi_{\gamma}\rangle$ and $|\psi_{\gamma'}\rangle$ with chiralities $\gamma,\gamma'\in\{+1,-1\}$, satisfying $\Gamma |\psi_{\gamma}\rangle=\gamma |\psi_{\gamma}\rangle$ and $\Gamma |\psi_{\gamma'}\rangle=\gamma' |\psi_{\gamma'}\rangle$. A symmetry-compatible perturbation $\delta H$ obeys $\Gamma \delta H=-\delta H \Gamma$, so
\begin{equation}
 (\gamma+\gamma')\langle\psi_\gamma|\delta H|\psi_{\gamma'}\rangle=0.
 \label{eq:endselection}
\end{equation}
Thus modes of the same chirality cannot couple within the zero-mode space, whereas opposite chiralities can. In the basis adapted to Eq.~\eqref{eq:enddecomposition}, an allowed effective end Hamiltonian is
\begin{equation}
 H_{\mathrm{end}}=
 \begin{pmatrix}0&A\\A^\dagger&0\end{pmatrix},
 \qquad A\in M_{n_+\times n_-}(\C).
 \label{eq:rectangular}
\end{equation}
Each nonzero singular value of $A$ produces an opposite-energy pair. Hence
\begin{equation}
 \dim_\C\ker H_{\mathrm{end}}
 =n_++n_--2\operatorname{rank}A\geq|\nu|.
 \label{eq:endbound}
\end{equation}
For unrestricted symmetry-compatible couplings between end states, the bound is attained by pairing as many opposite-chirality zero modes as possible. A pair with coupling matrix element $m$ acquires energies $\pm|m|$, while any unpaired modes remain at zero. Pairing therefore leaves the index obstruction unchanged.


\subsection{From removable pairs to a representation quotient}
We now express this argument in terms of symmetry representations. The group generated by the chiral operator is $G=\Z_2^\Gamma=\{1,\Gamma\}$. Its two irreducible representations (irreps) $W_\pm$ have $\Gamma=\pm1$, so the end zero-mode space carries
\begin{equation}
 Z\cong W_+^{\oplus n_+}\oplus W_-^{\oplus n_-}.
 \label{eq:endrepresentation}
\end{equation}
Removability depends on both the number of modes and their symmetry representations. We therefore ask which representations of $G$ admit an invertible Hamiltonian that preserves chiral symmetry. An invertible Hamiltonian $H_\Delta$ on a removable sector can be spectrally flattened to an operator $h$ satisfying
\begin{equation}
 h^\dagger=h,\qquad h^2=\id,\qquad \Gamma h=-h\Gamma.
 \label{eq:endflattening}
\end{equation}
Flattening continuously deforms the absolute value of each eigenvalue of $H_\Delta$ to $1$ while preserving its sign, giving $h=H_\Delta|H_\Delta|^{-1}$. Conversely, $\varepsilon h$ with $\varepsilon>0$ lifts every zero mode in that sector. Thus removability depends only on the existence of such an $h$. Section~\ref{sec:lifting} makes this equivalence precise.

Whether the operator representing an element of $G$ commutes or anticommutes with $h$ is recorded by 
\begin{equation}
	c: G \rightarrow \{+1, -1\},
\end{equation}
called the commutativity grading. For $G=\mathbb{Z}_2^\Gamma$, we have $c(1)=1$ and $c(\Gamma)=-1$.

To treat $h$ on the same footing as the symmetry, we adjoin it as an abstract generator and work with $G^h=G\times\Z_2^h=\Z_2^\Gamma\times\Z_2^h$. The commutation sign between the operators $\Gamma$ and $h$ is encoded in the factor system $\omega^c$, which specifies the phases in operator multiplication and is determined by the grading $c$. The abstract group law remains that of a direct product. With the operator ordering $\widetilde\rho(\Gamma^a,h^s)=\Gamma^a h^s$, anticommutation appears as the sign
\begin{equation}
 \omega^c\bigl((\Gamma^a,h^s),(\Gamma^b,h^t)\bigr)=(-1)^{sb},
 \label{eq:endfactor}
\end{equation}
where $a,b,s,t\in\{0,1\}$. The irreducible projective representation $V$ is unique up to equivalence and can be realized as
\begin{equation}
 \rho_V(\Gamma)=\sigma_3,\qquad \rho_V(h)=\sigma_1.
 \label{eq:endpauli}
\end{equation}
Forgetting $h$, or restricting to the original chiral group, gives
\begin{equation}
 \Res V=W_+\oplus W_-.
 \label{eq:endrestriction}
\end{equation}
This is exactly the opposite-chirality pairing rule, expressed as a relation between symmetry representations.

Zero-mode spaces combine by direct sum. Passing to formal integer combinations of representations gives the free abelian group generated by the irreps. Declaring the removable pair $W_+\oplus W_-$ to be trivial then gives the quotient
\begin{equation}
 \Qind_{\mathrm{AIII}}=
 \frac{\Z[W_+]\oplus\Z[W_-]}{\Z([W_+]+[W_-])}
 \cong\Z.
 \label{eq:aiii}
\end{equation}
The isomorphism sends the class of $n_+[W_+]+n_-[W_-]$ to $n_+-n_-$. In particular, $[W_-]=-[W_+]$ in the quotient: the negative generator is of the opposite physical chirality.

The ordinary chiral index is therefore a quotient of zero-mode representations by the representation content of gapped sectors. This observation motivates the general construction: enumerate the symmetry representations that may occur at zero energy, determine which extend after a flattened lifting Hamiltonian is adjoined, and quotient by their restrictions. The next section carries out these steps for general unitary or antiunitary projective symmetries.

\section{Symmetry data and the quotient index}
\label{sec:quotient}
We now generalize the quotient index introduced in the preceding example.
\subsection{Two symmetry gradings and projective corepresentations}
The constructions below apply to, or readily extend to, compact Lie groups. For simplicity, we take $G$ to be a finite symmetry group represented by unitary or antiunitary operators that commute or anticommute with the Hamiltonian $H$. We encode these properties by two gradings, the commutativity grading $c$ and the unitarity grading $\tau$, defined as homomorphisms
\begin{equation}
 c,\tau:G\longrightarrow\{+1,-1\}.
 \label{eq:gradings}
\end{equation}
The commutativity grading $c(g)=+1$ ($-1$) specifies that the symmetry operator commutes (anticommutes) with the Hamiltonian. The unitarity grading $\tau(g)=+1$ ($-1$) specifies a unitary (antiunitary) symmetry operator. An antiunitary symmetry is represented by a unitary operator followed by complex conjugation. Writing $\eps(g)=[1-\tau(g)]/2$, we therefore take
\begin{equation}
 \rho(g)=U_g K^{\eps(g)},\qquad
 \rho(g)H=c(g)H\rho(g),
 \label{eq:symmetry}
\end{equation}
where $K$ is complex conjugation. 

Following Wigner's formulation of symmetry actions on quantum states up to phase \cite{Wigner1939}, we allow a factor system $\omega$ in the multiplication law:
\begin{equation}
 \rho(g)\rho(k)=\omega(g,k)\rho(gk).\label{eq:projective}
\end{equation}
Nontrivial factor systems occur in projective crystalline symmetry algebras of lattice systems with gauge structures \cite{ZhaoProjectivePT2021,ShaoGauge2021,ChenGauge2023}. The same formalism also accommodates Clifford symmetry algebras: their $n$ generators represent the generators of $\Z_2^n$ projectively, with $\omega$ encoding their squares and mutual anticommutation relations. Associativity requires the $\tau$-twisted cocycle identity
\begin{equation}
 \omega(g,k)\omega(gk,l)=\omega(k,l)^{\tau(g)}\omega(g,kl).
 \label{eq:cocycle}
\end{equation}
For a phase $\omega$, the exponent $\tau(g)=-1$ implements complex conjugation, reflecting the antiunitarity of $\rho(g)$. Under a rephasing $\rho(g)\mapsto\beta(g)\rho(g)$, the factor system changes by the twisted coboundary
\begin{equation}
 \omega'(g,k)=
 \frac{\beta(g)\beta(k)^{\tau(g)}}{\beta(gk)}\omega(g,k).
 \label{eq:gauge}
\end{equation}

In a zero-mode problem, the commutativity grading $c$ is usually nontrivial. We also allow $c=1$ so that the preserving-only cases fit into the same framework; then there is no algebraic obstruction, because a Hamiltonian proportional to $\id$ is allowed on every sector. Throughout, ``representation'' includes projective corepresentations with the fixed data $(\tau,\omega)$---a corepresentation being a group action by linear and antilinear operators with the specified projective multiplication law. A representation equipped with a compatible $\Z_2$ grading is called a \emph{super representation}, or a \emph{super corepresentation} when antiunitary symmetries are included; an irreducible such object is called a \emph{super irrep}. The grading $c$ constrains a possible Hamiltonian but imposes no condition on the representation carried by states at exactly zero energy.

\subsection{Gappable modules and the flattened extension}
\label{sec:lifting}
Let $V$ be a finite-dimensional symmetry-invariant sector on which the unperturbed Hamiltonian vanishes. At the level of the algebraic obstruction, a perturbation on this sector is an operator $B$ satisfying
\begin{equation}
 B^\dagger=B,\qquad \rho(g)B=c(g)B\rho(g).
 \label{eq:liftingperturbation}
\end{equation}
It lifts the entire sector exactly when $B$ is invertible. We assume no particular matrix form, coupling pattern, or preferred perturbation direction.

This condition has a direct interpretation in degenerate perturbation theory. Extend $B$ by zero outside $V$ and consider $H_\lambda=H+\lambda B$ for small real $\lambda$. If $P$ projects onto $V$, then $H|_V=0$ gives the effective Hamiltonian
\begin{equation}
 H_{\mathrm{eff}}(\lambda)
 =\left.P(H+\lambda B)P\right|_V=\lambda B.
 \label{eq:perturbativeB}
\end{equation}
The zero-mode energies therefore become $\lambda b_j$, where $b_j$ are the eigenvalues of $B$. An invertible $B$ lifts every mode for any nonzero $\lambda$. Here the effective Hamiltonian is exact because the perturbation acts only within $V$. For a general symmetry-compatible perturbation $\delta H$ of an isolated zero-mode eigenspace $V$, degenerate perturbation theory instead gives $B=\left.P\delta H P\right|_V$ and $H_{\mathrm{eff}}(\lambda)=\lambda B+O(\lambda^2)$ for $H+\lambda\delta H$. Thus $B$ can also be viewed as the first-order perturbation that controls the splitting of the zero modes; if it is invertible, the sector is lifted for sufficiently small nonzero $\lambda$.

For any such invertible $B$, spectral flattening retains only its sign. Write $B=\sum_\lambda\lambda P_\lambda$, where $P_\lambda$ are its orthogonal spectral projectors and every eigenvalue $\lambda$ is real and nonzero. Its positive magnitude and inverse are
\begin{equation}
 \begin{aligned}
 |B|=(B^2)^{1/2}&=\sum_\lambda|\lambda|P_\lambda,\\
 |B|^{-1}&=\sum_\lambda|\lambda|^{-1}P_\lambda.
 \end{aligned}
 \label{eq:Bmodulus}
\end{equation}
The symmetry relation maps the $\lambda$ eigenspace to the $c(g)\lambda$ eigenspace, so $\rho(g)P_\lambda\rho(g)^{-1}=P_{c(g)\lambda}$. Since $|c(g)\lambda|=|\lambda|$, it follows that $\rho(g)|B|=|B|\rho(g)$, and likewise for $|B|^{-1}$. This argument also holds for antiunitary $\rho(g)$: the eigenvalues and the coefficients $|\lambda|^{\pm1}$ are real and hence unchanged by complex conjugation.

Define the flattened operator by
\begin{equation}
 h=B|B|^{-1}=\sum_\lambda\operatorname{sgn}(\lambda)P_\lambda.
 \label{eq:Bspectralflattening}
\end{equation}
The projector expansion gives $h^\dagger=h$ and $h^2=\sum_\lambda P_\lambda=\id$. Commutation of $|B|^{-1}$ with every symmetry then yields $\rho(g)h=c(g)h\rho(g)$. Thus
\begin{equation}
 h^\dagger=h,\qquad h^2=\id,\qquad
 \rho(g)h=c(g)h\rho(g).
 \label{eq:h}
\end{equation}
Moreover, for any energy scale $\varepsilon>0$, the path
\begin{equation}
 B_u=(1-u)B+u\varepsilon h,\qquad 0\leq u\leq1,
 \label{eq:flattenpath}
\end{equation}
preserves the symmetry and never closes the gap on $V$: an eigenvalue $\lambda\ne0$ evolves to $(1-u)\lambda+u\varepsilon\operatorname{sgn}\lambda$, which keeps its sign throughout. Conversely, every $h$ obeying Eq.~\eqref{eq:h} supplies the lifting perturbation $\varepsilon h$, which shifts all zero modes to $\pm\varepsilon$ with arbitrarily small perturbation norm. Therefore
\begin{equation}
 \begin{gathered}
 V\text{ admits a compatible invertible }B\\
 \Longleftrightarrow\quad
 V\text{ admits a compatible involution }h.
 \end{gathered}
 \label{eq:liftingequivalence}
\end{equation}
We call such a representation $V$ \emph{gappable} (equivalently, \emph{removable}). 

If $B$ lifts only part of $V$, then both $\ker B$ and its orthogonal complement are symmetry invariant. On $V_{\mathrm{lift}}=(\ker B)^\perp$, the restriction of $B$ is invertible, and the same construction yields $h^2=\id_{V_{\mathrm{lift}}}$. Extended by zero to the residual kernel, the flattened operator instead squares to the projector onto $V_{\mathrm{lift}}$, not to $\id_V$. 

Thus the involution condition characterizes complete lifting of each removable sector; partial lifting removes such a sector and leaves the zero-mode space to be classified. The Hermitian involution $h$ is a bookkeeping device representing \emph{every} symmetry-compatible perturbation that can lift zero modes after inessential energy scales have been removed. 



To determine which sectors are gappable by symmetry-preserving perturbations, we introduce the abstract group $G^h=G\times\Z_2^h$ and choose the operator convention
\begin{equation}
 \widetilde\rho(g,s)=\rho(g)h^s,\qquad s\in\{0,1\}.
 \label{eq:ordering}
\end{equation}
Direct multiplication fixes the lifted data:
\begin{align}
 \bar\tau(g,s)&=\tau(g),\nonumber\\
 \omega^c\bigl((g,s),(k,t)\bigr)
 &=\omega(g,k)c(k)^s.
 \label{eq:liftedcocycle}
\end{align}
The lifted factor satisfies the cocycle identity because $\omega$ does and the additional signs obey
$c(k)^s c(l)^{s+t}=c(l)^t c(kl)^s$, since $c$ is a homomorphism with values $\pm1$. The sign $c(k)^s$ records the commutation relation between $h^s$ and $\rho(k)$. Which factor is written first in Eq.~\eqref{eq:ordering} is a convention. Choosing $h^s\rho(g)$ instead changes the factor system by the twisted coboundary of the real one-cochain $c(g)^s$ and gives the equally valid $\omega(g,k)c(g)^t$.


The eigenspaces of $h$ define a compatible module grading,
\begin{equation}
 V=V^+\oplus V^-,\qquad
 \rho(g)V^\sigma\subseteq V^{c(g)\sigma}.
 \label{eq:modulegrading}
\end{equation}
Thus adjoining $h$ and supplying a compatible grading are equivalent operations. Treating $h$ as an additional group generator and replacing $\omega$ by $\omega^c$ allows us to apply Mackey's method for group extensions to construct all irreducible projective super representations of $G$.



\subsection{Definition of the quotient index}
Let $\Rep_{\tau,\omega}(G)$ be the group of formal integer combinations of finite-dimensional representations with the fixed data $(\tau,\omega)$, with direct sum as addition. Complete reducibility makes it free on the irreducible representations $W_i$:
\begin{equation}
 \Rep_{\tau,\omega}(G)=\bigoplus_i\Z[W_i].
 \label{eq:repgroup}
\end{equation}
This is an additive group and not a representation ring, because the tensor product of two representations does not in general carry the same factor system.

Similarly, let $\Rep^{\mathrm{gr}}_{c,\tau,\omega}(G)$ denote the super representation group: the group of formal integer combinations of finite-dimensional super representations with the fixed data $(c,\tau,\omega)$, using direct sum as addition and grading-preserving equivalence. The correspondence in Eqs.~\eqref{eq:ordering}--\eqref{eq:modulegrading} identifies a super representation $(V,\rho,h)$ of $G$ with a representation $\widetilde\rho(g,s)=\rho(g)h^s$ of $G^h$. This correspondence preserves direct sums and irreducibility, so
\begin{equation}
 \Rep^{\mathrm{gr}}_{c,\tau,\omega}(G)
 \cong\Rep_{\bar\tau,\omega^c}(G^h).
 \label{eq:superrepgroup}
\end{equation}
The grading $c$ specifies whether each symmetry preserves or exchanges the two graded subspaces, as in Eq.~\eqref{eq:modulegrading}.

Restricting a representation of $G^h$ to $G$ defines the map
\begin{equation}
 \Res:\Rep_{\bar\tau,\omega^c}(G^h)
       \longrightarrow\Rep_{\tau,\omega}(G).
 \label{eq:res}
\end{equation}
Under Eq.~\eqref{eq:superrepgroup}, this is precisely the homomorphism that forgets the module grading:
\begin{equation}
 \begin{gathered}
 F:\Rep^{\mathrm{gr}}_{c,\tau,\omega}(G)
 \longrightarrow\Rep_{\tau,\omega}(G),\\
 F([V,\rho,h])=[V,\rho].
 \end{gathered}
 \label{eq:forgetgrading}
\end{equation}
The zero-mode obstruction group can therefore be written equivalently as
\begin{equation}
 \boxed{\begin{aligned}
 \Qind(G;c,\tau,\omega)&=
 \frac{\Rep_{\tau,\omega}(G)}
 {\Res\!\left[\Rep_{\bar\tau,\omega^c}(G^h)\right]}\\
 &=\frac{\Rep_{\tau,\omega}(G)}
 {F\!\left[\Rep^{\mathrm{gr}}_{c,\tau,\omega}(G)\right]}.
 \end{aligned}}
 \label{eq:indexgroup}
\end{equation}

The numerator contains every possible zero-mode representation. The denominator contains the removable combinations and is the \emph{image} of the super representation group under $F$, rather than the super representation group itself. Distinct graded representations may have the same underlying ungraded representation. The quotient identifies zero-mode configurations that differ only by removable multiplets. In the finite semisimple setting, a representation has zero index if and only if it is itself gappable.


For actual representations, the quotient is equivalent to \emph{stable equivalence}: two zero-mode spaces are identified when they differ by removable sectors, that is, when
\begin{equation}
 V\oplus A\cong V'\oplus B
 \label{eq:stableequiv}
\end{equation}
for some gappable $A$ and $B$. 

\subsection{Index of a Hamiltonian and its stability}
For a finite-dimensional Hamiltonian obeying Eq.~\eqref{eq:symmetry}, the kernel $Z=\ker H$ carries a representation of $G$, and we define its index by
\begin{equation}
 \nu(H)=[\ker H]\quad\text{in }\Qind(G;c,\tau,\omega).
 \label{eq:kernelindex}
\end{equation}
The index depends only on the symmetry multiplicities of low-lying states and is independent of the details of a compatible Hamiltonian. States entering or leaving the kernel carry a gappable class and therefore do not change the index. A nonzero index obstructs the removal of all zero modes. 

The index $\nu$ is also additive: for two Hamiltonians with the same symmetry data and finite-dimensional zero-mode spaces,
\begin{equation}
 \nu(H_1\oplus H_2)=\nu(H_1)+\nu(H_2).
 \label{eq:additive}
\end{equation}



\subsection{Computation and Mackey construction}
\label{sec:quotientcomputation}
For finite $G$, Eq.~\eqref{eq:indexgroup} can be evaluated directly. Construct all irreps $W_i$ of $G$ and $V_\alpha$ of $G^h$, and form the restriction matrix
\begin{equation}
 \Res V_\alpha=\bigoplus_i W_i^{\oplus M_{i\alpha}},
 \qquad M:\Z^{n_h}\longrightarrow\Z^{n_G},
 \label{eq:matrixM}
\end{equation}
where $n_G$ and $n_h$ count the irreps of $G$ and $G^h$ with the specified factor systems. Then $\Qind=\Z^{n_G}/\im M$.


The following sections develop a systematic alternative using Mackey's construction of irreps of group extensions. This method constructs all super irreps for symmetry groups with or without antiunitary elements and organizes them into the two complex AZ types in the purely unitary case and the ten AZ types when antiunitary symmetries are included. We also formulate complete character criteria. 

Once the classification is established, we decompose the restriction of each super irrep into ungraded irreps, namely irreps of $G$. This decomposition depends only on the residual AZ symmetry type, or equivalently on the graded commutant type after the super-opposite conversion described in Sec.~\ref{sec:gradedcorrespondence}. Only paired, doubled, and single removable blocks occur. The complete indices are consequently integer differences and multiplicity parities; in particular, every nontrivial finite-order (torsion) factor is $\Z_2$. Thus the quotient index group has the general form
\begin{equation}
		\Qind(G;c,\tau,\omega)\cong
		 \mathbb{Z}_2^a\oplus\mathbb{Z}^b,
\end{equation}
with $a$ and $b$ nonnegative integers.

Section~\ref{sec:unitary} treats purely unitary symmetry groups, while Secs.~\ref{sec:tenfold} and \ref{sec:characters} treat groups containing antiunitary elements.



\section{Symmetry groups without antiunitary elements}
\label{sec:unitary}
Assume $\tau(g)=+1$ for every $g\in G$. Every symmetry operator is then complex-linear, although the symmetry representation may still be projective. Starting from the Hamiltonian-preserving subgroup $N$, we determine the orbits of its irreps and their little groups. We then construct the irreps of $G$ and $G^h$ by Mackey's method, compute their graded commutants, and evaluate the restriction quotients. We also formulate a complete set of character criteria to distinguish inequivalent irreps of $G^h$, or equivalently super irreps of $G$. 



\subsection{The preserving subgroup and its irreducible orbits}
\label{sec:complexorbits}
Set $N=\ker c$ and denote the nontrivial coset of $G/N$ by $\Gamma$. In the class-AIII end-mode example of Sec.~\ref{sec:motivation}, $N=\{1\}$. In general, $G$ has two cosets of $N$, and $G^h$ has four. The corresponding short exact sequences are
\begin{equation}
 \begin{gathered}
 1\longrightarrow N\longrightarrow G
   \longrightarrow\Z_2^\Gamma\longrightarrow1,\\
 1\longrightarrow N\longrightarrow G^h
   \longrightarrow\Z_2^\Gamma\times\Z_2^h\longrightarrow1.
 \end{gathered}
 \label{eq:unitaryextension}
\end{equation}
Here $\Gamma$ labels a coset, not necessarily an involution of $G$: a representative $v\in G\setminus N$ can square to a nontrivial element $v^2\in N$. Both factor systems, $\omega$ for $G$ and $\omega^c$ for $G^h$, restrict to the same $\omega|_N$ on $N$.

Let $\widehat N$ denote the equivalence classes of irreps of $N$ with this restricted multiplier. The reversing coset acts on $\widehat N$: conjugating an irrep by $v$ produces another irrep of $N$ and defines the natural action of $\Z_2^\Gamma$ on $\widehat N$:
\begin{equation}
 \Delta^\Gamma(n)=\xi(v,n)\Delta(v^{-1}nv),
  \label{eq:complexNorbit}
\end{equation}
for all $n\in N$, with
\begin{equation}
	\xi(v,n)=\frac{\omega(v^{-1},n)\omega(v^{-1}n,v)}
	{\omega(v^{-1},v)}.
\end{equation}
Here $\Delta$ represents an element of $\widehat N$, and the equivalence class of $\Delta^\Gamma$ is independent of the choice of coset representative $v$. Since $c(n)=1$ for $n\in N$, the added generator $h$ commutes with the action of $N$ and acts trivially on $\widehat N$. The groups $\Z_2^\Gamma$ and $\Z_2^\Gamma\times\Z_2^h$ therefore have the same orbits, and an orbit is either fixed or consists of two points:
\begin{equation}
 \begin{array}{ll}
 \text{(a) fixed irrep:}&\Delta^\Gamma\cong\Delta,\\
 \text{(b) two-point orbit:}&\Delta^\Gamma\not\cong\Delta.
 \end{array}
 \label{eq:complexorbitcases}
\end{equation}
Write $G_\Delta$ for the stabilizer of $\Delta$ and $B_\Delta=G_\Delta/N$ for the corresponding quotient, or little cogroup of the orbit. Adjoining $h$ enlarges them to $G_\Delta\times\Z_2^h$ and $B_\Delta\times\Z_2^h$, respectively. Distinct orbits lie in disjoint isotypic sectors---they have different irreducible content under $N$---and therefore cannot be mixed by a compatible Hamiltonian.

\subsection{Common Mackey construction}
\label{sec:unitaryconstruction}
Before treating the two orbit types, we give the construction common to the original group $G$ and its lifting extension $G^h$. Mackey's method \cite{Mackey1958} starts from the space $\widehat N$ of irreducible projective representations of $N$ with multiplier $\omega|_N$. The quotient $G/N$ acts on this space, and we choose one representative $\Delta$ from each orbit. Its stabilizer $G_\Delta$ is the little group, and $B_\Delta=G_\Delta/N$ is the little cogroup. Intertwiners realize the stabilizer action on $\Delta$ and determine the multiplier that must be removed from $\omega$ to obtain the residual factor system on $B_\Delta$. Tensoring a residual irrep with these intertwiners gives a representation of $G_\Delta$; inducing it to $G$ gives a full-group irrep. Varying the orbit and the residual irrep exhausts the irreps of $G$.

The same steps construct the irreps of $G^h$: the $N$-irrep orbits remain unchanged, while the little cogroup acquires the additional $h$ factor and its prescribed lifting multiplier. All operators in this subsection are complex-linear; Sec.~\ref{sec:construction} extends the construction to antiunitary symmetries.

\paragraph{Intertwiners and the residual multiplier.}
For $g\in G_\Delta$, let $\Delta^g$ be the conjugated projective representation defined by Eq.~\eqref{eq:complexNorbit} with $v$ replaced by $g$. Since $g$ stabilizes the equivalence class of $\Delta$, a unitary operator $L_g$ identifies the two $N$-actions:
\begin{equation}
 L_g\Delta^g(n)=\Delta(n)L_g,
 \qquad n\in N.
 \label{eq:unitaryintertwiner}
\end{equation}
We define the \emph{intertwiner} $L_g$ by the equivalent covariance relation with the original operators $\Delta(n)$:
\begin{equation}
 L_g\Delta(n)=
 \frac{\omega(g,n)}{\omega(gng^{-1},g)}
 \Delta(gng^{-1})L_g,
 \qquad n\in N.
 \label{eq:unitaryintertwinerCovariance}
\end{equation}
The phase ratio follows from the projective multiplication law and is unity for an ordinary representation. Equation~\eqref{eq:unitaryintertwinerCovariance} also serves as the definition in Sec.~\ref{sec:construction}, with $L_g$ denoting the whole semilinear operator when antiunitary symmetries are present. Here $L_g$ is complex-linear. We normalize $L_n=\Delta(n)$ for $n\in N$; irreducibility makes each unitary choice unique up to a phase. The product $L_gL_k$ and the intertwiner $L_{gk}$ implement the same action on $\Delta$, so Schur's lemma gives a scalar multiplier:
\begin{equation}
 L_gL_k=\lambda(g,k)L_{gk}.
 \label{eq:unitarylambda}
\end{equation}
With compatible choices of intertwiners on the cosets, the relative multiplier descends to the little cogroup:
\begin{equation}
 \alpha(\bar g,\bar k)=\frac{\omega(g,k)}{\lambda(g,k)},
 \qquad \bar g=gN.
 \label{eq:unitaryalpha}
\end{equation}
Thus factoring out the action on $\Delta$ leaves the multiplier $\alpha$, rather than the original $\omega$.

\paragraph{Original and enlarged representations.}
Let $D$ be an irreducible $\alpha$-representation of $B_\Delta$. On the reference component, the tensor product of $D$ with the intertwiners has multiplier $\alpha\lambda=\omega$. The stabilizer representation and its induction are
\begin{equation}
 \begin{gathered}
 \rho_{\Delta,D}(g)=D(\bar g)\otimes L_g,
 \qquad g\in G_\Delta,\\
 W_{\Delta,D}=\Ind_{G_\Delta}^{G}\rho_{\Delta,D}.
 \end{gathered}
 \label{eq:unitaryMackeyW}
\end{equation}
For $G^h$, choose the intertwiner on $\Delta$ to act trivially on the added $h$ factor. The ordering in Eq.~\eqref{eq:ordering} then fixes the residual multiplier on $B_\Delta\times\Z_2^h$:
\begin{equation}
 \alpha^c\bigl((b,s),(b',t)\bigr)
 =\alpha(b,b')c(b')^s.
 \label{eq:complexresidualfactor}
\end{equation}
Here $b,b'\in B_\Delta$ and $s,t\in\{0,1\}$. For an irreducible $\alpha^c$-representation $E$, the effective grading $h_E=E(1,1)$ obeys $h_E^2=\id$ and $E(b,0)h_E=c(b)h_EE(b,0)$. Tensoring and inducing give
\begin{equation}
 \begin{gathered}
 \widetilde\rho_{\Delta,E}(g,s)=E(\bar g,s)\otimes L_g,\\
 V_{\Delta,E}=\Ind_{G_\Delta\times\Z_2^h}^{G^h}
                    \widetilde\rho_{\Delta,E}.
 \end{gathered}
 \label{eq:unitaryMackeyV}
\end{equation}
These constructions exhaust the irreps of both groups as $\Delta$ ranges over one representative of each orbit and $D$ or $E$ ranges over the corresponding residual irreps. 
In each full-group irrep, the reference component is a multiplicity space tensored with $\Delta$, and the multiplicity space carries precisely the residual representation above.

\paragraph{Restriction and explicit induction.}
Since $h$ acts trivially on $\widehat N$, the original and enlarged constructions use the same coset representatives. Restriction is therefore determined entirely on the residual space:
\begin{equation}
 \begin{gathered}
 \Res_{B_\Delta}^{B_\Delta\times\Z_2^h}E
       \cong\bigoplus_D m_DD,\\
 \Res_G^{G^h}V_{\Delta,E}
       \cong\bigoplus_D m_DW_{\Delta,D}.
 \end{gathered}
 \label{eq:restrictioninduction}
\end{equation}
Both inductions have index $o=[G:G_\Delta]$, so the full-module dimension is the residual dimension multiplied by $o\dim_\C\Delta$. Thus the residual restriction multiplicities directly determine the full-group quotient relations.

For explicit matrices, write $L=G_\Delta$, which should not be confused with the intertwiners above. Let $A(l)$ be the reference stabilizer representation with multiplier $\omega|_L$, and choose left-coset representatives $q_i$ of $G/L$, including $q_1=1$. For each $g$ and $j$, decompose uniquely
\begin{equation}
 gq_j=q_i l,\qquad l\in L.
 \label{eq:cosetdecomposition}
\end{equation}
In the basis obtained by applying $\rho(q_i)$ to the reference basis, the only nonzero block in column $j$ is
\begin{equation}
 \rho(g)_{ij}=\frac{\omega(g,q_j)}{\omega(q_i,l)}A(l).
 \label{eq:unitaryinducedmatrix}
\end{equation}
This follows directly from $\rho(g)\rho(q_j)=\omega(g,q_j)\rho(q_i l)$ and retains the phases of a nonsplit extension. If the reference module has a compatible grading $h_L$, its induced block is
\begin{equation}
 h_i=c(q_i)h_L.
 \label{eq:unitaryinducedgrading}
\end{equation}
The same matrix prescription applies directly to $G^h$ with $L\times\Z_2^h$ and $\omega^c$. The following subsections evaluate these rules for the fixed and two-point orbits.

\subsection{Residual symmetry algebras and their lifting extensions}
\label{sec:complexfactors}
We now classify the residual factor systems for the two unitary orbit types and identify the lifting extension selected by Eq.~\eqref{eq:complexresidualfactor}.

\paragraph{Fixed $N$-irrep.}
The little cogroups for $G$ and $G^h$ are $\Z_2^\Gamma$ and $\Z_2^\Gamma\times\Z_2^h$. For a unitary cyclic group $\Z_2$, rephasing its generator removes its projective square, so its factor system is cohomologically trivial. For the enlarged little cogroup, let $\gamma$ and $h_E$ denote the effective generators, after absorbing the intertwiner multiplier. Independent rephasings normalize their squares, leaving
\begin{equation}
 \gamma^2=h_E^2=\id,\qquad
 h_E\gamma=\eta\gamma h_E,\qquad \eta=\pm1.
 \label{eq:complexfactorclasses}
\end{equation}
Indeed, commuting $h_E$ past $\gamma^2$ gives $\eta^2=1$, and rephasing either generator cannot change $\eta$. These give exactly two factor-system classes. Representatives in the ordering $\gamma^a h_E^s$ are
\begin{equation}
 \alpha_\eta\bigl((a,s),(b,t)\bigr)=\eta^{sb},
 \qquad a,b,s,t\in\{0,1\}.
 \label{eq:complexetacocycle}
\end{equation}
The commuting class $\eta=+1$ has four one-dimensional irreps. The anticommuting class $\eta=-1$ has twisted algebra $M_2(\C)$ and one two-dimensional irrep. Since $c(\Gamma)=-1$, Eq.~\eqref{eq:complexresidualfactor} selects the latter class: $h_E$ must anticommute with $\gamma$. The two cohomology classes are therefore alternatives in the classification of arbitrary factors, not an extra choice in the present lifting extension.

Before adjoining $h$, the residual symmetry algebra is generated by the odd involution $\gamma$. We write $\Cl_1^\C$ for the unital complex Clifford algebra with one odd generator squaring to $+1$; its even and odd parts are each one-dimensional over $\C$. With the commutativity grading inherited from $c$, the residual algebra is
\begin{equation}
 \mathscr A_\Delta=\C\id\oplus\C\gamma
 \cong_{\mathrm{gr}}\Cl_1^\C
 \qquad\text{(case a, AIII)}.
 \label{eq:complexresidualAIII}
\end{equation}
This graded algebra must be distinguished from the enlarged twisted algebra $M_2(\C)$.

\paragraph{Two-point orbit.}
Now $G_\Delta=N$, so the little cogroup for $G$ is trivial and that for $G^h$ is just $\Z_2^h$. Its factor system has only the trivial cohomology class: any projective square of a unitary generator can be removed by rephasing, and the prescribed lift already has $h_E^2=\id$. Its two irreps are $h_E=\pm1$. Before adjoining $h$, the residual symmetry algebra has no odd generator:
\begin{equation}
 \mathscr A_\Delta=\C
 \qquad\text{(case b, A; purely even)}.
 \label{eq:complexresidualA}
\end{equation}
Thus the two orbit types give the two complex residual symmetry types, AIII and A. Table~\ref{tab:complexsymmetry} summarizes their residual symmetry data, in parallel with Table~\ref{tab:Nsymmetry} for groups containing antiunitary elements.

\begin{table*}[tbp]
\caption{Complex residual symmetry algebras and their lifting extensions, the counterpart of Table~\ref{tab:Nsymmetry}. The AIII row comes from a fixed $N$-irrep, and the A row from a two-point $\Z_2^\Gamma$ orbit. The factor class listed is the one selected by Eq.~\eqref{eq:complexresidualfactor}: the commuting class also exists on $\Z_2^\Gamma\times\Z_2^h$ but is not the lifting extension. In both rows $h_E^2=\id$; in AIII, $\gamma^2=\id$. The algebra $\mathscr A_\Delta$ is the graded residual symmetry algebra before adjoining $h$, not the enlarged twisted algebra.}
\label{tab:complexsymmetry}
\begin{ruledtabular}
\begin{tabular}{cccccc}
AZ class & $B_\Delta$ & $B_\Delta\times\Z_2^h$ & Class of $\alpha^c$ & $\mathscr A_\Delta$ & Relation with $h_E$\\
A & $\{1\}$ & $\Z_2^h$ & trivial & $\C$ & none\\
AIII & $\Z_2^\Gamma$ & $\Z_2^\Gamma\times\Z_2^h$ & nontrivial, $\eta=-1$ & $\Cl_1^\C$ & $\{\gamma,h_E\}=0$\\
\end{tabular}
\end{ruledtabular}
\end{table*}

\subsection{Explicit original and enlarged irreps}
\label{sec:complexMackey}
We apply Sec.~\ref{sec:unitaryconstruction} to the residual modules just classified, starting from a $d$-dimensional $N$-irrep $\Delta$. The fixed orbit requires only tensoring, while the two-point orbit also requires induction. For a concrete matrix realization, it suffices to specify $\rho(n)$ for each $n\in N$, the matrix representing $v$, and, for $V$, the grading matrix $h$. The matrices of all other elements follow from
\begin{equation}
 \begin{gathered}
 \rho(nv)=\omega(n,v)^{-1}\rho(n)\rho(v),\\
 \widetilde\rho(g,s)=\rho(g)h^s.
 \end{gathered}
 \label{eq:complexallmatrices}
\end{equation}

\paragraph{Fixed $N$-irrep.}
The stabilizer is the full group, so no further induction is needed. Choose a unitary intertwiner $R$ implementing $v$ on $\Delta$. Triviality of the factor-system class on $\Z_2^\Gamma$ allows the normalization
\begin{equation}
 R^2=\omega(v,v)\Delta(v^2).
 \label{eq:complexRnormalization}
\end{equation}
The two residual irreps $\gamma=\pm1$ give two inequivalent $G$-irreps:
\begin{equation}
 \rho_{W_\pm}(n)=\Delta(n),\qquad
 \rho_{W_\pm}(v)=\pm R.
 \label{eq:complexWextensions}
\end{equation}
For $G^h$, choose the unique effective irrep $E$ of the anticommuting class as $\gamma=\sigma_3$ and $h_E=\sigma_1$. Tensoring with $\Delta$ gives
\begin{equation}
 \begin{gathered}
 V=E\otimes\Delta,\qquad
 \rho_V(n)=\begin{pmatrix}\Delta(n)&0\\0&\Delta(n)\end{pmatrix},\\
 \rho_V(v)=\begin{pmatrix}R&0\\0&-R\end{pmatrix},\qquad
 h=\begin{pmatrix}0&\id_d\\\id_d&0\end{pmatrix}.
 \end{gathered}
 \label{eq:unitarypair}
\end{equation}
The two inequivalent $W_\pm$ have dimension $d$, while the enlarged irrep $V$ has dimension $2d$.

\paragraph{Two-point orbit.}
The trivial little cogroup for $G$ yields a single irrep by projective induction:
\begin{equation}
 W=\Ind_N^G\Delta,\qquad
 \Res_N^G W\cong\Delta\oplus\Delta^\Gamma.
 \label{eq:unitaryinduction}
\end{equation}
Here $\Delta^\Gamma$ is defined in Eq.~\eqref{eq:complexNorbit}. Combining a basis of $\Delta$ with its image under $\rho(v)$ gives the explicit matrices
\begin{equation}
 \begin{gathered}
 \rho_W(n)=\begin{pmatrix}\Delta(n)&0\\0&\Delta^\Gamma(n)\end{pmatrix},\\
 \rho_W(v)=\begin{pmatrix}0&S_v\\\id_d&0\end{pmatrix},\qquad
 S_v=\omega(v,v)\Delta(v^2).
 \end{gathered}
 \label{eq:complexorbitmatrices}
\end{equation}
The lower-left identity sends the reference block to its partner, while the upper-right block applies $v$ a second time and includes its projective square. The second diagonal block is precisely the projective conjugate in Eq.~\eqref{eq:complexNorbit}. These matrices obey the full multiplication law. In general, $v^2\ne1$ and $S_v\ne\id_d$. Starting from $\Delta^\Gamma$ produces the same irrep $W$. 

For $G^h$, the two characters of $\Z_2^h$ extend $\Delta$ to the stabilizer $N\times\Z_2^h$ with $h=\pm\id_d$. Induction gives two inequivalent enlarged irreps
\begin{equation}
 V_\pm=\Ind_{N\times\Z_2^h}^{G^h}(\Delta\otimes E_\pm),
 \qquad E_\pm(h)=\pm1.
 \label{eq:complexinducedV}
\end{equation}
In the basis adapted to the two $N$ constituents, the induced grading is
\begin{equation}
 h_\pm=\pm\begin{pmatrix}\id_d&0\\0&-\id_d\end{pmatrix}.
 \label{eq:unitaryselfgap}
\end{equation}
For both $V_+$ and $V_-$, the matrices of every $g\in G$ are exactly those of $W$ in Eqs.~\eqref{eq:complexallmatrices} and \eqref{eq:complexorbitmatrices}; only $h_\pm$ differs. The sign on the second block follows because the reversing coset exchanges the blocks and anticommutes with $h$. All three modules $W,V_+,V_-$ have dimension $2d$. These two constructions exhaust the irreps of both groups, orbit by orbit.

\subsection{Graded division algebras and the complex correspondence}
\label{sec:complexcommutants}
For an enlarged irrep $(V,\rho,h)$, the degree-$p$ part of the graded commutant consists of complex-linear operators $M_p$ satisfying
\begin{align}
 M_p h&=(-1)^p hM_p,\nonumber\\
 M_p\rho(g)&=c(g)^p\rho(g)M_p,\qquad p=0,1.
 \label{eq:complexsupercommutant}
\end{align}
Even operators commute with $h$ and all symmetry operators. Odd operators anticommute with $h$ and the reversing symmetries, and commute with the preserving symmetries. Together they form the graded commutant $\mathscr D(V)=\mathscr D_0(V)\oplus\mathscr D_1(V)$. Because every symmetry is unitary, these constraints are complex-linear, so the commutant is a complex graded algebra. The kernel and image of a homogeneous endomorphism are invariant under $G$ and $h$; irreducibility therefore makes every nonzero homogeneous endomorphism invertible. The super Schur lemma consequently implies that $\mathscr D(V)$ is a graded \emph{division} algebra \cite{FreedMoore2013}. Up to graded isomorphism, there are two finite-dimensional complex graded division algebras: the purely even $\C$ and $\Cl_1^\C$ \cite{DeligneMorgan1999,FreedMoore2013}. We now use the Mackey construction to determine which algebra belongs to each residual symmetry type.

On the reference block, Schur's lemma reduces an operator that commutes with the irreducible action of $N$ to $m_p\otimes\id_\Delta$. The coset action then fixes its values on all other blocks. For representatives $q_i$ of $G/G_\Delta$,
\begin{equation}
 (M_p)_i=c(q_i)^p(m_p\otimes\id_\Delta),
 \label{eq:complexcommutanttransport}
\end{equation}
with no complex conjugation because $\tau=1$. Restriction to the reference block and extension by Eq.~\eqref{eq:complexcommutanttransport} are inverse operations that preserve multiplication and parity. Thus $\mathscr D(V)$ is isomorphic to $\mathscr D(E):=\End^{\mathrm{gr}}_{\mathscr A_\Delta}(E)$, the graded commutant of the residual symmetry algebra acting on the multiplicity space $E$, graded by $h_E$. This notation refers to the action on $E$, not to the vector space alone; multiplication is composition of endomorphisms. The explicit definition and the corresponding real isomorphism are given in Sec.~\ref{sec:realcommutants}.

\paragraph{Fixed $N$-irrep: AIII.}
Using Eq.~\eqref{eq:unitarypair}, a homogeneous endomorphism is a $2\times2$ multiplicity matrix tensored with $\id_d$. An even matrix commutes with both $\sigma_3$ and $\sigma_1$, while an odd matrix anticommutes with both. Hence
\begin{equation}
 \begin{gathered}
 \mathscr D_0(V)=\C\id_{2d},\qquad \mathscr D_1(V)=\C Q,\\
 Q=\sigma_2\otimes\id_d,\qquad Q^2=\id_{2d}.
 \end{gathered}
 \label{eq:complexpaircommutant}
\end{equation}
Because $Q^2=\id$, the assignment $\gamma\mapsto Q$ gives an explicit complex graded-algebra isomorphism from $\mathscr A_\Delta$ to $\mathscr D(V)$. Both have type $\Cl_1^\C$, or AIII.

\paragraph{Two-point orbit: A.}
The inequivalent $N$ constituents force an endomorphism commuting with $N$ to be block diagonal. Such a matrix cannot anticommute with $h_\pm$ unless it vanishes, so there is no odd part. For an even matrix, the relation with $v$ equates the two block scalars:
\begin{equation}
 \mathscr D_0(V_\pm)=\C\id_{2d},\qquad
 \mathscr D_1(V_\pm)=0.
 \label{eq:complexsinglecommutant}
\end{equation}
The map $z\mapsto z\id_{2d}$ identifies this purely even algebra with $\mathscr A_\Delta=\C$, or A.

The two calculations establish the correspondence
\begin{equation}
 \boxed{\mathscr D(V)\cong_{\mathrm{gr}}\mathscr A_\Delta
 \cong_{\mathrm{gr}}
 \begin{cases}
 \Cl_1^\C,&\Delta^\Gamma\cong\Delta,\\
 \C,&\Delta^\Gamma\not\cong\Delta.
 \end{cases}}
 \label{eq:complexAZcorrespondence}
\end{equation}
This is an isomorphism of complex graded algebras and hence also identifies their AZ labels. Both complex graded Morita classes are fixed by super opposition; the distinction needed in the real case is developed in Sec.~\ref{sec:gradedcorrespondence}. The displayed generator map is a choice; it concerns the residual algebra before adjoining $h$, not the enlarged little-cogroup algebra, and the resulting label characterizes this sector rather than the entire system. Table~\ref{tab:complexMackey} records the resulting two complex types.

\subsection{Restriction patterns and complete integer indices}
\label{sec:complexrestrictions}
We now forget $h$ and decompose each enlarged irrep into irreps of $G$. In case (a), the two $\sigma_3$ eigenspaces of Eq.~\eqref{eq:unitarypair} carry $W_+$ and $W_-$. Therefore
\begin{equation}
 \Res_G^{G^h}V\cong W_+\oplus W_-.
 \label{eq:complexpairedrestriction}
\end{equation}
Neither irrep admits a nonzero compatible Hamiltonian on its own. On $n_+W_+\oplus n_-W_-$, commutation with $N$ and reversal by $v$ give
\begin{equation}
 H=\begin{pmatrix}
 0&A\otimes\id_d\\A^\dagger\otimes\id_d&0
 \end{pmatrix},\qquad A\in M_{n_+\times n_-}(\C).
 \label{eq:unitaryblock}
\end{equation}
Invertibility requires equal multiplicities and an invertible $A$. Thus the paired restriction leaves the integer difference $n_+-n_-$.

In case (b), restricting an induced $V_\pm$ forgets the choice of the little-cogroup character $E_\pm$ and gives
\begin{equation}
 \Res_G^{G^h}V_+\cong\Res_G^{G^h}V_-\cong W.
 \label{eq:unitaryextensions}
\end{equation}
A single $W$ is therefore gappable by $h_\pm$. The two inequivalent enlarged irreps give the same restriction, not a doubled relation. This orbit has zero quotient, not a parity index.

Let $\mathscr S$ be the set of fixed $N$-irreps, and let $\mathscr O$ contain one representative from each two-point orbit. The complete restriction image is
\begin{equation}
 \begin{aligned}
 \im\Res={}&\bigoplus_{\Delta\in\mathscr S}
 \Z\bigl([W_{\Delta,+}]+[W_{\Delta,-}]\bigr)\\
 &\quad\oplus\bigoplus_{\Delta\in\mathscr O}\Z[W_\Delta].
 \end{aligned}
 \label{eq:unitaryimage}
\end{equation}
The construction exhausts all irreps, and the orbits have disjoint $N$-isotypic support. These are therefore all the independent relations, and
\begin{equation}
 \boxed{\begin{gathered}
 \Qind(G;c,1,\omega)\cong\bigoplus_{\Delta\in\mathscr S}\Z,\\
 \nu_\Delta=n_{\Delta,+}-n_{\Delta,-}.
 \end{gathered}}
 \label{eq:unitaryquotient}
\end{equation}
Only integer indices occur, including for nontrivial projective factors. Choosing the other extension as $W_{\Delta,+}$ reverses the corresponding integer. The AIII end-mode index of Sec.~\ref{sec:motivation} is the special case with a single fixed irrep. The graded commutants, restrictions, and indices are collected in Table~\ref{tab:complexMackey}, the complex counterpart of Table~\ref{tab:realMackey}.

\begin{table*}
\caption{Complex representation data, in parallel with Table~\ref{tab:realMackey}. By Eq.~\eqref{eq:complexAZcorrespondence}, the AZ label applies to both the residual symmetry algebra and the graded commutant. The even and odd pieces are complex vector spaces. In the A row, the two inequivalent enlarged irreps $V_\pm$ both restrict to the same $W$; this restriction is counted once and does not generate a parity relation. The integers $n_\pm$ count full $G$-irreps, not their complex dimensions.}
\label{tab:complexMackey}
\begin{ruledtabular}
\begin{tabular}{ccccccc}
AZ class & $\mathscr D_0$ & $\mathscr D_1$ & $\mathscr D$ & $\Res_G^{G^h}V$ & Index & Quotient\\
A & $\C$ & $0$ & $\C$ & $W$ & $0$ & $0$\\
AIII & $\C$ & $\C$ & $\Cl_1^\C$ & $W_+\oplus W_-$ & $n_+-n_-$ & $\Z$\\
\end{tabular}
\end{ruledtabular}
\end{table*}

\subsection{Character criteria and index formulas}
\label{sec:complexcharacters}
For finite $G$, projective-character orthogonality provides tests for an $N$-irrep, an enlarged irrep, and a full-group irrep. These parallel the normal-subgroup and super-irrep tests of Sec.~\ref{sec:characters} but require no antiunitary Wigner indicator.

\paragraph{Test on an $N$-irrep.}
Using Eq.~\eqref{eq:complexNorbit}, define
\begin{equation}
 X_\Gamma(\Delta)=\frac1{|N|}\sum_{n\in N}
 \chi_\Delta(n)^*\chi_{\Delta^\Gamma}(n).
 \label{eq:complexXGamma}
\end{equation}
Both characters have the same restricted multiplier, so orthogonality gives $X_\Gamma=1$ in case (a) and $0$ in case (b). Thus $X_\Gamma$ directly distinguishes the AIII and A sectors. The projective conjugation factor in Eq.~\eqref{eq:complexNorbit} is essential. Under a common rephasing of the projective representations, both characters acquire the same phase on $N$, which cancels in their inner product. Table~\ref{tab:complexindicators} summarizes this complex counterpart of the indicators in Table~\ref{tab:indicators}.

\begin{table}
\caption{Character tests on an $N$-irrep $\Delta$ and the constructed enlarged irrep $V$ for $\tau=1$, the counterpart of Table~\ref{tab:indicators}. The indicator $X_\Gamma(\Delta)$ determines the residual symmetry type. Here $o=[G:G_\Delta]$ and $m_\Delta$ is the multiplicity of each $N$-irrep in $V|_N$, so $R_N(V)=o m_\Delta^2$ is the complex dimension of the $N$-commutant, independently of $\dim_\C\Delta$. Both $V_+$ and $V_-$ in the A sector give the same row. These tests do not determine signed zero-mode multiplicities.}
\label{tab:complexindicators}
\begin{ruledtabular}
\begin{tabular}{cccccc}
Symmetry AZ & $\Z_2^\Gamma$ orbit & $X_\Gamma$ & $o$ & $m_\Delta$ & $R_N$\\
A & two-point & $0$ & $2$ & $1$ & $2$\\
AIII & fixed & $1$ & $1$ & $2$ & $4$\\
\end{tabular}
\end{ruledtabular}
\end{table}

\paragraph{Test on an enlarged irrep.}
Let $\chi_V$ be the character of $V|_N$. The constructions above give
\begin{equation}
 \chi_V=\begin{cases}
 2\chi_\Delta,&\text{case (a)},\\
 \chi_\Delta+\chi_{\Delta^\Gamma},&\text{case (b)}.
 \end{cases}
 \label{eq:complexNcharacters}
\end{equation}
Therefore
\begin{equation}
 R_N(V)=\frac1{|N|}\sum_{n\in N}|\chi_V(n)|^2
 =\begin{cases}4,&\mathrm{AIII},\\2,&\mathrm A.\end{cases}
 \label{eq:complexRN}
\end{equation}
This is the complex dimension of the $N$-commutant, not the graded commutant of the full symmetry action. The factor of two in the fixed case is the multiplicity of $\Delta$, whereas in the exchanged case there are two inequivalent constituents of multiplicity one. Writing $o$ for the orbit size and $m_\Delta$ for the multiplicity of each constituent gives $R_N=o m_\Delta^2$, as recorded alongside $X_\Gamma$ in Table~\ref{tab:complexindicators}, in parallel with Table~\ref{tab:indicators}.

\paragraph{Full-group test and signed index.}
For an irreducible $\omega$-representation $W$, its $c$-twist is
\begin{equation}
 \rho_{W^c}(g)=c(g)\rho_W(g),\qquad (W^c)^c\cong W.
 \label{eq:ctwist}
\end{equation}
The multiplier is unchanged. Character orthogonality gives
\begin{equation}
 \frac1{|G|}\sum_{g\in G}c(g)|\chi_W(g)|^2
 =\begin{cases}0,&W^c\not\cong W,\\1,&W^c\cong W.\end{cases}
 \label{eq:unitarytest}
\end{equation}
In case (a), $W_{\Delta,-}\cong W_{\Delta,+}^c$, and the test gives zero. In case (b), $W^c\cong W$, with $h_\pm$ as intertwiners, and it gives one. Equivalently, $G^h/G\cong\Z_2^h$ acts on the full-group irreps by this twist, since
\begin{equation}
 h\rho(g)h^{-1}=c(g)\rho(g).
 \label{eq:projectivehaction}
\end{equation}
A fixed $N$-irrep thus produces a two-point $c$-orbit of $G$-irreps, while a two-point $N$-orbit produces a fixed $G$-irrep. These are distinct notions of an orbit.

Finally, let $\chi_Z$ be the zero-mode character and choose $W=W_{\Delta,+}$ in a fixed sector. Using $\chi_{W^c}=c\chi_W$ to subtract the two character multiplicities gives
\begin{equation}
 \nu_\Delta=\frac1{|G|}\sum_{g\in G}
 [1-c(g)]\chi_W(g)^*\chi_Z(g).
 \label{eq:unitarycharindex}
\end{equation}
For a finite-dimensional Hamiltonian, $\chi_Z$ can be replaced by the full Hilbert-space character because every gappable restriction cancels in this difference. The orbit tests identify the structural type, while this formula determines the signed multiplicity imbalance.

\subsection{Summary}
For unitary symmetry groups, the Mackey construction exhausts the original and enlarged irreps using the orbits of $N=\ker c$. A fixed $N$-irrep gives residual type AIII and graded commutant $\Cl_1^\C$; its paired restriction leaves an integer multiplicity difference. A two-point orbit gives residual type A and graded commutant $\C$; its single restriction has trivial quotient. Thus the residual symmetry type agrees with the graded-commutant type, and the complete quotient is a direct sum of integer factors, with no torsion. The character criteria distinguish these structural types, while the character formula for $\nu_\Delta$ determines the signed index in each fixed sector. Table~\ref{tab:complexMackey} collects the commutants, restrictions, and indices.

\section{Symmetry groups with antiunitary elements and a reversing symmetry}
\label{sec:tenfold}
This section is structured in parallel to Sec.~\ref{sec:unitary}, except that the counterpart of Sec.~\ref{sec:complexcharacters} is separated as Sec.~\ref{sec:characters}. We now assume that both $\tau$ and $c$ are nontrivial, and consider the normal subgroup $N=\ker\tau\cap\ker c$, on which both gradings are trivial. As in Sec.~\ref{sec:unitary}, we build the irreps of $G$ and $G^h$ from the common normal subgroup $N$, retaining both gradings in the residual symmetry algebras \cite{Parthasarathy1969,ShawLever1974}. Throughout, $\Delta$ denotes an irreducible projective representation of $N$, $\mathcal W$ an irreducible corepresentation of $G$---written in script to distinguish it from the unitary representations $W$ in Sec.~\ref{sec:unitary}---and $V$ an irreducible corepresentation of $G^h$.

Table~\ref{tab:symbols} collects the symbols used in Sec.~\ref{sec:unitary} and in this section for the readers' reference.

\begin{table*}[tbp]
	\caption{Notation for the quotient construction and the reduction through $N$. Representations and corepresentations carry the specified projective factor systems.}
	\label{tab:symbols}
	\small
	\begin{ruledtabular}
		\begin{tabular}{lp{0.76\textwidth}}
			Symbols & Meanings \\
			\hline
			$G$ & Symmetry group \\
			$c,\,\tau$ & Commutativity and unitarity gradings, respectively \\
			$N$ & Unitary Hamiltonian-preserving subgroup, $\ker c\cap\ker\tau$ \\
			$B=G/N$ & Quotient group of symmetry gradings, $B\cong\im(c,\tau)$ \\
			$\Gamma,\,T,\,C$ & Unitary reversing, antiunitary preserving, and antiunitary reversing cosets \\
			$\Delta$ & Irreducible $N$-representation with multiplier $\omega|_N$ \\
			$G_\Delta$ & Stabilizer of the equivalence class $[\Delta]$ in $G$ \\
			$B_\Delta=G_\Delta/N$ & Little cogroup of $\Delta$ \\
			$o=[G:G_\Delta]$ & Size of the orbit of $\Delta$ \\
			$m_\Delta$ & Multiplicity of $\Delta$ in $V|_N$ \\
			$d=\dim_\C\Delta$ & Complex dimension of $\Delta$ \\
			$W,\,\mathcal W$ & Irreducible representation and corepresentation of $G$, respectively \\
			$V$ & Representation space; a $G^h$-(co)representation in the enlarged construction \\
			$E$ & Irreducible $\alpha^c$-(co)representation of $B_\Delta\times\Z_2^h$ \\
			$h,\,h_E$ & Compatible lifting involution on $V$ and residual involution on $E$, respectively \\
			$\alpha,\,\lambda$ & Residual little-cogroup multiplier and intertwiner multiplier, respectively \\
			$\mathscr A_\Delta$ & Graded residual symmetry algebra before adjoining $h_E$ \\
			$\mathscr B_V$ & Image of the original graded symmetry algebra on $V$, Eq.~\eqref{eq:representedalgebra} \\
			$\mathscr D(V)$ & Graded commutant of $V$, Eq.~\eqref{eq:supercommutant} \\
			$\Cl_r^\C$ & Complex Clifford algebra with $r$ mutually anticommuting odd generators, each squaring to $+1$; $\Cl_1^\C$ has one such generator \\
			$\Cl_{p,q}$ & Real Clifford algebra with $p+q$ mutually anticommuting odd generators: $p$ square to $+1$ and $q$ to $-1$ \\
			$\Qind$ & Representation-group quotient by the restriction image, Eq.~\eqref{eq:indexgroup} \\
			$X$ & Coset character indicators of an $N$-irrep \\
			$Y$ & Antiunitary coset character sums for an enlarged irrep \\
			$R_N$ & Squared $N$-character norm, equal to $\dim_\C\End_N(V)$ \\
		\end{tabular}
	\end{ruledtabular}
\end{table*}

\subsection{The preserving subgroup and its irreducible orbits}
\label{sec:unitarysubgroup}
The common starting subgroup consists of the elements that are unitary and preserve the Hamiltonian,
\begin{equation}
 N=\ker\tau\cap\ker c,\qquad
 B=G/N\cong\operatorname{im}(\tau,c).
 \label{eq:normal}
\end{equation}
Both gradings are trivial on $N$ and therefore descend to the quotient $B$, a subgroup of $\Z_2\times\Z_2$. Its nontrivial cosets are labeled $T$ (antiunitary, commuting), $C$ (antiunitary, anticommuting), and $\Gamma$ (unitary, anticommuting). Since both gradings are nontrivial, $B$ must contain an antiunitary coset and a reversing coset, which leaves only
\begin{equation}
 B\cong\begin{cases}
 \Z_2^C,&c=\tau,\\
 \Z_2^T\times\Z_2^C,&c\ne\tau.
 \end{cases}
 \label{eq:allowedB}
\end{equation}
In the two-element case the unique nontrivial coset has $(\tau,c)=(-1,-1)$ corresponding to an anticommuting antiunitary, so it must be called $C$ and not $T$ according to the AZ naming convention. In the four-element case $\Gamma=TC$. Note that a quotient $\Z_2^T$ would require $c=1$ and is excluded.

The two extensions to be analyzed are
\begin{equation}
 \begin{gathered}
 1\longrightarrow N\longrightarrow G\longrightarrow B\longrightarrow1,\\
 1\longrightarrow N\longrightarrow G^h
   \longrightarrow B\times\Z_2^h\longrightarrow1.
 \end{gathered}
 \label{eq:realextension}
\end{equation}
We treat both extensions simultaneously. Their factor systems are $\omega$ and $\omega^c$, which have the same restriction $\omega|_N$. Let $\widehat N$ be the set of equivalence classes of irreducible complex representations of $N$ with this restricted multiplier. An element $g\in G$ acts on such an irrep by conjugation and, when $g$ is antiunitary, complex conjugation:
\begin{align}
 \Delta^g(n)&=K^{\eps(g)}
    [\xi(g,n)\Delta(g^{-1}ng)]K^{\eps(g)},\nonumber\\
 \xi(g,n)&=\frac{\omega(g^{-1},n)\omega(g^{-1}n,g)}
                    {\omega(g^{-1},g)},\qquad n\in N.
 \label{eq:orbit}
\end{align}
On equivalence classes this action depends only on the coset in $B$. The stabilizer of $\Delta$ and its little group are
\begin{equation}
 G_\Delta=\{g\in G:\Delta^g\cong\Delta\},\qquad
 B_\Delta=G_\Delta/N,
 \label{eq:stabilizer}
\end{equation}
and one $\Delta$ from each orbit is enough. Distinct orbits occupy disjoint isotypic sectors---sets of $N$-irreps that no symmetry relates---so a compatible Hamiltonian, which commutes with $N$, cannot mix them.

The added generator $h$ acts trivially on $\widehat N$: $c(n)=1$ implies $h\rho(n)=\rho(n)h$. Consequently, $G$ and $G^h$ have exactly the same orbits, while their stabilizers and little cogroups are related by
\begin{equation}
 (G^h)_\Delta=G_\Delta\times\Z_2^h,\qquad
 (G^h)_\Delta/N=B_\Delta\times\Z_2^h.
 \label{eq:commonorbits}
\end{equation}
This separates the orbit construction from the lifting problem: adjoining $h$ changes the residual algebra, not the orbit of $\Delta$.


\subsection{Extension of the Mackey construction to antiunitary symmetries}
\label{sec:construction}
The orbit, tensor-product, induction, and restriction steps of Sec.~\ref{sec:unitaryconstruction} still apply. The new ingredient is semilinearity: an antiunitary stabilizer element acts antilinearly on both the reference irrep and the residual multiplicity space, and conjugates scalar phases when operators are composed. We therefore extend the unitary formulas below while retaining their orbit and restriction structure.

\paragraph{Semilinear intertwiners and residual factors.}
We use the same defining covariance relation, Eq.~\eqref{eq:unitaryintertwinerCovariance}, with $L_g$ now the whole operator on the space of $\Delta$. For $g\in G_\Delta$, choose $L_g=U_gK^{\eps(g)}$ with $U_g$ unitary, so $L_g$ is unitary for $\tau(g)=+1$ and antiunitary for $\tau(g)=-1$. In particular, $L_g(zv)=z^{\tau(g)}L_gv$ for $z\in\C$; this semilinearity accounts for scalar conjugation when composing operators. Normalize $L_n=\Delta(n)$ for $n\in N$. The defining relation has the same form in both cases, and Schur's lemma again gives
\begin{equation}
 L_gL_k=\lambda(g,k)L_{gk}.
 \label{eq:lambda}
\end{equation}
Associativity now makes $\lambda$ a $\tau$-twisted cocycle, just as for $\omega$ in Eq.~\eqref{eq:cocycle}. The quotient of multipliers still descends to $B_\Delta$ after compatible coset choices:
\begin{equation}
 \alpha(\bar g,\bar k)=\frac{\omega(g,k)}{\lambda(g,k)},
 \qquad\bar g=gN.
 \label{eq:alpha}
\end{equation}
Unlike a unitary square, an antiunitary square cannot be changed by rephasing its operator. Consequently, the intertwiner contribution in Eq.~\eqref{eq:alpha} must be retained when determining the residual antiunitary squares; they need not equal the squares of the bare representatives.

Let $D$ be an irreducible $\alpha$-corepresentation of $B_\Delta$, and write $D(\bar g)$ for its unitary matrix part. Tensoring with the intertwiner on $\Delta$ gives the little-group corepresentation
\begin{equation}
 \rho_{\Delta,D}(g)=[D(\bar g)\otimes U_g]K^{\eps(g)}.
 \label{eq:littlerep}
\end{equation}
There is one complex conjugation on the full tensor product, rather than two independently applied conjugations. Its multiplier is $\alpha\lambda=\omega$, and the same induction step gives
\begin{equation}
 \mathcal W_{\Delta,D}=\Ind_{G_\Delta}^{G}\rho_{\Delta,D}.
 \label{eq:inducedGcorep}
\end{equation}
For the enlarged group, the intertwiner on $\Delta$ again acts trivially on $h$. The residual lift is therefore still Eq.~\eqref{eq:complexresidualfactor}: $h$ is unitary and the extra factor $c(b')^s$ is real, so complex conjugation introduces no further sign. Let $E$ be an irreducible $\alpha^c$-corepresentation, writing $E(b,s)$ for its unitary matrix part and $E(b)=E(b,0)$. The effective grading relation now reads
\begin{equation}
 h_E^2=\id,\qquad
 E(b)K^{\eps(b)}h_E=c(b)h_EE(b)K^{\eps(b)}.
 \label{eq:effectivegrading}
\end{equation}
Tensoring with the same $L_g$ and inducing gives
\begin{align}
 \widetilde\rho_{\Delta,E}(g,s)
   &=[E(\bar g,s)\otimes U_g]K^{\eps(g)},\nonumber\\
 V_{\Delta,E}
   &=\Ind_{G_\Delta\times\Z_2^h}^{G^h}\widetilde\rho_{\Delta,E}.
 \label{eq:inducedGhcorep}
\end{align}
The exhaustiveness argument of Sec.~\ref{sec:unitaryconstruction} and the restriction identity \eqref{eq:restrictioninduction} extend to these corepresentations, with $W_{\Delta,D}$ replaced by $\mathcal W_{\Delta,D}$. In particular, the effective restriction multiplicities and the dimension factor $[G:G_\Delta]\dim_\C\Delta$ are unchanged by the presence of antiunitary elements.

\paragraph{Conjugation in the induced matrices.}
Use the coset decomposition \eqref{eq:cosetdecomposition} and the same transported basis as in Sec.~\ref{sec:unitaryconstruction}, now writing the reference stabilizer action as $A(l)K^{\eps(l)}$. The only change to Eq.~\eqref{eq:unitaryinducedmatrix} is conjugation of the reference matrix when $q_i$ is antiunitary. The nonzero block in column $j$ of the unitary matrix part is
\begin{equation}
 U(g)_{ij}=
 \frac{\omega(g,q_j)}{\omega(q_i,l)}
 K^{\eps(q_i)}A(l)K^{\eps(q_i)}.
 \label{eq:induced}
\end{equation}
The complete operator is $U(g)K^{\eps(g)}$. Setting $\eps=0$ recovers Eq.~\eqref{eq:unitaryinducedmatrix} exactly.

If the reference representation admits a compatible $h_L$, the corresponding block of the induced grading is
\begin{equation}
 h_i=c(q_i)K^{\eps(q_i)}h_LK^{\eps(q_i)}.
 \label{eq:inducedh}
\end{equation}
This extends Eq.~\eqref{eq:unitaryinducedgrading}: the commutativity grading supplies the sign $c(q_i)$, while the unitarity grading determines whether to conjugate $h_L$. Applying these formulas to the stabilizer module built from $E$ gives the full $V_{\Delta,E}$ and its grading. Section~\ref{sec:realexplicit} evaluates them for all ten residual types.

\subsection{Ten residual symmetry algebras}
\label{sec:horbitcoreps}
The classification now reduces to finite bookkeeping. The little cogroup $B_\Delta$ is a subgroup of $B\subseteq\Z_2\times\Z_2$ and therefore has one of five subgroup types. We count the possible \emph{projective symmetry algebras of the little cogroup $B_\Delta$}, using its inherited gradings and identifying factor systems up to twisted coboundaries. Adjoining $\Z_2^h$ introduces no additional factor-system choice: once $\alpha$ and $c|_{B_\Delta}$ are fixed, Eq.~\eqref{eq:complexresidualfactor} fixes $\alpha^c$, including $h_E^2=\id$ and every commutation sign involving $h_E$. We therefore do not classify arbitrary projective extensions of $B_\Delta\times\Z_2^h$. This smaller classification suffices because Eq.~\eqref{eq:restrictioninduction} identifies the effective restriction multiplicities with those entering the full-group quotient.

For the full four-element $B$, its five possible subgroups are
\begin{equation}
 \begin{gathered}
 \{1\},\quad \Z_2^\Gamma=\Z_2^{TC},\quad
 \Z_2^T,\quad \Z_2^C,\qquad \Z_2^T\times\Z_2^C.
 \end{gathered}
 \label{eq:fivecogroups}
\end{equation}
When $B=\Z_2^C$, only $\{1\}$ and $\Z_2^C$ are available. The factor-system classes for these subgroups are as follows.

\paragraph{Trivial or unitary little cogroup.}
The trivial group has one factor-system class. For the unitary group $\Z_2^\Gamma$, a normalized factor system is specified by $\alpha(\Gamma,\Gamma)=e^{i\theta}$. Rephasing the effective generator by $\Gamma\mapsto e^{-i\theta/2}\Gamma$ sets its square to $\id$. Thus every factor system on this unitary $\Z_2$ is \emph{cohomologically} trivial, although a given representative need not be identically one. These two subgroups contribute one projective symmetry type each, A and AIII.

\paragraph{One antiunitary generator.}
For $\Z_2^T$ or $\Z_2^C$, the twisted cocycle identity forces the normalized square to be real. Its value is a cohomology invariant, since rephasing an antiunitary operator $A$ gives $(e^{i\theta}A)^2=A^2$. Hence the two classes are specified by
\begin{equation}
 \kappa_T=\alpha(T,T)=\pm1
 \quad\text{or}\quad
 \kappa_C=\alpha(C,C)=\pm1.
 \label{eq:antiunitaryclasses}
\end{equation}
For a preserving generator these are AI and AII; for a reversing generator they are D and C. Each antiunitary $\Z_2$ therefore contributes two types.

\paragraph{Two antiunitary generators.}
For $\Z_2^T\times\Z_2^C$, the two independent cohomology invariants are $\kappa_T$ and $\kappa_C$. There is no further invariant in the mixed relation. Indeed, if the effective operators obey $TC=\zeta CT$, their rephasings by $e^{iu}$ and $e^{iv}$ change $\zeta$ to $e^{2i(u-v)}\zeta$ without changing either square. We may therefore choose commuting representatives. The four classes are
\begin{equation}
 (\kappa_T,\kappa_C)=(+1,+1),(+1,-1),(-1,+1),(-1,-1),
 \label{eq:fourfactorclasses}
\end{equation}
corresponding to BDI, CI, DIII, and CII, respectively. They include the trivial class $(+1,+1)$ and three nontrivial classes. The unitary product $TC$ can be phase-normalized to a chiral involution $\Gamma$; it supplies no additional factor-system invariant.

The five little cogroups therefore yield $1+1+2+2+4=10$ projective symmetry-algebra types, listed in Table~\ref{tab:Nsymmetry}. We call the type of $B_\Delta$ together with its factor class the orbit's \emph{residual type}, and the graded-division-algebra type of $\mathscr D(V)$ its \emph{commutant type}. Section~\ref{sec:gradedcorrespondence} relates their standard AZ labels by super opposition. We keep the residual symmetry label and the commutant label distinct. After identifying the residual type, we construct its original and enlarged irreps, compute their restriction, and determine the quotient index group.


For each residual type, the explicit constructions below exhaust the irreducible enlarged modules up to equivalence and the possible overall sign of $h_E$. Since the lift $\alpha^c$ is fixed by $\alpha$ and $c$, adjoining $h$ introduces no further factor-system choice.

\begin{table*}[tbp]
\caption{Residual symmetry algebras obtained from an $N$-irrep $\Delta$. The little cogroup is $B_\Delta=G_\Delta/N$, and the squares refer to effective operators after including the intertwiner multiplier in Eq.~\eqref{eq:alpha}. The unitarity and commutativity gradings are inherited from $(\tau,c)$ on $G$, not an auxiliary grading. A dash denotes an absent effective generator. When both $T$ and $C$ are present, $\Gamma$ is their normalized unitary product. The last column specifies the lifting relations in addition to $h_E^2=\id$. Only rows with $B_\Delta\subseteq B$ can occur. With both $c$ and $\tau$ nontrivial, $B$ has the two possibilities in Eq.~\eqref{eq:allowedB}; $B_\Delta=\Z_2^T$ can occur within the full quotient.}
\label{tab:Nsymmetry}
\begin{ruledtabular}
\begin{tabular}{cccccc}
Symmetry AZ & $B_\Delta$ & $T^2$ & $C^2$ & $\Gamma$ & Relations with $h_E$\\
A & $\{1\}$ & --- & --- & --- & none\\
AIII & $\Z_2^\Gamma$ & --- & --- & present & $\{\Gamma,h_E\}=0$\\
AI & $\Z_2^T$ & $+1$ & --- & --- & $[T,h_E]=0$\\
BDI & $\Z_2^T\times\Z_2^C$ & $+1$ & $+1$ & present & $[T,h_E]=0,\ \{C,h_E\}=0$\\
D & $\Z_2^C$ & --- & $+1$ & --- & $\{C,h_E\}=0$\\
DIII & $\Z_2^T\times\Z_2^C$ & $-1$ & $+1$ & present & $[T,h_E]=0,\ \{C,h_E\}=0$\\
AII & $\Z_2^T$ & $-1$ & --- & --- & $[T,h_E]=0$\\
CII & $\Z_2^T\times\Z_2^C$ & $-1$ & $-1$ & present & $[T,h_E]=0,\ \{C,h_E\}=0$\\
C & $\Z_2^C$ & --- & $-1$ & --- & $\{C,h_E\}=0$\\
CI & $\Z_2^T\times\Z_2^C$ & $+1$ & $-1$ & present & $[T,h_E]=0,\ \{C,h_E\}=0$\\
\end{tabular}
\end{ruledtabular}
\end{table*}

These effective types need not equal the AZ class of the full physical system. Antiunitary elements of $G$ may exchange inequivalent $N$-irreps and hence be absent from $B_\Delta$, leaving an effective A or AIII sector. If $B=\Z_2^C$, only D, C, and A occur, and only D has a nontrivial quotient. For the full four-element $B$, a little cogroup $B_\Delta=\Z_2^T$ remains possible, producing AI or AII even though $c$ is nontrivial on $G$. Excluding $B=\Z_2^T$ therefore does not remove the preserving-only rows from the tables. The unitary case of Sec.~\ref{sec:unitary} corresponds to $B=\Z_2^\Gamma$ or the trivial group.

\subsection{Explicit representations}
\label{sec:realexplicit}
We now give concrete matrices for the irreps $\mathcal W$ of $G$ and $V$ of $G^h$ for each residual type, so the construction can be used without the language of induced representations. Here induction places symmetry-related copies of the reference space in separate blocks; the symmetries then act within or exchange these blocks.

We use $d=\dim_\C\Delta$ and $J=-i\sigma_2$. Whenever the corresponding coset exists, let $t,u\in G$ represent the antiunitary cosets $T$ (preserving) and $C$ (reversing), and for the four-element quotient set $v=tu$, representing $\Gamma$. These representatives need not commute or square to one: their squares and products are absorbed into the intertwiners $U_l$ of Sec.~\ref{sec:construction}, whereas the canonical matrices on the residual multiplicity spaces have normalized squares.

We use $\rho_{\mathcal W}$ or $\rho_V$ for the restriction to $G$, and write $h$ separately for the enlarged irrep. Every antiunitary matrix below has one $K$ acting on the entire column vector, including all tensor factors. The lifting matrix $h$ is always complex-linear, never followed by $K$. The matrices of $N$ and of the displayed coset representatives generate the whole representation:
\begin{equation}
 \begin{gathered}
 \rho(ab)=\omega(a,b)^{-1}\rho(a)\rho(b),\\
 \widetilde\rho(g,s)=\rho_V(g)h^s.
 \end{gathered}
 \label{eq:concretereconstruction}
\end{equation}
Products in this equation are operator products, so an antiunitary factor conjugates any scalar to its right. The residual phase convention is chosen to match the canonical matrices in each case. In CI the displayed residual $T,C$ anticommute; rephasing one of them, for instance $C\mapsto iC$, gives the commuting convention of the classification above without changing their squares or type.

For the two-block cases choose an antiunitary representative $q\notin G_\Delta$, so the blocks are labeled by $1,q$ and $q^2\in N$. Two useful block matrices, fully specified by $\Delta$, are
\begin{equation}
 \begin{gathered}
 \mathsf N_q^{(m)}(n)=
 \begin{pmatrix}\id_m\otimes\Delta(n)&0\\
 0&\id_m\otimes\Delta^q(n)\end{pmatrix},\\
 S_q=\omega(q,q)\Delta(q^2),\qquad
 \mathsf S_q^{(m)}=
 \begin{pmatrix}0&\id_m\otimes S_q\\\id_{md}&0\end{pmatrix}.
 \end{gathered}
 \label{eq:twoblockbuilding}
\end{equation}
Thus $\rho(n)=\mathsf N_q^{(m)}(n)$ and $\rho(q)=\mathsf S_q^{(m)}K$. The superscript $m$ is the residual multiplicity, not an exponent. For $l\in G_\Delta$, define
\begin{equation}
 U_l^{[q]}=
 \frac{\omega(l,q)}{\omega(q,q^{-1}lq)}
 \overline{U_{q^{-1}lq}}.
 \label{eq:transportedintertwiner}
\end{equation}
The matrices $U_l$ here are defined on the whole stabilizer, as in Eq.~\eqref{eq:lambda}. In practice they can be obtained from the linear intertwining equations
\begin{equation}
 U_l\Delta(n)^{*\eps(l)}=
 \frac{\omega(l,n)}{\omega(lnl^{-1},l)}\Delta(lnl^{-1})U_l,
 \label{eq:intertwinerlinear}
\end{equation}
where $X^{*\eps}=X$ for $\eps=0$ and $X^{*\eps}=\overline X$ for $\eps=1$. After unitary normalization, their residual multiplier determines the appropriate canonical multiplicity matrices. Since $G/N$ is abelian, $q^{-1}lq$ and $l$ have the same little-cogroup label. All residual generator matrices $F_l$ used below are real. The complete matrix of a stabilizer generator is consequently
\begin{equation}
 \rho(l)=\begin{pmatrix}F_l\otimes U_l&0\\
 0&F_l\otimes U_l^{[q]}\end{pmatrix}K^{\eps(l)}.
 \label{eq:twoblockstabilizer}
\end{equation}
For $n\in N$, Eq.~\eqref{eq:transportedintertwiner} reduces to the projective-conjugate matrix $\Delta^q(n)$ of Eq.~\eqref{eq:orbit}. For a real reference lifting matrix $h_E$, the full lifting matrix is $\operatorname{diag}(h_E\otimes\id_d,c(q)h_E\otimes\id_d)$. These identities account for the block structure and the phases in every case, including the four-block case A treated below; no assumption of a split group extension is made.

We now present all ten constructions. Each case is fixed by three data: the little cogroup $B_\Delta$, which determines the number of blocks $o$; the residual multiplicity $m_\Delta$; and the lifting relations involving $h_E$, collected in Table~\ref{tab:Nsymmetry}. Each paragraph first states these data and then gives the matrices obtained from Eqs.~\eqref{eq:induced} and \eqref{eq:inducedh}. We treat DIII in greater detail because a single multiplet cannot be lifted and the required doubling must be exhibited explicitly.

\paragraph{A: four inequivalent $N$ constituents.}
No symmetry stabilizes $\Delta$, so the orbit has four points and induction from $N$ produces four blocks, one for each of $1,t,u,v$. The matrices of $t$ and $u$ follow from Eq.~\eqref{eq:induced}: each entry contains the phase of the coset decomposition and the transported reference matrix, and the lifting matrix is diagonal because $t$ preserves and $u$ reverses the Hamiltonian.

For $B_\Delta=\{1\}$ and the full four-element quotient, use the block order $(1,t,u,v)$ with $v=tu$. There is one $G$-irrep $\mathcal W$, with
\begin{equation}
 \rho_{\mathcal W}(n)=\operatorname{diag}
 \bigl(\Delta(n),\Delta^t(n),\Delta^u(n),\Delta^v(n)\bigr).
 \label{eq:concreteAn}
\end{equation}
Writing $I=\id_d$, the antiunitary generators are
\begin{equation}
 \begin{gathered}
 \rho_{\mathcal W}(t)=
 \begin{pmatrix}0&S_t&0&0\\I&0&0&0\\
 0&0&0&P_t\\0&0&\omega(t,u)I&0\end{pmatrix}K,\\
 \rho_{\mathcal W}(u)=
 \begin{pmatrix}0&0&S_u&0\\0&0&0&P_u\\
 I&0&0&0\\0&R_u&0&0\end{pmatrix}K,
 \end{gathered}
 \label{eq:concreteAexchange}
\end{equation}
where $S_t,S_u$ are defined in Eq.~\eqref{eq:twoblockbuilding} and
\begin{align}
 P_t&=\frac{\omega(t,v)}{\omega(u,u^{-1}tv)}
       \overline{\Delta(u^{-1}tv)},\nonumber\\
 P_u&=\frac{\omega(u,v)}{\omega(t,t^{-1}uv)}
       \overline{\Delta(t^{-1}uv)},\nonumber\\
 R_u&=\frac{\omega(u,t)}{\omega(v,v^{-1}ut)}
       \Delta(v^{-1}ut).
 \label{eq:concreteAphases}
\end{align}
All arguments of $\Delta$ in this formula lie in $N$. The two enlarged irreps $V_\pm$ have the same $G$ matrices and
\begin{equation}
 h_\pm=\pm\operatorname{diag}(I,I,-I,-I).
 \label{eq:concreteAh}
\end{equation}
The signs express preservation by $t$ and reversal by $u$. If instead $B=\Z_2^C$, the A orbit has only two blocks:
\begin{equation}
 \begin{gathered}
 \rho_{\mathcal W}(n)=\rho_{V_\pm}(n)=\mathsf N_u^{(1)}(n),\\
 \rho_{\mathcal W}(u)=\rho_{V_\pm}(u)=\mathsf S_u^{(1)}K,\\
 h_\pm=\pm\operatorname{diag}(I,-I).
 \end{gathered}
 \label{eq:concreteAreduced}
\end{equation} In either setting, $V_+$ and $V_-$ restrict to the same $\mathcal W$.

\paragraph{AIII: only the unitary reversing coset stabilizes $\Delta$.}
The unitary reversing coset $\Gamma$ stabilizes $\Delta$, whereas the antiunitary cosets exchange the two constituents; the orbit therefore has two points, represented by $1$ and $t$. The element $v$ acts inside each block with the intertwiner $U_v$ and its transported version $U_v^{[t]}$, and $h$ doubles the residual multiplicity; diagonalizing $\sigma_1$ in that multiplicity space splits $V$ into the two irreps $\mathcal W_\pm$.

Here $B_\Delta=\Z_2^\Gamma$. The two $G$-irreps are
\begin{equation}
 \begin{gathered}
 \rho_{\mathcal W_\pm}(n)=\mathsf N_t^{(1)}(n),\qquad
 \rho_{\mathcal W_\pm}(t)=\mathsf S_t^{(1)}K,\\
 \rho_{\mathcal W_\pm}(v)=
 \pm\begin{pmatrix}U_v&0\\0&U_v^{[t]}\end{pmatrix}.
 \end{gathered}
 \label{eq:concreteAIIIW}
\end{equation}
The enlarged irrep is
\begin{equation}
 \begin{gathered}
 \rho_V(n)=\mathsf N_t^{(2)}(n),\qquad
 \rho_V(t)=\mathsf S_t^{(2)}K,\\
 \rho_V(v)=\begin{pmatrix}\sigma_1\otimes U_v&0\\
 0&\sigma_1\otimes U_v^{[t]}\end{pmatrix},\\
 h=\operatorname{diag}(\sigma_3\otimes\id_d,\sigma_3\otimes\id_d).
 \end{gathered}
 \label{eq:concreteAIIIV}
\end{equation}
The missing generator $u=t^{-1}v$ is obtained from Eq.~\eqref{eq:concretereconstruction}. Diagonalizing $\sigma_1$ in the multiplicity space exposes the two restrictions $\mathcal W_\pm$. The different Pauli-axis choice from Sec.~\ref{sec:complexMackey} is a unitary change of basis.

\paragraph{AI: a preserving antiunitary stabilizer with residual square $+1$.}
The preserving antiunitary element $T$ stabilizes $\Delta$ while the reversing one exchanges the blocks, so the orbit has two points and the residual multiplicity stays one. Since $T$ preserves and $C$ reverses the Hamiltonian, $h_\pm$ is diagonal with opposite signs on the two blocks.

Now $B_\Delta=\Z_2^T$ and the two blocks are labeled by $1,u$. The original irrep has
\begin{equation}
 \begin{gathered}
 \rho_{\mathcal W}(n)=\mathsf N_u^{(1)}(n),\qquad
 \rho_{\mathcal W}(u)=\mathsf S_u^{(1)}K,\\
 \rho_{\mathcal W}(t)=\begin{pmatrix}U_t&0\\0&U_t^{[u]}\end{pmatrix}K.
 \end{gathered}
 \label{eq:concreteAI}
\end{equation}
Both enlarged irreps use these same $G$ matrices, with $h_\pm=\pm\operatorname{diag}(\id_d,-\id_d)$. The residual preserving symmetry fixes each block, while the physical reversing coset exchanges them.

\paragraph{BDI: both antiunitary cosets stabilize $\Delta$, with squares $(+1,+1)$.}
Both antiunitary cosets stabilize $\Delta$, so there is a single block. The squares $(+1,+1)$ give a two-dimensional effective module, in which the two original irreps differ only by the sign of the reversing generator; the enlarged irrep pairs them with $h=\sigma_3\otimes\id_d$.

There is only one orbit block. The two original irreps and the enlarged irrep are
\begin{equation}
 \begin{gathered}
 \rho_{\mathcal W_\pm}(n)=\Delta(n),\\
 \rho_{\mathcal W_\pm}(t)=U_tK,\qquad
 \rho_{\mathcal W_\pm}(u)=\pm U_uK,\\
 \rho_V(n)=\id_2\otimes\Delta(n),\qquad
 \rho_V(t)=(\id_2\otimes U_t)K,\\
 \rho_V(u)=(\sigma_1\otimes U_u)K,\qquad
 h=\sigma_3\otimes\id_d.
 \end{gathered}
 \label{eq:concreteBDI}
\end{equation}
The two $\sigma_1$ eigenspaces carry $\mathcal W_+$ and $\mathcal W_-$ when $h$ is forgotten.

\paragraph{D: a reversing antiunitary stabilizer with residual square $+1$.}
The reversing generator stabilizes $\Delta$ and its square $+1$ leaves a one-dimensional effective module. A Hermitian matrix that reverses sign under $C$ must vanish, so the lifting needs a second copy; with residual multiplicity two, $h=\operatorname{diag}(\sigma_3\otimes\id_d,\sigma_3\otimes\id_d)$ acts inside each block.

For the full quotient, $B_\Delta=\Z_2^C$ and the blocks are labeled by $1,t$. The original irrep is
\begin{equation}
 \begin{gathered}
 \rho_{\mathcal W}(n)=\mathsf N_t^{(1)}(n),\qquad
 \rho_{\mathcal W}(t)=\mathsf S_t^{(1)}K,\\
 \rho_{\mathcal W}(u)=\begin{pmatrix}U_u&0\\0&U_u^{[t]}\end{pmatrix}K.
 \end{gathered}
 \label{eq:concreteDW}
\end{equation}
A lifting matrix needs a doubled reference space:
\begin{equation}
 \begin{gathered}
 \rho_V(n)=\mathsf N_t^{(2)}(n),\qquad
 \rho_V(t)=\mathsf S_t^{(2)}K,\\
 \rho_V(u)=\begin{pmatrix}\sigma_1\otimes U_u&0\\
 0&\sigma_1\otimes U_u^{[t]}\end{pmatrix}K,\\
 h=\operatorname{diag}(\sigma_3\otimes\id_d,\sigma_3\otimes\id_d).
 \end{gathered}
 \label{eq:concreteDV}
\end{equation}
Its restriction is two copies of $\mathcal W$, rather than two inequivalent irreps: the two one-dimensional antiunitary actions $K$ and $-K$ are equivalent by rephasing. For $B=\Z_2^C$, there is no $t$ block:
\begin{equation}
 \begin{gathered}
 \rho_{\mathcal W}(n)=\Delta(n),\qquad \rho_{\mathcal W}(u)=U_uK,\\
 \rho_V(n)=\id_2\otimes\Delta(n),\\
 \rho_V(u)=(\sigma_1\otimes U_u)K,\qquad h=\sigma_3\otimes\id_d.
 \end{gathered}
 \label{eq:concreteDreduced}
\end{equation}

\paragraph{DIII: a full stabilizer with squares $(-1,+1)$.}
Here $B_\Delta=B$ and the orbit is a single point, so there is one block and the effective module is the whole residual space. Take $T_E=JK$ and $C_E=K$ on $\C^2$. On a single effective multiplet, the condition $C_Eh_E=-h_EC_E$ with $C_E=K$ makes a Hermitian $h_E$ purely imaginary, hence $h_E=a\sigma_2$ with $a\in\mathbb R$. Since $T_E\sigma_2T_E^{-1}=-\sigma_2$ for $T_E=JK$, preservation by $T_E$ forces $a=0$. Thus one multiplet admits no lifting involution.

On two copies the grading exists, but its form depends on the basis in which the copies are written. In the basis of Eq.~\eqref{eq:concreteDIIIV}, where the reversing generator acts as $C_E=(\sigma_1\otimes\id_2)K$ rather than as $K$ on each copy separately, $h_E=\sigma_3\otimes\id_2$ commutes with $T_E=(\id_2\otimes J)K$ and anticommutes with $C_E$. In the direct-sum basis of the two copies, where $C_E=\id_4K$, the same grading reads $h_E=\sigma_2\otimes\sigma_1$, as in Eq.~\eqref{eq:doubledDIII}. Tensoring with $\Delta$ and applying Eq.~\eqref{eq:inducedh} with a single coset then gives Eq.~\eqref{eq:concreteDIIIV} below.

On the single orbit block, the original irrep is
\begin{equation}
 \begin{gathered}
 \rho_{\mathcal W}(n)=\id_2\otimes\Delta(n),\\
 \rho_{\mathcal W}(t)=(J\otimes U_t)K,\qquad
 \rho_{\mathcal W}(u)=(\id_2\otimes U_u)K.
 \end{gathered}
 \label{eq:concreteDIIIW}
\end{equation}
The enlarged irrep requires an additional two-dimensional factor:
\begin{equation}
 \begin{gathered}
 \rho_V(n)=\id_2\otimes\id_2\otimes\Delta(n),\\
 \rho_V(t)=(\id_2\otimes J\otimes U_t)K,\\
 \rho_V(u)=(\sigma_1\otimes\id_2\otimes U_u)K,\\
 h=\sigma_3\otimes\id_2\otimes\id_d.
 \end{gathered}
 \label{eq:concreteDIIIV}
\end{equation}
Forgetting $h$ gives two equivalent copies of $\mathcal W$; in particular, the two residual modules with $(T,C)=(JK,K)$ and $(JK,-K)$ are equivalent. One explicit intertwiner between them is $i\sigma_1$.

\paragraph{AII: a preserving antiunitary stabilizer with residual square $-1$.}
The preserving generator stabilizes $\Delta$ with square $-1$, which forces a two-dimensional effective module and brings in the quaternionic factor $J$; the reversing coset exchanges the two blocks, and $h_\pm$ is again diagonal with opposite signs.

The two blocks are again labeled by $1,u$, but the residual multiplicity is two:
\begin{equation}
 \begin{gathered}
 \rho_{\mathcal W}(n)=\mathsf N_u^{(2)}(n),\qquad
 \rho_{\mathcal W}(u)=\mathsf S_u^{(2)}K,\\
 \rho_{\mathcal W}(t)=\begin{pmatrix}J\otimes U_t&0\\
 0&J\otimes U_t^{[u]}\end{pmatrix}K.
 \end{gathered}
 \label{eq:concreteAII}
\end{equation}
Both $V_\pm$ have these $G$ matrices and $h_\pm=\pm\operatorname{diag}(\id_{2d},-\id_{2d})$. Each underlying irrep is already gappable.

\paragraph{CII: a full stabilizer with squares $(-1,-1)$.}
Both cosets stabilize $\Delta$, with squares $(-1,-1)$; the two original irreps differ by the sign of the reversing generator, and the enlarged irrep pairs them on a four-dimensional effective module.

The two original irreps differ by the sign of the reversing generator:
\begin{equation}
 \begin{gathered}
 \rho_{\mathcal W_\pm}(n)=\id_2\otimes\Delta(n),\\
 \rho_{\mathcal W_\pm}(t)=(J\otimes U_t)K,\\
 \rho_{\mathcal W_\pm}(u)=\pm(J\otimes U_u)K.
 \end{gathered}
 \label{eq:concreteCIIW}
\end{equation}
They are paired in the enlarged irrep
\begin{equation}
 \begin{gathered}
 \rho_V(n)=\id_2\otimes\id_2\otimes\Delta(n),\\
 \rho_V(t)=(\id_2\otimes J\otimes U_t)K,\\
 \rho_V(u)=(\sigma_1\otimes J\otimes U_u)K,\\
 h=\sigma_3\otimes\id_2\otimes\id_d.
 \end{gathered}
 \label{eq:concreteCIIV}
\end{equation}
Diagonalizing the first $\sigma_1$ identifies the two inequivalent original irreps.

\paragraph{C: a reversing antiunitary stabilizer with residual square $-1$.}
The reversing generator stabilizes $\Delta$ with square $-1$, so a single two-dimensional effective multiplet already admits $h=\sigma_3$: the matrix is real, and $J\sigma_3=-\sigma_3J$ supplies the reversal.

For the full quotient use blocks $1,t$. No extra multiplicity is needed to lift the original irrep:
\begin{equation}
 \begin{gathered}
 \rho_{\mathcal W}(n)=\rho_V(n)=\mathsf N_t^{(2)}(n),\\
 \rho_{\mathcal W}(t)=\rho_V(t)=\mathsf S_t^{(2)}K,\\
 \rho_{\mathcal W}(u)=\rho_V(u)=
 \begin{pmatrix}J\otimes U_u&0\\0&J\otimes U_u^{[t]}\end{pmatrix}K,\\
 h=\operatorname{diag}(\sigma_3\otimes\id_d,\sigma_3\otimes\id_d).
 \end{gathered}
 \label{eq:concreteC}
\end{equation}
For $B=\Z_2^C$ there is just one block:
\begin{equation}
 \begin{gathered}
 \rho_{\mathcal W}(n)=\rho_V(n)=\id_2\otimes\Delta(n),\\
 \rho_{\mathcal W}(u)=\rho_V(u)=(J\otimes U_u)K,\qquad
 h=\sigma_3\otimes\id_d.
 \end{gathered}
 \label{eq:concreteCreduced}
\end{equation} Thus a single full irrep is gappable in either setting.

\paragraph{CI: a full stabilizer with squares $(+1,-1)$.}
Both cosets stabilize $\Delta$, with squares $(+1,-1)$. One two-dimensional effective multiplet already admits $h=\sigma_1$, which is real and therefore preserved by $T$, while $J\sigma_1=-\sigma_1J$ gives reversal by $C$.

One irreducible corepresentation already supports the lifting matrix:
\begin{equation}
 \begin{gathered}
 \rho_{\mathcal W}(n)=\rho_V(n)=\id_2\otimes\Delta(n),\\
 \rho_{\mathcal W}(t)=\rho_V(t)=(\sigma_1\otimes U_t)K,\\
 \rho_{\mathcal W}(u)=\rho_V(u)=(J\otimes U_u)K,\qquad
 h=\sigma_1\otimes\id_d.
 \end{gathered}
 \label{eq:concreteCI}
\end{equation}
Here the real lifting matrix commutes with the preserving generator and anticommutes with the reversing one.

All ten constructions can be verified by ordinary block multiplication, with complex conjugation applied to antiunitary factors. The block matrices in Eqs.~\eqref{eq:twoblockbuilding} and \eqref{eq:twoblockstabilizer} retain the projective phases and nonsplit squares that would be lost if every coset exchange were replaced by a bare Pauli matrix; in the four-block A case, the phases $P_t$, $P_u$, and $R_u$ appear explicitly. These matrices display the paired, doubled, and single restrictions. Section~\ref{sec:realcommutants} computes their graded commutants, Sec.~\ref{sec:gradedcorrespondence} identifies the associated AZ labels through graded Morita equivalence, and Sec.~\ref{sec:realrestrictions} derives the complete quotient indices.

\subsection{Super division algebras of the constructed super irreps}
\label{sec:realcommutants}
For each of the ten residual symmetry classes of an $N$-irrep $\Delta$, we determine the graded commutant of the constructed super irrep $V=V_{\Delta,E}$. We first establish the super division property, then reduce the calculation to the residual multiplicity module $E$, and finally compute its even and odd parts and their multiplication. Throughout this subsection, class names refer to the residual symmetry class of $\Delta$. The standard AZ label of the resulting super division algebra is a separate question, addressed in Sec.~\ref{sec:gradedcorrespondence}.

The graded commutant $\mathscr D(V)=\mathscr D_0(V)\oplus\mathscr D_1(V)$ consists of complex-linear homogeneous endomorphisms satisfying
\begin{align}
 M_p h&=(-1)^p hM_p,\nonumber\\
 M_p\rho(g)&=c(g)^p\rho(g)M_p,\qquad p=0,1.
 \label{eq:supercommutant}
\end{align}
It is a real graded algebra because multiplication by $i$ need not preserve the antiunitary constraint. The kernel and image of a homogeneous $M_p$ are invariant under both $G$ and $h$. Irreducibility of $V$ therefore makes every nonzero homogeneous element invertible, establishing the graded division property. This is the kernel-and-image argument of the super Schur lemma \cite{FreedMoore2013}. It remains to determine the resulting algebra in each Mackey sector.

\paragraph{Reduction to the effective module.}
Here $E$ is the residual multiplicity space in the reference $N$-isotypic component $E\otimes\Delta$. It carries the residual symmetry action $\rho_E$ of $\mathscr A_\Delta$ and the grading involution $h_E$. We use the abbreviation
\begin{equation}
 \mathscr D(E):=\End^{\mathrm{gr}}_{\mathscr A_\Delta}(E)
 =\mathscr D_0(E)\oplus\mathscr D_1(E)
 \label{eq:effectivecommutantdefinition}
\end{equation}
for the graded commutant of this action. Explicitly, $m\in\mathscr D_p(E)$ is a real-linear endomorphism satisfying
\begin{align}
 mh_E&=(-1)^p h_Em,\nonumber\\
 m\rho_E(a)&=(-1)^{p|a|}\rho_E(a)m
 \quad(a\in\mathscr A_\Delta\text{ homogeneous}).
 \label{eq:effectivecommutantabstract}
\end{align}
Since $\mathscr A_\Delta$ includes the even complex structure $i\id_E$, every such $m$ is automatically complex-linear. Multiplication in $\mathscr D(E)$ is composition. Thus $\mathscr D(E)$ depends on the residual action and the grading, not merely on the vector space $E$. The residual module is graded irreducible, so the super Schur lemma makes its graded commutant a real super division algebra. The involution $h_E$ specifies parity; it is not an additional generator of $\mathscr A_\Delta$.

Because $N$ is unitary and preserving, every $M_p$ commutes with $N$ and preserves its isotypic subspaces. On the reference block $E\otimes\Delta$, Schur's lemma gives
\begin{equation}
 M_p|_{E\otimes\Delta}=m_p\otimes\id_\Delta.
 \label{eq:commutantreference}
\end{equation}
Inserting the little-group action of Eq.~\eqref{eq:inducedGhcorep} into Eq.~\eqref{eq:supercommutant} cancels the intertwiner $U_g$ and leaves exactly
\begin{align}
 m_p h_E&=(-1)^p h_E m_p,\nonumber\\
 m_p E(b)&=c(b)^p E(b)K^{\eps(b)}m_pK^{\eps(b)},
 \label{eq:effectivecommutant}
\end{align}
for $b\in B_\Delta$. These are precisely the defining conditions of Eq.~\eqref{eq:effectivecommutantabstract} written for the residual generators, where $E(b)$ is their matrix part and $E(b)K^{\eps(b)}$ is the full semilinear operator. Thus $m_p\in\mathscr D_p(E)$.

Conversely, any such $m_p$ extends uniquely to the other $N$-isotypic blocks. In the coset basis of Sec.~\ref{sec:construction}, its block on $q_i(E\otimes\Delta)$ is
\begin{equation}
 (M_p)_i=c(q_i)^p K^{\eps(q_i)}
 (m_p\otimes\id_\Delta)K^{\eps(q_i)}.
 \label{eq:inducedcommutant}
\end{equation}
The sign follows from $M_p\rho(q_i)=c(q_i)^p\rho(q_i)M_p$. The effective relations ensure consistency with the little group, while the scalar projective factors cancel under conjugation. Together with Eq.~\eqref{eq:inducedh}, this establishes all required relations on the induced module. Restriction to the reference block and extension by Eq.~\eqref{eq:inducedcommutant} are inverse maps. They preserve multiplication and parity because the coset signs satisfy $c(q_i)^pc(q_i)^q=c(q_i)^{p+q}$. Hence
\begin{equation}
 \mathscr D(V_{\Delta,E})\cong_{\mathrm{gr}}\mathscr D(E)
 \label{eq:commutantinduction}
\end{equation}
as real graded algebras. In particular, neither $\dim_\C\Delta$ nor the orbit size changes the graded division-algebra type. When the effective algebra is complex, its complex structure is transported by Eq.~\eqref{eq:inducedcommutant}; it need not be scalar multiplication by $i$ on the full induced module.

\paragraph{Effective graded commutants.}
We identify $\mathscr D(E)$ directly from the residual symmetry data. In the following calculations, AZ names denote the residual symmetry types of Table~\ref{tab:Nsymmetry}; Sec.~\ref{sec:gradedcorrespondence} assigns standard AZ labels to the resulting commutant algebras. Write $F=\mathscr D_0(E)$. For each type with a nonzero odd part, choose an invertible odd element $j$ and write
\begin{equation}
 \begin{gathered}
 \mathscr D(E)=F\oplus Fj,\qquad j^2=s,\\
 jx=\varphi(x)j\quad(x\in F).
 \end{gathered}
 \label{eq:realgradedpresentation}
\end{equation}
We determine $F$, $\varphi$, and $s$ directly from Eq.~\eqref{eq:effectivecommutant}. Below $J=-i\sigma_2$ and $K$ conjugates the entire effective space.

We denote by $\Cl_{p,q}$ the unital real Clifford algebra generated by $p+q$ mutually anticommuting odd elements, with $p$ generators squaring to $+1$ and $q$ squaring to $-1$. We retain this grading when identifying the commutant algebras below.

For A, AI, and AII, an irreducible enlarged module has $h_E=\pm\id$, so its odd commutant vanishes. The ordinary commutants of the effective modules with no generator, $T=K$, and $T=JK$ are $\C$, $\R$, and $\HH$, respectively. For AIII choose $h_E=\sigma_3$ and $\Gamma=\sigma_1$. A matrix commuting with both is scalar, while a matrix anticommuting with both is proportional to $\sigma_2$. Thus $F=\C$ and $j=\sigma_2$ gives $\varphi=\operatorname{id}$ and $s=+1$, namely $\Cl_1^\C$.

For D and C choose $h_E=\sigma_3$, with $C=\sigma_1K$ and $C=JK$, respectively. The even constraint in either case gives
\begin{equation}
 F=\{\operatorname{diag}(z,\bar z):z\in\C\}.
 \label{eq:DCeven}
\end{equation}
The odd constraint gives $Fj$ with $j=i\sigma_1$ in D and $j=\sigma_1$ in C. Both exchange the two diagonal entries, so $\varphi$ is complex conjugation, while $s=-1$ in D and $s=+1$ in C. If $I_F=\operatorname{diag}(i,-i)$, the two real odd generators $j$ and $I_Fj$ anticommute and both square to $s$. Their algebras are therefore $\Cl_{0,2}$ and $\Cl_{2,0}$.

For BDI take $(h_E,T,C)=(\sigma_3,K,\sigma_1K)$. Its even commutant is $\R\id$ and its odd commutant is $\R i\sigma_2$, giving $s=-1$ and $\Cl_{0,1}$. For CI take $(h_E,T,C)=(\sigma_1,\sigma_1K,JK)$. The corresponding spaces are $\R\id$ and $\R\sigma_2$, giving $s=+1$ and $\Cl_{1,0}$. In both cases $\varphi=\operatorname{id}$.

For the two remaining types, use four-dimensional effective modules:
\begin{equation}
 \begin{gathered}
 h_E=\sigma_3\otimes\id_2,\qquad
 T=(\id_2\otimes J)K,\\
 C=\begin{cases}
 (\sigma_1\otimes\id_2)K,&\mathrm{DIII},\\
 (\sigma_1\otimes J)K,&\mathrm{CII}.
 \end{cases}
 \end{gathered}
 \label{eq:quaternionmodels}
\end{equation}
Represent the quaternions by
\begin{equation}
 \mathbb H_J=\{A\in M_2(\C):AJ=J\bar A\}\cong\HH.
 \label{eq:quaternionmatrices}
\end{equation}
Commutation with $h_E$ makes an even endomorphism block diagonal. The $T$ constraint puts each block in $\mathbb H_J$, and the $C$ constraint relates the two blocks. For an odd endomorphism the same calculation starts with off-diagonal blocks. The solutions are
\begin{align}
 F_{\mathrm{DIII}}&=\{\operatorname{diag}(A,\bar A):A\in\mathbb H_J\},\nonumber\\
 j_{\mathrm{DIII}}&=(i\sigma_2)\otimes J,\qquad j_{\mathrm{DIII}}^2=+\id,\nonumber\\
 F_{\mathrm{CII}}&=\{\operatorname{diag}(A,A):A\in\mathbb H_J\},\nonumber\\
 j_{\mathrm{CII}}&=(i\sigma_2)\otimes\id_2,\qquad j_{\mathrm{CII}}^2=-\id.
 \label{eq:quaternioncommutants}
\end{align}
In each case the full odd space is $Fj$, and $j$ commutes with $F$. Thus $\varphi=\operatorname{id}$, with $s=+1$ for DIII and $s=-1$ for CII. Quaternion units $I_1,I_2,I_3\in F$ now yield three anticommuting odd generators:
\begin{align}
 (I_\mu j)^2&=-s,\nonumber\\
 (I_\mu j)(I_\nu j)&=-(I_\nu j)(I_\mu j)\quad(\mu\ne\nu).
 \label{eq:realquaternionodd}
\end{align}
Together with their products these span $F\oplus Fj$, giving $\Cl_{0,3}$ for DIII and $\Cl_{3,0}$ for CII. The displayed modules exhaust the effective enlarged irreps up to equivalence and the possible choices $h_E\mapsto-h_E$. Those choices do not change Eq.~\eqref{eq:effectivecommutant} or its algebra.

Table~\ref{tab:realMackey} summarizes these results for all ten residual symmetry classes in its $\mathscr D_0$, $\mathscr D_1$, and $\mathscr D$ columns. By Eq.~\eqref{eq:commutantinduction}, these are also the even and odd parts and the graded division-algebra types of the full constructed super irreps. The separate commutant AZ labels in that table are explained in Sec.~\ref{sec:gradedcorrespondence}.

\subsection{Super Morita classes and the AZ correspondence}
\label{sec:gradedcorrespondence}
\begin{table}[tbp]
\caption{AZ labels of the ten graded Morita classes and their unique real super division-algebra representatives. The algebras $\R$, $\C$, and $\HH$ are purely even; the Clifford generators are odd. Complex algebras are also regarded as real algebras. This labels each algebra's own Morita class, independently of its realization as a commutant.}
\label{tab:AZMorita}
\begin{ruledtabular}
\begin{tabular}{ccc}
AZ label & Division representative & Graded center\\
A & $\C$ & $\C$\\
AIII & $\Cl_1^\C$ & $\C$\\
AI & $\R$ & $\R$\\
BDI & $\Cl_{1,0}$ & $\R$\\
D & $\Cl_{2,0}$ & $\R$\\
DIII & $\Cl_{3,0}$ & $\R$\\
AII & $\HH$ & $\R$\\
CII & $\Cl_{0,3}$ & $\R$\\
C & $\Cl_{0,2}$ & $\R$\\
CI & $\Cl_{0,1}$ & $\R$\\
\end{tabular}
\end{ruledtabular}
\end{table}
The AZ correspondence follows from a general algebra--commutant relation: a graded-simple algebra and its graded commutant on an irreducible graded module have super-opposite Morita classes. After recalling the classification that assigns AZ labels to these classes, we recall this relation and apply it to the symmetry algebra represented on the Mackey super irrep. The residual algebra of $\Delta$ has the same Morita type as this represented algebra, so its AZ label determines the opposite label of the super commutant.

\paragraph{Super Morita equivalence and division representatives.}
All algebras in this discussion are finite-dimensional, associative, unital real $\Z_2$-graded algebras. A superalgebra $\mathscr A=\mathscr A_0\oplus\mathscr A_1$ is \emph{graded simple} if it has no nonzero proper graded two-sided ideal. It is a \emph{super division algebra} if every nonzero homogeneous element is invertible; invertibility of arbitrary inhomogeneous elements is not required. Graded simplicity is likewise weaker than simplicity after forgetting the grading.

Two superalgebras are \emph{super Morita equivalent}, denoted $\mathscr A\sim_{\mathrm{grMorita}}\mathscr B$, when their graded module categories are equivalent compatibly with parity and graded intertwiners. This identifies the representation structure while allowing matrix multiplicities to change. For example, tensoring with the graded matrix algebra $\End_{\R}(U)$ of a nonzero finite-dimensional graded vector space $U$ leaves the class unchanged. The tensor product here is graded: interchanging homogeneous factors of parities $p$ and $q$ contributes $(-1)^{pq}$. The graded analogue of the division-algebra reduction for simple algebras gives exactly one super division-algebra representative, up to graded isomorphism, in each graded Morita class of graded-simple algebras \cite{Wall1964,DeligneMorgan1999,DeligneSpinors1999}.

There are exactly ten such real super division algebras, hence ten graded Morita classes of graded-simple real algebras. Eight have graded center $\R$ and two have graded center $\C$. The eight real-centered classes are represented by real Clifford algebras, with class determined by $p-q$ modulo eight for $\Cl_{p,q}$. The two complex-centered classes are represented by complex Clifford algebras of even and odd rank, regarded as real superalgebras. Thus real Clifford algebras alone account for eight classes; including the two complex-centered classes gives ten \cite{Wall1964,DeligneSpinors1999,FreedMoore2013}.

\paragraph{$CT$ algebras and the AZ labeling.}
The ten AZ symmetry classes supply another complete set of representatives of these same ten graded Morita classes. Their $CT$ algebras are generated by the symmetries present, together with the even complex structure $I=i\id$, using the grading $c$ and the specified antiunitary squares. Class A has only the complex scalars, and AIII has an independent odd unitary generator. Proposition B.4 and Appendix C of Freed and Moore identify their graded Morita classes \cite{FreedMoore2013}. Consequently, each super division algebra receives a unique AZ label: the label of the $CT$ algebra in its graded Morita class. This assignment is intrinsic to the algebra and does not depend on the representation in whose commutant it occurs. Table~\ref{tab:AZMorita} gives the dictionary used here.

\paragraph{The algebra--commutant relation.}
Write $\operatorname{AZ}(\mathscr A)$ for the label of the graded Morita class of $\mathscr A$, using Table~\ref{tab:AZMorita}. For a superalgebra $\mathscr D$, its super opposite $\mathscr D^{\mathrm{sop}}$ has the same graded vector space and multiplication
\begin{equation}
 d\star e=(-1)^{|d||e|}ed
 \qquad(d,e\text{ homogeneous}).
 \label{eq:superoppositeproduct}
\end{equation}
In particular, super opposition reverses the square of an odd element.

The following is a standard consequence of the graded Schur and double-centralizer theorems, recalled here to fix the super-opposite convention. A direct reference is Deligne's \emph{Notes on spinors}, Sec.~3.3, Eq.~(3.3.3), and Sec.~3.4, p.~114 \cite{DeligneSpinors1999}. For a simple graded module with commutant $\mathscr D$, Sec.~3.4 gives the graded matrix decomposition
\begin{equation}
 \mathscr A\cong_{\mathrm{gr}}
 M_{r|s}(k)\,\widehat\otimes_k\,\mathscr D^{\mathrm{sop}}.
 \label{eq:Delignecommutant}
\end{equation}
Here $k$ is the graded center, $k=\R$ or $\C$, and $M_{r|s}(k)=\End_k(k^{r|s})$ is a graded matrix algebra. Applying Deligne's statement over this center covers both kinds of real superalgebra in Table~\ref{tab:AZMorita}. The matrix factor is Morita trivial, which gives the relation below.

Specifically, let $\mathscr A$ be a finite-dimensional graded-simple real algebra and $W$ a nonzero finite-dimensional irreducible graded left $\mathscr A$-module. Its graded commutant $\mathscr D=\End^{\mathrm{gr}}_{\mathscr A}(W)$ is a real super division algebra, and
\begin{equation}
 \mathscr A\sim_{\mathrm{grMorita}}\mathscr D^{\mathrm{sop}},
 \qquad
 \mathscr D\sim_{\mathrm{grMorita}}\mathscr A^{\mathrm{sop}}.
 \label{eq:generalcommutantMorita}
\end{equation}
Thus their graded Morita classes are super opposite.

For completeness, the super Schur lemma makes $\mathscr D$ a super division algebra. For homogeneous $w\in W$ and $d\in\mathscr D$, the right action $w\cdot d=(-1)^{|w||d|}d(w)$ makes $W$ an $\mathscr A$--$\mathscr D^{\mathrm{sop}}$ bimodule. The graded double-centralizer theorem yields Eq.~\eqref{eq:Delignecommutant}, so removing the graded matrix factor gives $\mathscr A\sim_{\mathrm{grMorita}}\mathscr D^{\mathrm{sop}}$. Taking super opposites gives the second relation. This is the standard argument of Ref.~\cite{DeligneSpinors1999}, Sec.~3.4.

Let $\iota$ be the involution on AZ labels induced by super opposition. Equation~\eqref{eq:generalcommutantMorita} implies
\begin{equation}
 \operatorname{AZ}(\mathscr D)=\iota\!\left(\operatorname{AZ}(\mathscr A)\right).
 \label{eq:generalcommutantAZ}
\end{equation}
The relation $\Cl_{p,q}^{\mathrm{sop}}\cong_{\mathrm{gr}}\Cl_{q,p}$ and Table~\ref{tab:AZMorita} give
\begin{equation}
 \begin{gathered}
 \mathrm{BDI}\longleftrightarrow\mathrm{CI},\qquad
 \mathrm D\longleftrightarrow\mathrm C,\qquad
 \mathrm{DIII}\longleftrightarrow\mathrm{CII},\\
 \iota(X)=X\quad(X=\mathrm A,\mathrm{AIII},\mathrm{AI},\mathrm{AII}).
 \end{gathered}
 \label{eq:AZopposite}
\end{equation}
Here ``opposite AZ labels'' means related by $\iota$; the four fixed labels need not change. This explains the coincident complex labels in Sec.~\ref{sec:complexcommutants}.

\paragraph{Application to the Mackey super irreps.}
For the constructed irreducible graded module $V=V_{\Delta,E}$, define the represented symmetry algebra
\begin{equation}
 \mathscr B_V=
 \operatorname{Alg}_{\R}\langle\id,I,\rho_V(g):g\in G\rangle
 \subseteq\End_{\R}(V),\qquad I=i\id.
 \label{eq:representedalgebra}
\end{equation}
This is the image of the original symmetry algebra on the chosen sector. It is graded by $c$, with $I$ even, and is graded simple because $V$ is graded irreducible. The involution $h$ supplies the module grading and is not adjoined as an additional generator. Its graded commutant is exactly $\mathscr D(V)$: inclusion of $I$ ensures that all its elements are complex-linear. The relation in Eq.~\eqref{eq:generalcommutantMorita} therefore applies to $\mathscr B_V$ and $\mathscr D(V)$.

The residual symmetry algebra $\mathscr A_\Delta$ includes complex scalars and the effective symmetry generators, with parity inherited from $c$, before adjoining $h_E$. It is a graded-simple real algebra acting irreducibly on the graded multiplicity module $E$. Applying the same relation to this action and using the induction isomorphism of Eq.~\eqref{eq:commutantinduction} gives
\begin{equation}
 \boxed{\begin{gathered}
 \mathscr A_\Delta\sim_{\mathrm{grMorita}}
 \mathscr D(E)^{\mathrm{sop}}
 \cong_{\mathrm{gr}}\mathscr D(V)^{\mathrm{sop}}
 \sim_{\mathrm{grMorita}}\mathscr B_V,\\
 \operatorname{AZ}(\mathscr B_V)
 =\operatorname{AZ}(\mathscr A_\Delta)
 =\operatorname{AZ}_{\mathrm{sym}}(\Delta).
 \end{gathered}}
 \label{eq:representedMorita}
\end{equation}
Thus tensoring with $\Delta$ and induction preserve the symmetry algebra's graded Morita type. The residual AZ label is the label of the full represented symmetry algebra of that sector.

The AZ correspondence now follows directly from the general algebra--commutant relation:
\begin{equation}
 \boxed{\begin{aligned}
 \operatorname{AZ}[\mathscr D(V_{\Delta,E})]
 &=\iota\!\left(\operatorname{AZ}(\mathscr B_{V_{\Delta,E}})\right)\\
 &=\iota\!\left(\operatorname{AZ}_{\mathrm{sym}}(\Delta)\right).
 \end{aligned}}
 \label{eq:AZcommutantlabel}
\end{equation}
Equivalently,
\begin{equation}
 \begin{gathered}
 \mathscr A_\Delta\sim_{\mathrm{grMorita}}
 \mathscr D(V_{\Delta,E})^{\mathrm{sop}},\\
 \operatorname{AZ}[\mathscr D(V_{\Delta,E})^{\mathrm{sop}}]
 =\operatorname{AZ}_{\mathrm{sym}}(\Delta).
 \end{gathered}
 \label{eq:AZcorrespondence}
\end{equation}
The calculations of Sec.~\ref{sec:realcommutants} give explicit division-algebra realizations of this relation for all ten residual types; equality of the symmetry and commutant labels is neither assumed nor required.

Class D gives a concrete distinction. Here $C$ is odd and antiunitary, while $I$ is even, so
\begin{equation}
 \begin{gathered}
 I^2=-\id,\qquad C^2=\id,\qquad CI=-IC,\\
 (IC)^2=\id,\qquad C(IC)=-(IC)C.
 \end{gathered}
 \label{eq:Dsymmetryalgebra}
\end{equation}
Thus $C$ and $IC$ are two anticommuting odd generators with positive square. By contrast, the effective commutant calculated above has odd generators $j=i\sigma_1$ and $I_Fj$, both with negative square. Consequently,
\begin{equation}
 \mathscr A_{\mathrm D}\cong_{\mathrm{gr}}\Cl_{2,0},
 \qquad
 \mathscr D(E_{\mathrm D})\cong_{\mathrm{gr}}\Cl_{0,2}.
 \label{eq:Dsymmetrycommutant}
\end{equation}
These superalgebras are super opposites and belong to different graded Morita classes: their standard AZ labels are D and C, respectively. Their underlying ungraded algebras are $M_2(\R)$ and $\HH$. The residual symmetry remains class D; its commutant has class C under the same Morita-class labeling.

The AZ correspondence is independent of the orbit size, the dimension of $\Delta$, and the choice of effective enlarged irrep $E$. It uses the inherited commutativity grading and the fixed lifting extension in Eq.~\eqref{eq:complexresidualfactor}; it makes no claim about arbitrary projective extensions or the AZ label of the full physical system.

Table~\ref{tab:realMackey} summarizes this correspondence and the representation data. The restriction rules and resulting zero-mode indices are derived next in Sec.~\ref{sec:realrestrictions}.

\begin{table*}[tbp]
\caption{Representation data for the residual symmetry algebras of Table~\ref{tab:Nsymmetry}. The symmetry and commutant columns use the same graded Morita-class labeling; they are related by the super-opposite involution in Eq.~\eqref{eq:AZopposite}. Restrictions and indices are organized by the residual symmetry label. $\Delta$ is an $N$-irrep, $\mathcal W$ and $\mathcal W_\pm$ are full $G$-corepresentations, and $V$ is a $G^h$-corepresentation. Equivalent underlying restrictions are counted once, even if there are inequivalent choices of $h$. The odd pieces are specified as real vector spaces, except that the AIII row is complex; their multiplication is fixed in the text. The indices count full $G$-multiplets, not their complex dimensions or the number of $N$ constituents.}
\label{tab:realMackey}
\begin{ruledtabular}
\begin{tabular}{cccccccc}
Symmetry AZ & Commutant AZ & $\mathscr D_0$ & $\mathscr D_1$ & $\mathscr D$ & $\Res_G^{G^h}V$ & Index & Quotient\\
A & A & $\C$ & $0$ & $\C$ & $\mathcal W$ & $0$ & $0$\\
AIII & AIII & $\C$ & $\C$ & $\Cl_1^\C$ & $\mathcal W_+\oplus\mathcal W_-$ & $n_+-n_-$ & $\Z$\\
AI & AI & $\R$ & $0$ & $\R$ & $\mathcal W$ & $0$ & $0$\\
BDI & CI & $\R$ & $\R$ & $\Cl_{0,1}$ & $\mathcal W_+\oplus\mathcal W_-$ & $n_+-n_-$ & $\Z$\\
D & C & $\C$ & $\R^2$ & $\Cl_{0,2}$ & $2\mathcal W$ & $n\bmod2$ & $\Z_2$\\
DIII & CII & $\HH$ & $\HH$ & $\Cl_{0,3}$ & $2\mathcal W$ & $n\bmod2$ & $\Z_2$\\
AII & AII & $\HH$ & $0$ & $\HH$ & $\mathcal W$ & $0$ & $0$\\
CII & DIII & $\HH$ & $\HH$ & $\Cl_{3,0}$ & $\mathcal W_+\oplus\mathcal W_-$ & $n_+-n_-$ & $\Z$\\
C & D & $\C$ & $\R^2$ & $\Cl_{2,0}$ & $\mathcal W$ & $0$ & $0$\\
CI & BDI & $\R$ & $\R$ & $\Cl_{1,0}$ & $\mathcal W$ & $0$ & $0$\\
\end{tabular}
\end{ruledtabular}
\end{table*}

\subsection{Restriction patterns and complete indices}
\label{sec:realrestrictions}

By Eq.~\eqref{eq:restrictioninduction}, the effective restriction rules determine those of the full irreps. After duplicate restrictions are identified, the three elementary forms are
\begin{align}
 \text{paired:}\quad &\Res V=\mathcal W_+\oplus\mathcal W_-,
      \quad\mathcal W_+\not\cong\mathcal W_-,\nonumber\\
 \text{doubled:}\quad &\Res V=\mathcal W\oplus\mathcal W,\nonumber\\
 \text{single:}\quad &\Res V=\mathcal W.
 \label{eq:threepatterns}
\end{align}
Equivalent underlying restrictions generate the same subgroup, even when their enlarged irreps are inequivalent. A single already gappable irrep contributes the relation $[\mathcal W]=0$ and therefore no surviving invariant. We enumerate the distinct restriction patterns orbit by orbit. In AIII the two irreps of the effective chiral generator are paired by $h_E$. In BDI and CII the two inequivalent effective corepresentations are likewise distinguished by the normalized chiral product and have equal complex dimensions; the enlarged module pairs them. In all three cases the minimal restriction is $D_+\oplus D_-$, giving $\mathcal W_+\oplus\mathcal W_-$ after tensoring and induction.

The D and DIII rows instead have one effective ungraded irrep that cannot carry a lifting involution on a single copy. For D, take $C=K$ on $\C$: the only Hermitian scalar compatible with reversal is zero, whereas two copies admit $h_E=\sigma_2$. For DIII, take $T=JK$ and $C=K$ on $\C^2$, where $J=-i\sigma_2$. Reversal permits only a Hamiltonian proportional to $\sigma_2$, and preservation by $T$ forces its coefficient to vanish. Two copies admit
\begin{equation}
 \begin{gathered}
 T=(\id_2\otimes J)K,\qquad C=\id_4K,\\
 h_E=\sigma_2\otimes\sigma_1.
 \end{gathered}
 \label{eq:doubledDIII}
\end{equation}
Thus the doubled restriction holds for one- and two-dimensional effective multiplets, respectively. The protected parity counts full $G$-corepresentations after induction, not individual complex states.

The A, AI, and AII rows admit a scalar $h_E=\pm\id$ because every effective generator preserves the Hamiltonian. In C, the irreducible effective corepresentation $C=JK$ admits $h_E=\sigma_3$. In CI, $T=\sigma_1K$ and $C=JK$ admit $h_E=\sigma_1$. These five rows therefore restrict to a single already gappable irrep. Different choices of the grading can have the same restriction and must not be counted as independent relations.

Because distinct orbits carry disjoint irreducible content, the quotient can be assembled orbit by orbit. Within each orbit, the restrictions just listed exhaust the ungraded irreps and contribute either one paired relation, one doubled relation, or a single removable irrep. The three cases give
\begin{align}
 &(\Z\oplus\Z)/\Z(1,1)\cong\Z,
 &&\nu=n_+-n_-,\nonumber\\
 &\Z/2\Z\cong\Z_2,
 &&\nu=n\pmod2,\nonumber\\
 &\Z/\Z=0.
 \label{eq:threeindices}
\end{align}
If there are $a$ independent paired sectors and $b$ independent doubled sectors, then
\begin{equation}
 \boxed{\Qind\cong\Z^a\oplus(\Z_2)^b.}
 \label{eq:mainclassification}
\end{equation}
The two members of a pair can be interchanged, reversing the sign of its integer index. A parity can equivalently be written $(-1)^n$. Table~\ref{tab:realMackey} gives these indices and quotients for each residual symmetry type.

The converse makes the index complete. If every component vanishes, equal multiplicities can be paired, even multiplicities can be grouped into doubled blocks, and every single block is already removable. This decomposes the zero-mode representation into restrictions of enlarged modules and constructs a lifting involution block by block; every gappable representation arises in this way. Hence, in the finite semisimple setting, vanishing of the complete index is both necessary and sufficient for algebraic lifting.

If $d_j$ is the complex dimension of either $G$-corepresentation in paired sector $j$, and $d_k$ is that of the irreducible $G$-corepresentation in doubled sector $k$, then
\begin{equation}
 \dim_\C\ker H\ \geq\
 \sum_{j\in\mathrm{paired}}d_j|\nu_j|
 +\sum_{k\in\mathrm{doubled}}d_k\bar\nu_k,
 \label{eq:bound}
\end{equation}
where $\bar\nu_k\in\{0,1\}$. These dimensions include both $\Delta$ and its orbit multiplicity. Equality is attainable for arbitrary symmetry-compatible finite-dimensional perturbations; locality or spatial restrictions can prevent attaining it in a specific microscopic model.

\subsection{Summary}
Including antiunitary symmetries extends the Mackey construction through semilinear intertwiners and twisted residual factors. The five possible little cogroups and their factor classes yield ten residual AZ types. For every induced super irrep, the super opposite of its graded commutant represents the graded Morita class of the residual symmetry algebra, independently of the reference-irrep dimension and orbit size. The commutant label itself is obtained by Eq.~\eqref{eq:AZopposite}. In residual symmetry labels, the explicit constructions give paired restrictions in AIII, BDI, and CII, doubled restrictions in D and DIII, and single restrictions in the remaining types. These produce integer, parity, and trivial quotient contributions, respectively, as collected in Table~\ref{tab:realMackey}. The independent orbit contributions form the complete quotient, whose vanishing is equivalent to algebraic lifting in the finite semisimple setting. Section~\ref{sec:characters} gives character tests for identifying these structural types without constructing the intertwiners.

\section{Character criteria}
\label{sec:characters}
An orbit's type fixes its contribution to the quotient, so the practical task is to determine that type without constructing the intertwiners. The tests below implement the little-group construction for groups containing antiunitary elements; Sec.~\ref{sec:complexcharacters} gave their complex counterparts. All AZ names in this section are residual symmetry labels. The standard label of the graded commutant follows by applying $\iota$ in Eq.~\eqref{eq:AZopposite}.

\subsection{Tests on a normal-subgroup irrep}
\begin{table*}[tbp]
\caption{Character tests on an $N$-irrep $\Delta$ and the constructed super irrep $V$ when all four grading cosets exist. All AZ labels refer to residual symmetry types. The $X$ indicators act on $\Delta$, whereas $Y_T$, $Y_C$, and $R_N$ use the character of $V|_N$. The orbit size $o=[G:G_\Delta]$ and constituent multiplicity $m_\Delta$ give $Y_T=o m_\Delta X_T$, $Y_C=o m_\Delta X_C$, and $R_N=o m_\Delta^2$. The latter is the complex dimension of the $N$-commutant. The indicator $X_\Gamma$ distinguishes AIII from A when both antiunitary indicators vanish; it is zero for AI, AII, D, and C because a stabilizer containing only $T$ or only $C$ cannot also contain their product.}
\label{tab:indicators}
\begin{ruledtabular}
\begin{tabular}{ccccccccc}
 & \multicolumn{3}{c}{$N$-irrep $\Delta$} & \multicolumn{2}{c}{Mackey multiplicities} & \multicolumn{3}{c}{Super irrep $V$}\\
\cline{2-4}\cline{5-6}\cline{7-9}
Symmetry AZ & $X_T$ & $X_C$ & $X_\Gamma$ & $o$ & $m_\Delta$ & $Y_T$ & $Y_C$ & $R_N$\\
A & $0$ & $0$ & $0$ & $4$ & $1$ & $0$ & $0$ & $4$\\
AIII & $0$ & $0$ & $1$ & $2$ & $2$ & $0$ & $0$ & $8$\\
AI & $+1$ & $0$ & $0$ & $2$ & $1$ & $2$ & $0$ & $2$\\
BDI & $+1$ & $+1$ & $1$ & $1$ & $2$ & $2$ & $2$ & $4$\\
D & $0$ & $+1$ & $0$ & $2$ & $2$ & $0$ & $4$ & $8$\\
DIII & $-1$ & $+1$ & $1$ & $1$ & $4$ & $-4$ & $4$ & $16$\\
AII & $-1$ & $0$ & $0$ & $2$ & $2$ & $-4$ & $0$ & $8$\\
CII & $-1$ & $-1$ & $1$ & $1$ & $4$ & $-4$ & $-4$ & $16$\\
C & $0$ & $-1$ & $0$ & $2$ & $2$ & $0$ & $-4$ & $8$\\
CI & $+1$ & $-1$ & $1$ & $1$ & $2$ & $2$ & $-2$ & $4$\\
\end{tabular}
\end{ruledtabular}
\end{table*}

For finite $G$, the type can be determined without constructing intertwiners. The two antiunitary indicators below are character averages over a coset and take values $+1$, $-1$, or $0$; the unitary overlap indicator takes values $0$ or $1$. Let $\chi_\Delta(n)=\tr\Delta(n)$. When the corresponding cosets exist, choose $t$ antiunitary and preserving and $u$ antiunitary and reversing, and define the twisted Wigner indicators \cite{Sumita2019}
\begin{align}
 X_T(\Delta)&=\frac{1}{|N|}\sum_{b\in Nt}
                      \omega(b,b)\chi_\Delta(b^2),\nonumber\\
 X_C(\Delta)&=\frac{1}{|N|}\sum_{b\in Nu}
                      \omega(b,b)\chi_\Delta(b^2).
 \label{eq:XTC}
\end{align}
A zero value means that the antiunitary coset does not stabilize $\Delta$; a nonzero value is the residual square with which it does. The three cases follow from the orthogonality of corepresentations: averaged over the coset, the antiunitary intertwining form either vanishes, or is symmetric, or is antisymmetric. The average is normalized by the size of the coset, not by the order of $G$.

For a unitary reversing representative $v$, define the character overlap \cite{Sumita2019}
\begin{equation}
 X_\Gamma(\Delta)=\frac{1}{|N|}\sum_{n\in N}
          \chi_\Delta(n)^*\chi_{\Delta^v}(n).
 \label{eq:XGamma}
\end{equation}
Character orthogonality gives $X_\Gamma=1$ if $\Delta^v\cong\Delta$ and zero otherwise. Using Eq.~\eqref{eq:orbit}, this expression includes the projective conjugation factor automatically. In the full four-coset case, take $v=tu$.

These tests are independent of the projective phase convention. For antiunitary $b$, the transformation in Eq.~\eqref{eq:gauge} gives
\begin{equation}
 \omega'(b,b)\chi'_\Delta(b^2)=\omega(b,b)\chi_\Delta(b^2).
 \label{eq:indicatorgauge}
\end{equation}
Thus the Wigner indicators are gauge invariant. The common rephasing of $\chi_\Delta$ and $\chi_{\Delta^v}$ likewise cancels in the character inner product of Eq.~\eqref{eq:XGamma}.

Table~\ref{tab:indicators} determines the ten types. An absent coset is not the same as a vanishing indicator for a present coset: if the coset is absent, the corresponding indicator is undefined rather than zero. For a subgroup $B$ of the four-element group, use only the tests whose cosets exist. Indicators determine the \emph{type} of an orbit, not the signed multiplicities in a physical zero-mode space. In a paired sector one must still distinguish the full-group corepresentations $\mathcal W_+$ and $\mathcal W_-$, for example by the normalized effective chiral operator; in a doubled sector one counts full irreducible corepresentations modulo two.

\subsection{Tests on an already constructed super irrep}
The preceding tests act on a single irrep of $N$. When applied to an entire super irrep, the same traces also acquire factors from the orbit size and the multiplicity of each constituent, changing both their magnitudes and their pattern. For an irreducible super representation $\widetilde\rho$ of a full four-coset group, let $\chi=\tr\widetilde\rho|_N$ and set
\begin{align}
 Y_T&=\frac{1}{|N|}\sum_{b\in Nt}\omega(b,b)\chi(b^2),\nonumber\\
 Y_C&=\frac{1}{|N|}\sum_{b\in Nu}\omega(b,b)\chi(b^2),\nonumber\\
 R_N&=\frac{1}{|N|}\sum_{n\in N}|\chi(n)|^2.
 \label{eq:Ytests}
\end{align}
If $\widetilde\rho|_N=\bigoplus_\delta m_\delta\delta$, then $R_N=\sum_\delta m_\delta^2$. This is the complex dimension of the $N$-commutant, not in general the dimension of the graded commutant of $G$.

The complete values are shown in Table~\ref{tab:indicators}. They obey the exact relations
\begin{equation}
 \begin{aligned}
 Y_T&=o m_\Delta X_T(\Delta),\\
 Y_C&=o m_\Delta X_C(\Delta),\qquad
 R_N=o m_\Delta^2.
 \end{aligned}
 \label{eq:Yrelations}
\end{equation}
These relations follow by summing the constituent characters: each of the $o$ orbit constituents occurs $m_\Delta$ times, and its antiunitary indicators agree with those of $\Delta$. The norm $R_N$ is the dimension of the $N$-commutant and takes the values $2$, $4$, $8$, or $16$ in Table~\ref{tab:indicators}. For a fixed value of $R_N$, the signed pair $(Y_T,Y_C)$ resolves the remaining possibilities, for example D, C, and AII when $R_N=8$. A and AIII instead have the same vanishing signed pair and are distinguished by $R_N=4$ and $8$, respectively. Thus the triple $(Y_T,Y_C,R_N)$ distinguishes all ten real structural types in the full four-coset setting. For a quotient with only an antiunitary preserving coset, the corresponding $(Y_T,R_N)$ values are $(1,1)$, $(-2,4)$, and $(0,2)$ for the real, quaternionic, and complex types. For a quotient with only an antiunitary reversing coset, $Y_C=2,-2,0$ identifies D, C, and A. These reduced-group normalizations differ from the four-coset entries.

The orbit tests identify the structural type but do not determine the signed multiplicities in a physical zero-mode space. In a paired sector, distinguish $\mathcal W_+$ from $\mathcal W_-$, for example by the normalized effective chiral operator, and take their multiplicity difference. In a doubled sector, count complete irreducible corepresentations modulo two. Table~\ref{tab:realMackey} records these indices; neither an orbit indicator nor the raw complex kernel dimension replaces the required multiplet count.

\section{Application to end zero modes of one-dimensional systems}
\label{sec:examples}
The class-AIII end modes of Sec.~\ref{sec:motivation} illustrate the complex quotient developed in Sec.~\ref{sec:unitary}. For the eight real AZ classes considered here, there are no additional symmetry sectors, and the unitary Hamiltonian-preserving subgroup is $N=\{1\}$. Its unique irrep is stabilized by the full symmetry group, so the residual symmetry algebra is the physical end-mode symmetry algebra. The quotient indices therefore follow directly from the general restriction rules in Table~\ref{tab:realMackey}. AI and AII are preserving-only classes, for which the quotient vanishes identically.

The calculations below illustrate the general theory: they identify the physical multiplets and the elementary perturbations that remove them. Table~\ref{tab:endsummary} collects the removable content, the quotient, and the form of the protected end zero modes. The end-mode forms are consistent with the classification in Ref.~\cite{ZhaoWang2014}. AI and AII have no Hamiltonian-reversing symmetry; the remaining six classes do.

The identification with bulk topology is not algebraic. The end quotient is computed from the symmetry representation alone; its agreement with the stable bulk classes is additional physical input supplied by bulk--boundary correspondence, not a consequence of representation theory.

\begin{table*}[tbp]
\caption{End-mode applications for the eight real AZ classes with $N=\{1\}$. The quotient follows directly from the general theory; the explicit calculations in this section identify the physical multiplets and their lifting operators. The protected content is the minimal remainder after removing all symmetry-allowed blocks at one decoupled end; accidental additional zero modes are not counted. For integer indices, the sign specifies the surviving chirality; $\bar\nu=\nu\bmod2\in\{0,1\}$ in the parity cases. Majorana terminology refers to BdG realizations. The CII integer counts Kramers doublets, so the conventional complex winding is $2\nu$. AI and AII admit scalar lifting, while C and CI admit lifting within each complete two-dimensional multiplet.}
\label{tab:endsummary}
\begin{ruledtabular}
\begin{tabular}{clcl}
AZ class & Elementary removable content & Quotient & Protected end zero modes\\
\hline
AI & One multiplet (scalar lifting) & $0$ & None\\
BDI & Opposite-chirality one-dimensional multiplets & $\Z$ & $|\nu|$ modes of one chirality\\
D & Two Majorana modes & $\Z_2$ & $\bar\nu$ Majorana modes\\
DIII & Two Majorana Kramers pairs & $\Z_2$ & $\bar\nu$ Majorana Kramers pairs\\
AII & One Kramers doublet (scalar lifting) & $0$ & None\\
CII & Opposite-chirality Kramers doublets & $\Z$ & $|\nu|$ doublets of one chirality\\
C & One complete two-dimensional multiplet & $0$ & None\\
CI & One complete two-dimensional multiplet & $0$ & None\\
\end{tabular}
\end{ruledtabular}
\end{table*}

\subsection{End-mode symmetry algebras}
As in Sec.~\ref{sec:motivation}, consider one end of a semi-infinite system, or the limit of a finite chain in which the two ends decouple. Let $Z$ be its finite-dimensional zero-mode space. There are no additional crystalline or internal symmetry sectors in this application. When present, the antiunitary generators satisfy
\begin{equation}
 \begin{gathered}
 TH=HT,\qquad CH=-HC,\\
 T^2=\kappa_T,\qquad C^2=\kappa_C.
 \end{gathered}
 \label{eq:endTC}
\end{equation}
The four chiral classes have both generators, with square signs BDI $(+,+)$, DIII $(-,+)$, CII $(-,-)$, and CI $(+,-)$. D and C have only the reversing generator, with $C^2=+1$ and $-1$, respectively. AI and AII have only the preserving generator, with $T^2=+1$ and $-1$. In the four chiral classes, choose commuting representatives for $T$ and $C$ and define
\begin{equation}
 \Gamma=\begin{cases}
 TC,&\mathrm{BDI},\\
 -TC,&\mathrm{CII},\\
 iTC,&\mathrm{DIII},\ \mathrm{CI}.
 \end{cases}
 \label{eq:endGamma}
\end{equation}
It obeys $\Gamma^2=\id$ and $\Gamma H=-H\Gamma$. In BDI and CII, $T$ and $C$ preserve each chiral eigenspace; in DIII and CI they exchange them. This difference is why the latter two classes do not have an integer chirality imbalance.

Since $N=\{1\}$, no further induction is needed: we exhibit the irreducible end-mode corepresentations and the lifting extensions that realize the general restriction rules. In superconducting realizations, $Z$ is the complex Bogoliubov--de Gennes (BdG) kernel; particle-hole redundancy must be retained when translating its multiplets into physical Majorana modes.

\paragraph{Preserving-only classes AI and AII.}
Neither class has a symmetry anticommuting with $H$. On every end multiplet the scalar involution $h=\id$ is allowed, so an arbitrarily small $\varepsilon\id$ moves the entire multiplet away from zero without breaking the symmetry. In AII the Kramers degeneracy survives, but it is not pinned to zero energy. Thus every representation is already in the restriction image:
\begin{equation}
 \Qind_{\mathrm{AI}}=\Qind_{\mathrm{AII}}=0.
 \label{eq:endAIAII}
\end{equation}
These quotients are formally defined but carry no zero-mode obstruction. We next illustrate the general restriction rules for each of the remaining six symmetry algebras.

\subsection{Integer quotients in BDI and CII}
\paragraph{BDI.}
The two inequivalent irreducible zero-mode corepresentations $\mathcal W_\pm$ are complex one-dimensional. They can be represented as
\begin{equation}
 T=K,\qquad C=\pm K,\qquad \Gamma=\pm\id.
 \label{eq:endBDIirreps}
\end{equation}
A compatible Hermitian Hamiltonian on either irrep is zero: preservation by $T$ makes a scalar real, and reversal by $C$ forces it to vanish. On $\mathcal W_+\oplus\mathcal W_-$, however, $T=K$ and $C=\sigma_3K$ admit $h=\sigma_1$. Thus the elementary restriction is $\Res V=\mathcal W_+\oplus\mathcal W_-$, giving
\begin{equation}
 \begin{gathered}
 \Qind_{\mathrm{BDI}}=
 \frac{\Z[\mathcal W_+]\oplus\Z[\mathcal W_-]}
 {\Z([\mathcal W_+]+[\mathcal W_-])}\cong\Z,\\
 \nu_{\mathrm{BDI}}=n_+-n_-.
 \end{gathered}
 \label{eq:endBDIquotient}
\end{equation}
Opposite chiralities can be removed in pairs; an imbalance leaves at least $|\nu_{\mathrm{BDI}}|$ end zero modes. In a BdG realization these are unpaired Majorana modes, with the sign recording their chiral symmetry character.

\paragraph{CII.}
Let $J=-i\sigma_2$, so $J$ is real and $J^2=-\id_2$. The two irreducible corepresentations now have complex dimension two:
\begin{equation}
 T=JK,\qquad C=\pm JK,\qquad \Gamma=\pm\id_2.
 \label{eq:endCIIirreps}
\end{equation}
Each is a Kramers doublet of definite chirality. A Hermitian Hamiltonian on one doublet that is preserved by $T$ is scalar, and reversal by $C$ again forces it to vanish. An opposite-chirality pair admits
\begin{equation}
 \begin{gathered}
 T=(\id_2\otimes J)K,\qquad
 C=(\sigma_3\otimes J)K,\\
 \Gamma=\sigma_3\otimes\id_2,\qquad
 h=\sigma_1\otimes\id_2.
 \end{gathered}
 \label{eq:endCIIlifting}
\end{equation}
The restriction is therefore paired, just as in BDI:
\begin{equation}
 \begin{gathered}
 \Qind_{\mathrm{CII}}=
 \frac{\Z[\mathcal W_+]\oplus\Z[\mathcal W_-]}
 {\Z([\mathcal W_+]+[\mathcal W_-])}\cong\Z,\\
 \nu_{\mathrm{CII}}=n_+-n_-.
 \end{gathered}
 \label{eq:endCIIquotient}
\end{equation}
Here $n_\pm$ count Kramers doublets, not complex states. Hence
\begin{equation}
 \dim_\C Z_+-\dim_\C Z_-=2\nu_{\mathrm{CII}},
 \label{eq:endCIIcount}
\end{equation}
and at least $2|\nu_{\mathrm{CII}}|$ complex end states remain. This normalization explains why the conventional complex winding number in class CII takes values in $2\Z$, although the quotient written in elementary-multiplet units is $\Z$ \cite{Ryu2010}.

\subsection{Parity quotients in D and DIII}
\paragraph{D.}
The only antiunitary generator is $C$, with $C^2=+1$. Its unique irreducible zero-mode corepresentation $\mathcal W$ is complex one-dimensional and can be chosen as $C=K$. A Hermitian scalar cannot reverse sign under complex conjugation unless it vanishes. Two copies, however, admit $C=\id_2K$ and the purely imaginary Hermitian involution $h=\sigma_2$. Hence the elementary restriction is $\Res V=2\mathcal W$, and
\begin{equation}
 \begin{gathered}
 \Qind_{\mathrm D}=\frac{\Z[\mathcal W]}{2\Z[\mathcal W]}
 \cong\Z_2,\\
 \nu_{\mathrm D}=n\pmod2=\dim_\C Z\pmod2.
 \end{gathered}
 \label{eq:endDquotient}
\end{equation}
More generally, in a basis with $C=K$ an allowed Hermitian end Hamiltonian is $H_{\mathrm{end}}=iA$, where $A$ is real and antisymmetric. An odd-dimensional $A$ is singular, while an even-dimensional one can be invertible. In a BdG realization, each one-dimensional zero-mode multiplet is a Majorana mode: pairs can hybridize to nonzero opposite energies, but an odd number leaves one protected end Majorana \cite{Kitaev2001}. No time-reversal or unitary chiral symmetry is needed for this parity obstruction.

\paragraph{DIII.}
There is one irreducible zero-mode corepresentation $\mathcal W$, of complex dimension two, with
\begin{equation}
 T=JK,\qquad C=K,\qquad \Gamma=\sigma_2.
 \label{eq:endDIIIirrep}
\end{equation}
Since $T$ exchanges the two chiralities, they occur equally often in this multiplet. Nevertheless, a single multiplet is protected. The $C$ constraint makes a Hermitian $2\times2$ lifting Hamiltonian proportional to $\sigma_2$, and the $T$ constraint sets its coefficient to zero. Two copies admit the involution in Eq.~\eqref{eq:doubledDIII}, so the elementary restriction is $\Res V=2\mathcal W$. Consequently,
\begin{equation}
 \begin{gathered}
 \Qind_{\mathrm{DIII}}=
 \frac{\Z[\mathcal W]}{2\Z[\mathcal W]}\cong\Z_2,\\
 \nu_{\mathrm{DIII}}=n\pmod2
 =\tfrac12\dim_\C Z\pmod2.
 \end{gathered}
 \label{eq:diii}
\end{equation}
One nontrivial end multiplet is a Majorana Kramers pair in the superconducting realization \cite{ZhangKaneMele2013}. Two such pairs can be lifted, whereas an odd number leaves one protected pair. The parity of the raw complex kernel dimension is always zero and would miss this obstruction.

\subsection{Trivial quotients in C and CI}
\paragraph{C.}
There is only a reversing antiunitary generator, now with $C^2=-1$. Its unique irreducible corepresentation $\mathcal W$ has complex dimension two and can be chosen as $C=JK$. It already admits $h=\sigma_3$, since $J\sigma_3=-\sigma_3J$ and $h$ is real. Thus $\Res V=\mathcal W$, rather than $2\mathcal W$, and
\begin{equation}
 \Qind_{\mathrm C}=\frac{\Z[\mathcal W]}{\Z[\mathcal W]}=0.
 \label{eq:endCquotient}
\end{equation}
Although the zero-mode multiplet is even-dimensional, this degeneracy does not protect it: an allowed perturbation splits it to opposite nonzero energies. Unlike class D, class C has no end-mode parity obstruction.

\paragraph{CI.}
A single complex two-dimensional irreducible corepresentation can be chosen as
\begin{equation}
 T=K,\qquad C=JK,\qquad \Gamma=\sigma_2.
 \label{eq:endCIirrep}
\end{equation}
Unlike DIII, it already admits the lifting involution $h=\sigma_3$: it is real and therefore preserved by $T$, while $J\sigma_3=-\sigma_3J$ gives reversal by $C$. Thus $\Res V=\mathcal W$ and
\begin{equation}
 \Qind_{\mathrm{CI}}=
 \frac{\Z[\mathcal W]}{\Z[\mathcal W]}=0.
 \label{eq:endCIquotient}
\end{equation}
There is no symmetry-protected end zero-mode content in this class. Accidental zero modes can occur, but a symmetry-compatible perturbation can remove each complete multiplet.

\subsection{Agreement with bulk topology}
Let $\mathcal K_X^{(1)}$ denote the reduced stable classification of one-dimensional gapped free-fermion bulks in class $X$, relative to a fixed trivial reference. In the conventional real-class order $X=(\mathrm{AI},\mathrm{BDI},\mathrm D,\mathrm{DIII},\mathrm{AII},\mathrm{CII},\mathrm C,\mathrm{CI})$, the periodic table and the end quotients give \cite{Schnyder2008,Ryu2010,Kitaev2009}
\begin{equation}
 \begin{aligned}
 (\mathcal K_X^{(1)})_X
 &\cong(0,\Z,\Z_2,\Z_2,0,\Z,0,0)\\
 &\cong(\Qind_X)_X.
 \end{aligned}
 \label{eq:1Dbulkquotients}
\end{equation}
The CII entry is written in Kramers-multiplet units; multiplication by two identifies it with the conventional $2\Z$ winding classification. The AIII quotient of Sec.~\ref{sec:motivation} likewise agrees with its one-dimensional $\Z$ bulk classification.

This agreement is more than an equality of abstract groups. Fix an end orientation and a symmetry-compatible termination against the trivial vacuum, and permit boundary stabilization only by gappable multiplets. The one-dimensional bulk--boundary index correspondence then identifies the stable bulk class with the quotient class of its end kernel:
\begin{equation}
 \begin{aligned}
 \partial_X:\mathcal K_X^{(1)}&\xrightarrow{\ \cong\ }\Qind_X,\\
 [H_{\mathrm{bulk}}]&\longmapsto[\ker H_{\mathrm{end}}].
 \end{aligned}
 \label{eq:1Dbulkboundary}
\end{equation}
These boundary data are part of the definition of $\partial_X$. Adding an isolated nongappable boundary multiplet would change the end quotient without changing the bulk Hamiltonian. Such an addition changes the boundary realization and is not an allowed stabilization in Eq.~\eqref{eq:1Dbulkboundary}.
In AIII and BDI, the winding invariant equals the signed end chirality imbalance, after fixing the orientation convention. In CII the conventional winding is twice the signed doublet imbalance. In D the bulk $\Z_2$ invariant is the parity of the end Majorana modes, whereas in DIII it is the parity of the end Majorana Kramers pairs. C and CI have trivial stable bulk classifications and no protected end multiplet. AI and AII likewise have trivial one-dimensional bulk classifications, consistent with the absence of a zero-energy obstruction when all symmetries preserve the Hamiltonian.

Physically, symmetry-preserving perturbations can only remove the paired, doubled, or single blocks exhibited above, so a nontrivial bulk invariant enforces a nonzero end quotient class while stacking inverse bulk phases supplies mutually removable end content. The correspondence is between stable bulk classes and stable end-mode classes, under the assumptions of local gapped free-fermion systems, a decoupled end, the fixed compatible termination above, and stabilization by gappable boundary degrees of freedom.

\section{Application to boundary sigma-model topological terms}
\label{sec:application}
Topological terms in boundary nonlinear sigma models encode obstructions to Anderson localization of disordered boundary fermions \cite{Schnyder2008,Ryu2010}. For minimal boundary Dirac models in the eight real AZ classes, Ref.~\cite{ZhaoSigma2026} relates these terms to the kernel index of an interpolation cylinder and evaluates the index by directly extending its real Clifford symmetry algebra. Here we derive the cylinder indices from the general representation theory developed above. We identify the unitary Hamiltonian-preserving subgroup, determine its irreducible orbits and residual symmetry algebras, and use the Mackey--AZ correspondence and restriction rules to obtain the quotient. The texture-index theorem and fermion-phase calculation then supply the connection to the sigma-model topological terms.

\subsection{The interpolation cylinder and its grading}
Let $Q_0$ and $Q_1$ be endpoint sigma-field textures in the normalized target manifold. A symmetry-compatible interpolation $Q_t$ may leave this target and range over the larger Hubbard--Stratonovich field space, provided it preserves the real-linear symmetry constraints of the fermionic kernel; for example, one may take $Q_t=(1-t)Q_0+tQ_1$. Allowing the interpolation to leave the target is essential when the endpoints represent distinct homotopy classes. Let $\mathcal D_t$ be the associated boundary fermionic operator. An elliptic cylinder representative is
\begin{equation}
 \widehat{\mathcal D}=
 \begin{pmatrix}\partial_t&\mathcal D_t\\
 -\mathcal D_t^\dagger&-\partial_t\end{pmatrix},
 \qquad \widehat H=i\widehat{\mathcal D}.
 \label{eq:cylinder}
\end{equation}
The auxiliary coordinate $t$ is not physical time. The construction requires invertible asymptotic endpoint operators and a symmetry-compatible Fredholm domain. The additional matrix factor ensures ellipticity in the Euclidean conventions of Ref.~\cite{ZhaoSigma2026}.

For the minimal boundary modules used there, lifting the physical boundary intertwiners to the cylinder, including the additional cylinder and auxiliary factors, produces a real internal symmetry algebra $\mathcal C^{(r)}$. This is the symmetry algebra of the full allowed cylinder family, not the boundary algebra obtained by merely renaming the physical symmetry operators. In particular, lifting can change the effective antiunitary squares. Commutation with $\widehat H$ grades the cylinder algebra:
\begin{equation}
 a\widehat H=(-1)^{|a|}\widehat Ha,\qquad
 r=s-D-1\pmod8.
 \label{eq:cylindergrading}
\end{equation}
Here $D$ is the boundary spatial dimension, and the physical real classes are indexed by $s=0,\ldots,7$ in the order AI, BDI, D, DIII, AII, CII, C, CI. This induced grading is \emph{not} necessarily the canonical grading that declares every Clifford generator odd. Keeping it is indispensable: identical ungraded matrix algebras can have different zero-mode quotients.

The algebra $\mathcal C^{(r)}$ realizes the symmetry data denoted by $G$ in the general construction, and adjoining a compatible lifting involution realizes $G^h$. More precisely, $G$ is a group of homogeneous symmetry operators, whereas $\mathcal C^{(r)}$ is their real operator algebra; we specify a finite projective generating group below. Its preserving subgroup generates $\mathcal C_0^{(r)}$, the algebraic counterpart of $\ker c$. Writing $\mathcal M(\mathcal C)$ for the additive group of finite-dimensional real modules, the quotient is
\begin{align}
 \mathcal C_{\mathrm{ext}}
 &=\langle\mathcal C,h:\ h^2=1,\ ha=(-1)^{|a|}ah\rangle,
 \nonumber\\
 \Qind_r&=\coker\left[
 \mathcal M(\mathcal C_{\mathrm{ext}}^{(r)})
 \xrightarrow{\Res}\mathcal M(\mathcal C^{(r)})\right].
 \label{eq:cliffordquotient}
\end{align}
The numerator modules have no chosen module grading, while the algebra retains its cylinder grading. Since the even algebra contains the complex structure $I$, the added $h$ is complex-linear. Thus Eq.~\eqref{eq:cliffordquotient} implements exactly the lifting condition of Sec.~\ref{sec:lifting}, and the class of $\ker\widehat H$ is the cylinder index.

Table~\ref{tab:cylinder} lists the generators of the cylinder algebra, its preserving subalgebra, and the unitary preserving subgroup. Together with the relations below, these are the input data for the orbit and residual-symmetry calculation.

\begin{table*}[tbp]
\caption{Generators for the cylinder subgroup calculation, obtained from the lifted symmetries of Ref.~\cite{ZhaoSigma2026}. The first two generator columns specify unital real algebras; the identity is implicit. The $N^{(r)}$ column uses the projective convention of Eq.~\eqref{eq:cylinderN}, with scalar phases, including $I$, absorbed into the multiplier; a dash denotes the trivial subgroup. Only $Q_x,Q_y$ are needed because $Q_z=-Q_xQ_y$. The last column gives $TPT^{-1}=\eta_rP$; a dash means $P$ is absent. The remaining generator relations are given in Eqs.~\eqref{eq:cylinderIT}--\eqref{eq:cylinderseeds}.}
\label{tab:cylinder}
\begin{ruledtabular}
\begin{tabular}{ccccc}
$r$ & Generators of $\mathcal C^{(r)}$ & Generators of $\mathcal C_0^{(r)}$ & Generators of $N^{(r)}$ & $\eta_r$\\
\hline
$0$ & $I,T$ & $I,T$ & --- & ---\\
$1$ & $I,T,P$ & $I,T$ & --- & $-1$\\
$2$ & $I,T,L,P$ & $I,T,L$ & $L$ & $-1$\\
$3$ & $I,T,Q_x,Q_y,P$ & $I,T,Q_x,Q_y$ & $Q_x,Q_y$ & $+1$\\
$4$ & $I,T,Q_x,Q_y$ & $I,T,Q_x,Q_y$ & $Q_x,Q_y$ & ---\\
$5$ & $I,T,Q_x,Q_y,P$ & $I,T,Q_x,Q_y$ & $Q_x,Q_y$ & $-1$\\
$6$ & $I,T,L,P$ & $I,T,L$ & $L$ & $+1$\\
$7$ & $I,T,P$ & $I,T$ & --- & $+1$\\
\end{tabular}
\end{ruledtabular}
\end{table*}

\subsection{The unitary preserving subgroup and its irreducible orbits}
The lifted cylinder generators provide both gradings required by the general theory. In every case they include the complex structure $I$ and a preserving antiunitary $T$, with
\begin{equation}
 I^2=T^2=-\id,\qquad IT=-TI,\qquad
 [I,\widehat H]=[T,\widehat H]=0.
 \label{eq:cylinderIT}
\end{equation}
The square $T^2=-\id$ includes the cylinder and auxiliary factors. It is the same in all eight cases, irrespective of the square of the original boundary antiunitary. In the following, $T$ denotes this concrete cylinder operator. It must be distinguished from the abstract coset label $T$ in $B=G/N$: the operator represents the antiunitary preserving coset, but $T^2=-\id$ is an operator relation, whereas the coset has order two.

Except at $r=0,4$, there is also a complex-linear reversing involution $P$. The remaining complex-linear preserving generators are an involution $L$ at $r=2,6$, or quaternion units $Q_x,Q_y,Q_z$ at $r=3,4,5$. Their relations, obtained from the cylinder lift in Ref.~\cite{ZhaoSigma2026}, are
\begin{align}
 &P^2=\id,\quad [I,P]=0,\quad
 P\widehat H=-\widehat H P,\nonumber\\
 &TPT^{-1}=\eta_r P,\qquad
 \eta_r=\begin{cases}
 -1,&r=1,2,5,\\
 +1,&r=3,6,7,
 \end{cases}
 \label{eq:cylinderTP}
\end{align}
and, when the indicated generators exist,
\begin{equation}
 \begin{gathered}
 L^2=\id,\quad TLT^{-1}=-L,\quad PL=-LP,\\
 Q_a^2=-\id,\quad Q_xQ_y=-Q_yQ_x=-Q_z,\\
 [T,Q_a]=[P,Q_a]=0.
 \end{gathered}
 \label{eq:cylinderseeds}
\end{equation}
Here $L$ and every $Q_a$ commute with $I$ and $\widehat H$; relations involving $P$ are omitted when $P$ is absent.

To apply the finite-group formulation literally, take the group generated by these operators and identify scalar phases with the factor system $\omega_r$. In particular, $I$ acts as the fixed scalar $i$ and is absorbed into the projective multiplier. Denote the resulting finite projective symmetry group by $G^{(r)}$. Its operator algebra, including the scalar complex structure, is $\mathcal C^{(r)}$, and its lifted projective representation realizes $\mathcal C_{\mathrm{ext}}^{(r)}$. The unitary preserving subgroup is
\begin{equation}
 \begin{gathered}
 N^{(r)}=\ker\tau_r\cap\ker c_r\subseteq G^{(r)},\\
 N^{(r)}\cong
 \begin{cases}
 \{1\},&r=0,1,7,\\
 \Z_2^L,&r=2,6,\\
 \Z_2^x\times\Z_2^y,&r=3,4,5.
 \end{cases}
 \end{gathered}
 \label{eq:cylinderN}
\end{equation}
The last line is projective: the cosets of $Q_x,Q_y$ commute, but their matrices anticommute. Before scalar phases are removed, these subgroups are generated by $I$, by $I,L$, or by $I,Q_x,Q_y$, respectively. Thus Eq.~\eqref{eq:cylinderN} singles out the complex-linear part of the preserving symmetry algebra in precisely the sense of Sec.~\ref{sec:unitarysubgroup}.

Let $\Delta_a^{(r)}\in\widehat{N^{(r)}}$ denote an irrep with the prescribed restricted multiplier. The three possibilities in Eq.~\eqref{eq:cylinderN} have, respectively, one one-dimensional irrep, two characters $\Delta_\pm(L)=\pm1$, and one two-dimensional projective irrep. For the last case we may choose
\begin{equation}
 \Delta(Q_x)=i\sigma_1,\qquad
 \Delta(Q_y)=i\sigma_2,\qquad
 \Delta(Q_z)=i\sigma_3.
 \label{eq:cylinderDeltaQ}
\end{equation}
These matrices generate $M_2(\C)$, proving uniqueness of the projective irrep. Consequently, the complex-linear preserving algebras generated by $N^{(r)}$ are $\C$, $\C\oplus\C$, and $M_2(\C)$ in the three cases, respectively.

For $r=0,1,7$ the unique irrep is fixed by every symmetry. For $r=3,4,5$ uniqueness likewise makes the orbit a single point; the antiunitary action is implemented by
\begin{equation}
 R=JK,\qquad J=-i\sigma_2,\qquad R^2=-\id_2,
 \label{eq:cylinderR}
\end{equation}
which commutes with all three matrices in Eq.~\eqref{eq:cylinderDeltaQ}. At $r=2,6$, both $T$ and $P$ exchange $\Delta_+$ and $\Delta_-$, whereas $TP$ stabilizes each. There is therefore one two-point orbit and its little cogroup is the reversing antiunitary group $\Z_2^C$, represented by $C=TP$.

Single-orbit support is a consequence of irreducibility. For an irreducible $\mathcal C^{(r)}$-module $\mathcal W$, the sum of the $N^{(r)}$-isotypic components belonging to any one $G^{(r)}$-orbit is invariant under the full cylinder symmetry algebra. Hence the irreps occurring in $\mathcal W|_{N^{(r)}}$ must form a single orbit, with equal multiplicities. This general statement concerns the irreps appearing in $\mathcal W$; it does not require the entire admissible space $\widehat{N^{(r)}}$ to be one orbit. For the present cylinder algebras, the explicit action above establishes this stronger property as well. In particular, at $r=2,6$ both characters occur in each irreducible cylinder module and contribute one orbit sector, not two independent indices.

\subsection{Residual AZ types, graded commutants, and quotient indices}
We now divide out the action on $\Delta_a^{(r)}$ as in Eq.~\eqref{eq:alpha}. For the one-dimensional fixed irreps at $r=0,1,7$, the reference antiunitary intertwiner can be chosen as $K$, with square $+1$. The residual preserving antiunitary therefore has $T_E^2=-1$. At $r=3,4,5$, the reference intertwiner in Eq.~\eqref{eq:cylinderR} instead has square $-1$, so
\begin{equation}
 T_E^2=\frac{T^2}{R^2}=+1
 \qquad(r=3,4,5).
 \label{eq:cylinderresidualT}
\end{equation}
This is the residual-factor-system correction: the physical cylinder square is divided by the intertwiner square on the reference irrep.

When $P$ is present in a fixed orbit, its reference intertwiner is the identity. The residual odd unitary $P_E$ has square $+1$ and obeys $T_EP_E=\eta_rP_ET_E$. Defining the residual reversing antiunitary $C_E=T_EP_E$ gives
\begin{equation}
 C_E^2=\eta_rT_E^2
 \qquad(r=1,3,5,7).
 \label{eq:cylinderresidualC}
\end{equation}
Hence $r=1$ has residual squares $(-1,+1)$ and type DIII; $r=7$ has $(-1,-1)$ and type CII. After the quaternionic intertwiner is removed, $r=3$ has $(+1,+1)$ and type BDI, while $r=5$ has $(+1,-1)$ and type CI. The preserving-only cases $r=0,4$ have types AII and AI, respectively.

For the exchanged characters at $r=2,6$, there is no preserving antiunitary in the little cogroup. Since $\Delta_\pm$ are real one-dimensional characters, the intertwiner for $TP$ can be chosen as $K$ and has square $+1$. The residual square is therefore
\begin{equation}
 C_E^2=(TP)^2=\eta_rT^2=-\eta_r
 =\begin{cases}+1,&r=2,\\-1,&r=6.\end{cases}
 \label{eq:cylinderexchangedsquares}
\end{equation}
These are precisely residual types D and C. The full cylinder family has the preserving antiunitary $T$, but it exchanges the two reference sectors and is absent from their little cogroup.

Tensoring each effective enlarged module $E$ with $\Delta_a^{(r)}$ and inducing constructs the super irrep $V_{a,E}^{(r)}$. Equations~\eqref{eq:commutantinduction} and \eqref{eq:AZcorrespondence} now apply directly:
\begin{equation}
 \begin{gathered}
 \mathscr D(V_{a,E}^{(r)})\cong_{\mathrm{gr}}\mathscr D(E),\\
 \operatorname{AZ}[\mathscr D(V_{a,E}^{(r)})^{\mathrm{sop}}]
 =\operatorname{AZ}_{\mathrm{sym}}(\Delta_a^{(r)}).
 \end{gathered}
 \label{eq:cylinderMackeyAZ}
\end{equation}
Table~\ref{tab:cylinderMackey} gives the resulting commutants and restrictions. The symmetry labels refer to the residual cylinder sectors, and the commutant labels follow by super opposition; the original boundary class enters through $r=s-D-1$.

\begin{table*}[tbp]
\caption{Derivation of the cylinder indices by the general theory. Scalar phases are included in the fixed multiplier, as in Eq.~\eqref{eq:cylinderN}. The two-dimensional irrep for $N^{(r)}=\Z_2^x\times\Z_2^y$ is projective. Here $d=\dim_\C\Delta_a^{(r)}$, $o$ is its orbit size, and $B_\Delta$ is its little cogroup. Each row has one orbit of admissible $N^{(r)}$-irreps. The residual symmetry label determines the restriction by Table~\ref{tab:realMackey}; the separate commutant label is obtained by Eq.~\eqref{eq:AZopposite}. $\mathcal W$ denotes a full cylinder multiplet after tensoring and induction.}
\label{tab:cylinderMackey}
\small
\begin{ruledtabular}
\begin{tabular}{cccccccccc}
$r$ & $N^{(r)}$ & $(d,o)$ & $B_\Delta$ & Sym. AZ & Comm. AZ & $\mathscr D(V)$ & $\Res V$ & Index & $\Qind_r$\\
\hline
$0$ & $\{1\}$ & $(1,1)$ & $\Z_2^T$ & AII & AII & $\HH$ & $\mathcal W$ & $0$ & $0$\\
$1$ & $\{1\}$ & $(1,1)$ & $\Z_2^T\times\Z_2^C$ & DIII & CII & $\Cl_{0,3}$ & $2\mathcal W$ & $n\bmod2$ & $\Z_2$\\
$2$ & $\Z_2^L$ & $(1,2)$ & $\Z_2^C$ & D & C & $\Cl_{0,2}$ & $2\mathcal W$ & $n\bmod2$ & $\Z_2$\\
$3$ & $\Z_2^x\times\Z_2^y$ & $(2,1)$ & $\Z_2^T\times\Z_2^C$ & BDI & CI & $\Cl_{0,1}$ & $\mathcal W_+\oplus\mathcal W_-$ & $n_+-n_-$ & $\Z$\\
$4$ & $\Z_2^x\times\Z_2^y$ & $(2,1)$ & $\Z_2^T$ & AI & AI & $\R$ & $\mathcal W$ & $0$ & $0$\\
$5$ & $\Z_2^x\times\Z_2^y$ & $(2,1)$ & $\Z_2^T\times\Z_2^C$ & CI & BDI & $\Cl_{1,0}$ & $\mathcal W$ & $0$ & $0$\\
$6$ & $\Z_2^L$ & $(1,2)$ & $\Z_2^C$ & C & D & $\Cl_{2,0}$ & $\mathcal W$ & $0$ & $0$\\
$7$ & $\{1\}$ & $(1,1)$ & $\Z_2^T\times\Z_2^C$ & CII & DIII & $\Cl_{3,0}$ & $\mathcal W_+\oplus\mathcal W_-$ & $n_+-n_-$ & $\Z$\\
\end{tabular}
\end{ruledtabular}
\end{table*}

The general restriction theorem now determines each quotient. At $r=1,2$, the residual DIII and D modules require doubling, leaving the parity of the number of full cylinder multiplets. At $r=3,7$, residual BDI and CII pair two inequivalent irreps, leaving their multiplicity difference. At $r=0,4,5,6$, residual AII, AI, CI, and C each admit lifting of one complete multiplet. Thus
\begin{equation}
 (\Qind_0,\ldots,\Qind_7)
 =(0,\Z_2,\Z_2,\Z,0,0,0,\Z).
 \label{eq:cylinderpattern}
\end{equation}
This reproduces the direct extension calculation of Ref.~\cite{ZhaoSigma2026} from the orbit, residual-factor, and restriction data of our general theory. In particular, the trivial quotients at $r=5,6$ follow from their residual CI and C types even though each full cylinder multiplet has an even complex dimension.

The normalization also follows from induction. At $r=1$, the effective ungraded DIII multiplet has complex dimension two and $(d,o)=(1,1)$. At $r=2$, the effective D multiplet has dimension one and $(d,o)=(1,2)$. Both full multiplets therefore have complex dimension two, giving
\begin{equation}
 \nu_{\mathrm{cyl}}=
 \tfrac12\dim_\C\ker\widehat H\pmod2.
 \label{eq:cylparity}
\end{equation}
At $r=3$, a one-dimensional residual BDI multiplet is tensored with the two-dimensional $\Delta$; at $r=7$, the residual CII multiplet is itself two-dimensional. The integers in both cases count full multiplets of the two inequivalent types. These counts illustrate why the reference-irrep dimension and orbit size must be retained when translating the quotient into a kernel dimension.

In these two cases, $r=3,7$, the cylinder Clifford algebra has two inequivalent ungraded irreducible modules $\mathcal W_\pm$. Their restrictions to $N^{(r)}$ have the same irrep content, so either starting module yields the same residual AZ type and quotient-index group. The lifting relation $[\mathcal W_+]+[\mathcal W_-]=0$ in the quotient nevertheless makes them opposite generators of $\Z$. Thus exchanging $\mathcal W_+$ and $\mathcal W_-$ reverses the orientation of the integer index $\nu=n_+-n_-$, without changing its type or the obstruction to lifting.

\subsection{Texture topology and the fermion phase}
The subgroup calculation gives the abstract group pattern $\Qind_r\cong KO^{r-3}(\mathrm{pt})$, rather than the physical bulk pattern $KO^{-r}(\mathrm{pt})$. For stable auxiliary rank and $D>0$, the compact localization targets $\mathcal M_s\simeq R_{4-s}$ satisfy \cite{Ryu2010}
\begin{equation}
 \pi_D(\mathcal M_s)\cong KO^{r-3}(\mathrm{pt})
 \cong\Qind_r.
 \label{eq:target}
\end{equation}
Agreement of groups does not fix the physical map. Reference~\cite{ZhaoSigma2026} establishes, under its minimal-module, slowly varying endpoint, and Fredholm hypotheses, the normalized correspondence
\begin{align}
 \Phi_{s,D}:\pi_D(\mathcal M_s)&\longrightarrow\Qind_r,\nonumber\\
 [Q_1]-[Q_0]&\longmapsto[\ker\widehat H[Q_t]].
 \label{eq:texturemap}
\end{align}
The normalization is fixed there by deformation and excision to a Clifford defect whose Gaussian oscillator kernel carries one elementary index. This analytic identification is a separate input beyond the algebraic quotient.

At $r=1,2$, a field-independent congruence, a common regularization, and real Pfaffian spectral flow further give the relative phase \cite{ZhaoSigma2026,RyuMudryObuseFurusaki2007}
\begin{equation}
 \mathcal P[Q_1,Q_0]=(-1)^{\nu_{\mathrm{cyl}}[Q_1,Q_0]}.
 \label{eq:pfaffian}
\end{equation}
Combining Eqs.~\eqref{eq:texturemap} and \eqref{eq:pfaffian} yields, relative to a fixed reference texture, the minimal-model topological weight
\begin{equation}
 \begin{aligned}
 e^{-S_{\mathrm{top}}[Q]}&=(-1)^{\nu_{\mathrm{hom}}[Q]},\\
 S_{\mathrm{top}}&=i\pi\nu_{\mathrm{hom}}\pmod{2\pi i}.
 \end{aligned}
 \label{eq:theta}
\end{equation}
Independent physical copies add their parities. For instance, a two-dimensional AII boundary has $s=4$ and $r=1$, so the cylinder quotient is $\Z_2$ and the stated minimal theory has the nontrivial theta character \cite{RyuMudryObuseFurusaki2007}. A two-dimensional DIII boundary instead has $r=0$ and no same-dimensional cylinder obstruction; its Wess--Zumino (WZ) term \cite{Schnyder2008} is detected by an integer texture problem one dimension higher.

More generally, the integer cases $r=3,7$ normalize closed textures in dimension $D$. A WZ action lives in dimension $n=D-1$ and requires the complex-parent anomaly calculation as an additional input. For that $n$-dimensional fermion theory, $r_n=r+1=4,0$ modulo eight. The quotient fixes the integer period, while the anomaly determines the local phase variation. The unit spherical WZ level reported in Ref.~\cite{ZhaoSigma2026} uses both ingredients; it does not follow from the integer kernel class alone.

The application therefore has three logically distinct steps. The representation quotient determines which cylinder zero modes are removable; a normalized texture-index theorem identifies their class with relative texture topology; and regulated spectral flow or an anomaly calculation determines the fermion phase. Exceptional auxiliary rank, $D=0$ component conventions, rapidly varying endpoints, microscopic ultraviolet phases, and restricted local perturbations require additional analysis. The quotient does not remove these qualifications.

\section{Discussion and conclusions}
We have developed a representation-theoretic approach to the stability of zero modes in systems with symmetries that commute or anticommute with the Hamiltonian. The central object is the quotient index group: the additive group of ungraded symmetry representations or corepresentations modulo the image obtained by forgetting the grading of compatible super representations. Spectral flattening gives this quotient a direct physical meaning. A zero-mode sector admits an invertible symmetry-compatible perturbation precisely when it admits a compatible grading, so the restriction image records the representation content that can be removed from zero energy. The quotient retains the obstruction after all such removable content is identified.

Computing this group led us to a systematic investigation of graded corepresentations by Mackey's construction. Starting from the unitary Hamiltonian-preserving normal subgroup $N=\ker\tau\cap\ker c$, we organize its irreducible projective representations into symmetry orbits. Within each orbit, the stabilizer intertwiners separate the action on the reference irrep from a residual symmetry problem on its multiplicity space. Solving that smaller problem and inducing to the full group constructs all irreducible super corepresentations, with the projective phases and antiunitary actions retained. The possible little cogroups and residual factor systems yield two complex structural types when all symmetries are unitary and ten real structural types when antiunitary symmetries are included. Their residual symmetry algebras are the corresponding AZ $CT$ algebras. A given symmetry group can contain several inequivalent sectors of the same type and need not realize every type.

This construction makes the relation between the symmetry and representation versions of the tenfold way explicit. The residual graded module and the induced super irrep have isomorphic graded commutants, independently of the reference-irrep dimension and orbit size. The general algebra--commutant relation then identifies the graded Morita class of the residual symmetry algebra with that of the full symmetry algebra represented on the induced module. Their commutant has the super-opposite class. Thus the established classification of super division algebras becomes a concrete classification of the super irreps constructed from each residual AZ sector. With a common Morita-class convention, the symmetry and commutant labels are related by super opposition, as summarized in Tables~\ref{tab:complexMackey} and \ref{tab:realMackey}. We also derived character criteria that determine these structural types either from the starting $N$-irrep or from a completed super irrep, retaining the orbit and multiplicity factors in the latter case.

The full quotient index group follows by exhaustively enumerating these super irreps and their restrictions, or equivalently by summing the contributions from all independent $N$-irrep orbits. Each orbit contributes $0$, $\Z$, or $\Z_2$. These possibilities correspond to removing one already gappable multiplet, pairing two inequivalent multiplets, or combining two copies of the same multiplet. The nontrivial indices are consequently multiplicity differences and multiplicity parities; only the former occur in the purely unitary case. Equivalent restrictions are counted once even when they arise from inequivalent gradings. In the finite semisimple setting, these relations are complete: all index components vanish if and only if a symmetry-compatible perturbation can lift the entire zero-mode sector. The construction therefore determines both the obstruction and the elementary pairing or doubling processes that remove unprotected modes.

The two applications illustrate how this algebraic information enters physical index problems. For one-dimensional end states, the quotient gives chiral imbalances, Majorana parities, and Majorana Kramers-pair parities, with the stable bulk invariants supplied by bulk--boundary correspondence. For interpolation cylinders of disordered boundary Dirac theories, the subgroup and residual-symmetry analysis determines the protected cylinder-kernel class and its normalization in full multiplet units. A texture-index theorem and a fermion-phase or anomaly calculation then relate that class to sigma-model theta or Wess--Zumino terms \cite{ZhaoSigma2026}.

The appendices examine alternative starting points for the Mackey construction. Appendix~\ref{app:complexalternative} starts from the full unitary group $G$ and organizes its irreps by the twist generated by the added lifting involution. This exchanges the auxiliary residual labels A and AIII relative to the physical commutant labels. Appendix~\ref{app:realalternative} starts from the full unitary subgroup $G_0=\ker\tau$, constructs the enlarged corepresentations, and calculates the corresponding permutation of the ten residual labels. In both cases, keeping the auxiliary grading distinct from the physical commutativity grading recovers the same full super irreps, graded commutants, and quotient indices as the construction through $N$. The residual labels depend on how the representation problem is organized, while the protected zero-mode content does not.

Together, the quotient construction and the Mackey analysis provide an explicit way to determine which zero-mode multiplets are protected by a given projective symmetry group. The result concerns symmetry-compatible perturbations on the zero-mode representation. Incorporating microscopic locality, momentum-dependent families, or interactions requires additional structure, providing directions for extending this framework beyond the algebraic lifting problem.

\begin{acknowledgments}
GPT 6 Astra assisted with language polishing throughout the manuscript and the verification of derivations. This work was supported by the General Research Fund of Hong Kong (Grant Nos.~17302525 and 17301224).
\end{acknowledgments}

\appendix
\section{Alternative complex construction from the full group}
\label{app:complexalternative}
Section~\ref{sec:unitary} starts from $N=\ker c$. An equivalent construction starts instead from the full unitary group $G$ and the extension
\begin{equation}
 1\longrightarrow G\longrightarrow G^h
 \longrightarrow\Z_2^h\longrightarrow1.
 \label{eq:altcomplexextension}
\end{equation}
This approach begins with the original irreps already known; only their extension or pairing under $h$ remains to be determined. Its residual AZ labels differ from those of the $N$-based residual algebras. We specify below the grading responsible for this difference.

\subsection{Action on full-group irreps and the two orbit types}
Let $W$ be an irreducible $\omega$-representation of $G$. Projective conjugation by $h$ acts on $\widehat G$ as the twist by the commutativity grading,
\begin{equation}
 W\longmapsto W^c,\qquad
 \rho_{W^c}(g)=c(g)\rho_W(g).
 \label{eq:altcomplexaction}
\end{equation}
Thus an orbit has one or two elements. In contrast to the construction from $N$, it is now the added generator itself that can exchange the reference irreps.

\paragraph{Two-point orbit: $W^c\not\cong W$.}
The stabilizer is $G$, with trivial little cogroup. Its residual algebra is the purely even $\C$, which has residual type A. There is a single enlarged irrep for the unordered pair $\{W,W^c\}$:
\begin{equation}
 \begin{gathered}
 \rho_V(g)=\begin{pmatrix}\rho_W(g)&0\\0&c(g)\rho_W(g)\end{pmatrix},\\
 h=\begin{pmatrix}0&\id_d\\\id_d&0\end{pmatrix},
 \end{gathered}
 \label{eq:altcomplexpairmatrices}
\end{equation}
where $d=\dim_\C W$. To calculate the graded commutant, a commuting endomorphism is scalar on each irreducible block; an odd endomorphism must intertwine the $c$-twisted actions. Imposing its additional anticommutation with $h$ gives
\begin{equation}
 \mathscr D_0(V)=\C\id_{2d},\qquad
 \mathscr D_1(V)=\C(\sigma_2\otimes\id_d).
 \label{eq:altcomplexpaircommutant}
\end{equation}
Hence the commutant has type $\Cl_1^\C$, or AIII, although the residual type is A.

\paragraph{Fixed irrep: $W^c\cong W$.}
Choose a unitary intertwiner $S$ with $S\rho_W(g)=c(g)\rho_W(g)S$. Schur's lemma makes $S^2$ scalar, and a phase choice sets $S^2=\id$. The stabilizer is all of $G^h$, with little cogroup $\Z_2^h$ and a cohomologically trivial factor system. There are two inequivalent extensions:
\begin{equation}
 \rho_{V_\pm}(g)=\rho_W(g),\qquad h_\pm=\pm S.
 \label{eq:altcomplexfixedmatrices}
\end{equation}
The even commutant is $\C\id_d$. Every candidate odd intertwiner is proportional to $S$, and cannot anticommute with $h_\pm=\pm S$ unless it vanishes. Therefore
\begin{equation}
 \mathscr D(V_\pm)=\C\qquad\text{(purely even, class A)}.
 \label{eq:altcomplexfixedcommutant}
\end{equation}
For the residual classification, declare $h$ odd using the quotient character $\eta(g,s)=(-1)^s$. The residual algebra is then generated by one odd involution, namely $\Cl_1^\C$ of type AIII. This $\eta$ is an auxiliary grading on the enlarged group, not the commutativity grading $c$ on $G$.

\subsection{Reversed labels and agreement with the preserving-subgroup construction}
The two cases yield the label reversal
\begin{equation}
 \boxed{\mathrm A_{\mathrm{res}}\longleftrightarrow
 \mathrm{AIII}_{\mathrm{comm}},\qquad
 \mathrm{AIII}_{\mathrm{res}}\longleftrightarrow
 \mathrm A_{\mathrm{comm}}.}
 \label{eq:altcomplexreversal}
\end{equation}
Once the reversal in Eq.~\eqref{eq:altcomplexreversal} is established, the restriction, index, and quotient follow from the row of Table~\ref{tab:complexMackey} with the paired \emph{graded-commutant label}; no new derivation is needed. A fixed $N$-irrep in Sec.~\ref{sec:unitary} produces the two $G$-irreps exchanged by $h$ here, whereas a two-point $N$-orbit produces the single $c$-invariant $G$-irrep. Both starting subgroups therefore give the same enlarged irreps and physical graded commutants. Only the residual algebra to which the preliminary AZ label is attached changes.

\section{Alternative real construction from the unitary subgroup}
\label{app:realalternative}
This alternative construction shares its subgroup-based induction and block structure with the antilinear representation theory developed by Rumynin and Taylor \cite{RumyninTaylor2021}. We use that established framework for the enlarged projective symmetry problem. The purpose here is to determine how the resulting auxiliary residual labels map to the physical graded commutants and to verify agreement with the quotient computed from $N$ in the main text.

For a group containing antiunitary elements, let $G_0=\ker\tau$ be its unitary subgroup. This is generally larger than $N=\ker\tau\cap\ker c$: a unitary reversing symmetry belongs to $G_0$ but not to $N$. The extension used here is
\begin{equation}
 \begin{gathered}
 1\longrightarrow G_0\longrightarrow G^h
 \longrightarrow B_0\longrightarrow1,\\
 B_0\cong\Z_2^T\times\Z_2^h.
 \end{gathered}
 \label{eq:altrealextension}
\end{equation}
Choose any antiunitary representative $a\in G$. The label $T$ in $B_0$ denotes $aG_0$; it does \emph{not} require $c(a)=+1$. In particular, this construction also applies when $c=\tau$ and every antiunitary symmetry reverses the Hamiltonian. Write $\Omega=\omega^c$ and $\bar\tau(g,s)=\tau(g)$ on $G^h$.

The construction can also be organized in two stages: first construct projective representations of the full unitary subgroup $G_0^h=G_0\times\Z_2^h$ with multiplier $\Omega|_{G_0^h}$, then extend or induce across the antiunitary coset. The second stage is the projective counterpart of the index-two construction in Ref.~\cite{RumyninTaylor2021}. Below we organize both stages directly through $G_0$-irrep orbits, retaining the additional action of $h$ and distinguishing its auxiliary grading from $c$.

The residual AZ classification uses
\begin{equation}
 \begin{gathered}
 \eta(g,s)=(-1)^s,\\
 (\bar\tau,\eta)(T)=(-1,+1),\qquad
 (\bar\tau,\eta)(h)=(+1,-1).
 \end{gathered}
 \label{eq:altauxgrading}
\end{equation}
Thus $h$ plays the role of the odd unitary residual generator, and $Th$ that of an odd antiunitary generator. Both gradings are trivial on $G_0$, so this residual classification is well defined on the little cogroups. The auxiliary $\eta$ must not be substituted for the physical commutativity grading $c$ in the graded commutant of $V$.

With the unitarity grading $\bar\tau$ and the auxiliary grading $\eta$, the construction of Sec.~\ref{sec:tenfold} proceeds unchanged, with the same five little cogroups and hence the same ten types. Only the label bookkeeping changes because the residual algebra is now computed with $(\bar\tau,\eta)$ rather than $c$; the physical object is unaffected. Every irreducible corepresentation $V$ of $G^h$ still has a graded commutant with respect to the commutativity grading $c$, and its type is one of the ten graded division algebras
\[
\C,\ \Cl_1^\C,\ \R,\ \Cl_{1,0},\ \Cl_{2,0},\ \Cl_{3,0},\ \HH,\ \Cl_{0,3},\ \Cl_{0,2},\ \Cl_{0,1},
\]
whose Freed--Moore AZ labels are, respectively, A, AIII, AI, BDI, D, DIII, AII, CII, C, and CI. These labels refer to the algebras' own graded Morita classes, not to the auxiliary residual symmetry types that produce them. The commutant algebra therefore labels the physical structural type, independently of the residual label used in this appendix. This is the graded analogue of Dyson's threefold way, in which Wigner's three types of corepresentations are labelled by the three real division algebras $\R$, $\HH$, and $\C$: the commutativity grading replaces those three by the ten graded division algebras just listed.

\subsection{Orbits, little cogroups, and residual multipliers}
Let $\Delta$ be an irrep of $G_0$ with multiplier $\omega|_{G_0}$. The two generators act on $\widehat{G_0}$ as
\begin{align}
 \Delta^h(k)&=c(k)\Delta(k),\nonumber\\
 \Delta^T(k)&=\overline{\xi(a,k)\Delta(a^{-1}ka)},
 \qquad k\in G_0,
 \label{eq:altrealactions}
\end{align}
where $\xi$ is defined in Eq.~\eqref{eq:orbit}. These actions commute on equivalence classes, and $(Th)\cdot[\Delta]=T\cdot[h\cdot\Delta]$. Their orbits have sizes $o=4,2,1$. Let $S_\Delta\subseteq G^h$ be the stabilizer and $K_\Delta=S_\Delta/G_0$ its little cogroup. The five possibilities are
\begin{equation}
 K_\Delta=\{1\},\ \langle h\rangle,\ \langle T\rangle,
 \ \langle Th\rangle,\ B_0.
 \label{eq:altlittlegroups}
\end{equation}
Choose semilinear intertwiners $L_x=U_xK^{\eps(x)}$ on $\Delta$ for $x\in S_\Delta$, with $L_k=\Delta(k)$ for $k\in G_0$. As before, their multiplier $\lambda$ removes the reference-space contribution:
\begin{equation}
 L_xL_y=\lambda(x,y)L_{xy},\qquad
 \beta(\bar x,\bar y)=\frac{\Omega(x,y)}{\lambda(x,y)}.
 \label{eq:altresidualmultiplier}
\end{equation}
Compatible coset choices make $\beta$ a multiplier on $K_\Delta$. When $h\in S_\Delta$, choose $U_h^2=\id$; this is possible because $h^2=1$ and the remaining scalar square of its unitary intertwiner can be removed by a phase. The effective $h_E$ can then also be normalized to square to $\id$. The residual antiunitary squares are those of $\beta$, not necessarily the bare squares of $a$ or $ah$.

The trivial subgroup gives A. The unitary subgroup $\langle h\rangle$ has one factor class, with its odd generator normalized to square to $+1$, giving AIII. The antiunitary subgroups $\langle T\rangle$ and $\langle Th\rangle$ each have two square signs: they give AI/AII and D/C, respectively. For the full little cogroup normalize $h_E^2=\id$ and write
\begin{equation}
 T_E^2=\kappa_T,\qquad (T_Eh_E)^2=\kappa_{Th}.
 \label{eq:altresidualsquares}
\end{equation}
The two signs specify the four factor classes BDI, DIII, CII, and CI. Equivalently, $T_Eh_E=\zeta h_ET_E$ with $\zeta=\kappa_{Th}/\kappa_T=\pm1$. Thus the five little cogroups give $1+1+2+2+4=10$ residual projective symmetry types.

These ten possibilities form a universal list; they are not ten sectors that must occur for every $G$. For example, if $c|_{G_0}=1$, the $h$ action on $\widehat{G_0}$ is trivial. Every stabilizer then contains $h$, so only $\langle h\rangle$ and $B_0$ can occur. The given physical multiplier further constrains the possible factor classes.

\subsection{Explicit residual modules and full enlarged corepresentations}
Table~\ref{tab:altresidualmodules} gives every irreducible residual module $E$ and records separately its auxiliary residual symmetry label, the resulting physical graded commutant and its Freed--Moore label, and the label of the full represented symmetry algebra $\mathscr B_V$. Write $J=-i\sigma_2$. A dash means that the generator is absent from the little cogroup, not that it acts by zero. The independent signs of $h_E$ in AIII, BDI, and CII distinguish two residual irreps. In DIII and CI the two signs are equivalent. All entries use one complex conjugation $K$ on the full effective space.

\begin{table*}[tbp]
\caption{Residual modules for the construction from $G_0$. The auxiliary residual label uses $(\bar\tau,\eta)$. The last three columns use the physical grading $c$: the commutant algebra, its Freed--Moore AZ label, and the AZ label of the represented original symmetry algebra $\mathscr B_V$. The last label is $\iota(\operatorname{AZ}[\mathscr D(V)])$ and equals the $N$-based residual symmetry label of Sec.~\ref{sec:tenfold}. $o=[B_0:K_\Delta]$ is the orbit size; $e=\dim_\C E$ and the full enlarged irrep has dimension $oe\dim_\C\Delta$. For the full little cogroup the odd antiunitary is $T_Eh_E$. Each independent sign in the $h_E$ column gives a different enlarged irrep after induction.}
\label{tab:altresidualmodules}
\small
\begin{ruledtabular}
\begin{tabular}{cccccccccc}
Aux. AZ & $K_\Delta$ & $o$ & $e$ & $T_E$ & $(Th)_E$ & $h_E$ & $\mathscr D(V)$ & Comm. AZ & $\operatorname{AZ}(\mathscr B_V)$\\
\hline
A & $\{1\}$ & $4$ & $1$ & --- & --- & --- & $\Cl_1^\C$ & AIII & AIII\\
AIII & $\langle h\rangle$ & $2$ & $1$ & --- & --- & $\pm1$ & $\C$ & A & A\\
AI & $\langle T\rangle$ & $2$ & $1$ & $K$ & --- & --- & $\Cl_{0,1}$ & CI & BDI\\
BDI & $B_0$ & $1$ & $1$ & $K$ & $T_Eh_E$ & $\pm1$ & $\R$ & AI & AI\\
D & $\langle Th\rangle$ & $2$ & $1$ & --- & $K$ & --- & $\Cl_{1,0}$ & BDI & CI\\
DIII & $B_0$ & $1$ & $2$ & $JK$ & $T_Eh_E$ & $\sigma_2$ & $\Cl_{2,0}$ & D & C\\
AII & $\langle T\rangle$ & $2$ & $2$ & $JK$ & --- & --- & $\Cl_{3,0}$ & DIII & CII\\
CII & $B_0$ & $1$ & $2$ & $JK$ & $T_Eh_E$ & $\pm\id_2$ & $\HH$ & AII & AII\\
C & $\langle Th\rangle$ & $2$ & $2$ & --- & $JK$ & --- & $\Cl_{0,3}$ & CII & DIII\\
CI & $B_0$ & $1$ & $2$ & $\sigma_1K$ & $T_Eh_E$ & $\sigma_3$ & $\Cl_{0,2}$ & C & D\\
\end{tabular}
\end{ruledtabular}
\end{table*}

The reference stabilizer module is obtained directly from these matrices and the intertwiner on $\Delta$:
\begin{equation}
 A_E(x)K^{\eps(x)}
 =[E(\bar x)\otimes U_x]K^{\eps(x)},\qquad x\in S_\Delta,
 \label{eq:altlittlematrix}
\end{equation}
where $E(\bar x)$ denotes the matrix part of the listed residual operator. Its factor is $\beta\lambda=\Omega$. The desired enlarged irrep is
\begin{equation}
 V_{\Delta,E}=\Ind_{S_\Delta}^{G^h}(E\otimes L).
 \label{eq:altinduction}
\end{equation}
Varying one $\Delta$ per $B_0$ orbit and the residual modules in Table~\ref{tab:altresidualmodules} gives all irreducible corepresentations of $G^h$.

For a matrix realization, choose representatives $q_i$ of $G^h/S_\Delta$, taking the first to be the identity. Whenever $xq_j=q_i l$ with $l\in S_\Delta$, the only nonzero block in column $j$ is
\begin{equation}
 \rho_V(x)_{ij}=\frac{\Omega(x,q_j)}{\Omega(q_i,l)}
 K^{\eps(q_i)}A_E(l)K^{\eps(q_i)}K^{\eps(x)}.
 \label{eq:altinducedmatrix}
\end{equation}
The final $K^{\eps(x)}$ denotes a single conjugation of the entire block vector. In particular, setting $x=h$ gives the physical lifting involution whether or not $h$ belongs to the reference stabilizer. When $h$ exchanges different $G_0$-isotypic blocks, the table's $h_E$ cannot be used as the full lifting matrix.

For clarity, the two-block case has a simple closed form. Choose $q\notin S_\Delta$, so $q^2\in G_0$. For $l\in S_\Delta$ and $l'=q^{-1}lq$, define
\begin{equation}
 A_E^{[q]}(l)=\frac{\Omega(l,q)}{\Omega(q,l')}
 K^{\eps(q)}A_E(l')K^{\eps(q)}.
 \label{eq:alttransportedblock}
\end{equation}
Then the complete generating matrices are
\begin{equation}
 \begin{gathered}
 \rho_V(l)=\begin{pmatrix}A_E(l)&0\\0&A_E^{[q]}(l)\end{pmatrix}K^{\eps(l)},\\
 \rho_V(q)=\begin{pmatrix}0&\Omega(q,q)A_E(q^2)\\\id&0\end{pmatrix}K^{\eps(q)}.
 \end{gathered}
 \label{eq:alttwoblock}
\end{equation}
One may take $q=a$ for $K_\Delta=\langle h\rangle$, and $q=h$ for either antiunitary little cogroup. For the trivial little cogroup use four blocks $(1,a,h,ah)$ in Eq.~\eqref{eq:altinducedmatrix}; for the full little cogroup no induction is needed. These prescriptions retain nontrivial projective phases and nonsplit antiunitary squares.

\subsection{The graded commutant and the map of real AZ labels}
The commutant of the physical super representation remains defined by Eq.~\eqref{eq:supercommutant}, using $c$. It cannot be identified by simply assigning the residual AZ label in Table~\ref{tab:altresidualmodules}. To determine the relation, set
\begin{equation}
 A_p=h^pM_p.
 \label{eq:altcommutanttransform}
\end{equation}
The two anticommutation signs cancel on $G$, whereas the sign on $h$ remains:
\begin{equation}
 A_p\rho(g)=\rho(g)A_p,\qquad
 A_ph=(-1)^p hA_p.
 \label{eq:altcommutantconditions}
\end{equation}
In particular $A_p$ commutes with $G_0$, so on the reference isotypic block it is $m_p\otimes\id_\Delta$. The residual constraints become
\begin{equation}
 m_pE(b)=\eta(b)^p E(b)K^{\eps(b)}m_pK^{\eps(b)},
 \qquad b\in K_\Delta.
 \label{eq:altresidualcommutant}
\end{equation}
Denote the degree-tagged solution spaces by $\mathscr E_p(E)$. They can coincide as ordinary matrix spaces when no odd residual generator is present, but they remain distinct homogeneous spaces: Eq.~\eqref{eq:altcommutanttransport} gives them different signs on the exchanged blocks. On another block the extension is uniquely fixed as
\begin{equation}
 (A_p)_i=\eta(q_i)^p K^{\eps(q_i)}
 (m_p\otimes\id_\Delta)K^{\eps(q_i)}.
 \label{eq:altcommutanttransport}
\end{equation}
This reduces the calculation to the small matrices in Table~\ref{tab:altresidualmodules}, independently of the dimension of $\Delta$ or the orbit size.

The transformation in Eq.~\eqref{eq:altcommutanttransform} changes multiplication of two odd elements. In fact,
\begin{equation}
 \begin{gathered}
 (h^pA_p)(h^qA_q)=(-1)^{pq}h^{p+q}A_pA_q,\\
 m_p\star m_q=(-1)^{pq}m_pm_q.
 \end{gathered}
 \label{eq:altsigntwist}
\end{equation}
Consequently, $\mathscr D(V)$ is isomorphic to $\mathscr E(E)$ equipped with this sign-twisted multiplication, but not generally to $\mathscr E(E)$ with ordinary matrix multiplication. In particular, the square of an odd generator changes sign. This sign change produces the residual-to-commutant permutation.

Here is an explicit calculation for all ten cases. Put $F=\mathscr E_0(E)$ and use $J=-i\sigma_2$ as above. For residual AIII, BDI, and CII, $h_E=\pm\id$ forces $\mathscr E_1=0$, leaving $F=\C,\R,\HH$, respectively. Their physical commutants therefore have labels A, AI, and AII.

For residual A, AI, and AII the only residual constraints are absent or preserving antiunitary ones. Thus $\mathscr E_0=\mathscr E_1=F$ with $F=\C,\R,\HH$, and one can choose the ordinary odd matrix $j=\id$. After Eq.~\eqref{eq:altsigntwist}, $j\star j=-\id$. The three physical graded algebras are $\Cl_1^\C$, $\Cl_{0,1}$, and $\Cl_{3,0}$, with standard AZ labels AIII, CI, and DIII. In the complex case multiplying the odd generator by $i$ changes its square to $+1$; in the quaternionic case the three odd elements $I_\mu j$ have positive square.

For residual D and C, respectively, the odd antiunitary is $K$ or $JK$. The even algebra is $F=\R$ or $\HH$, and the odd space is $Fj$ with $j=i\id$. Its ordinary square is $-\id$, and $j$ commutes with $F$. The sign twist makes its square positive, giving $\Cl_{1,0}$ and $\Cl_{0,3}$, or standard commutant labels BDI and CII.

The remaining full-little-cogroup cases are distinguished by a complex even algebra:
\begin{equation}
 \begin{aligned}
 F_{\mathrm{DIII}}&=\operatorname{span}_{\R}\{\id_2,J\},\\
 j_{\mathrm{DIII}}&=i\sigma_1,\qquad j_{\mathrm{DIII}}^2=-\id_2,\\
 F_{\mathrm{CI}}&=\operatorname{span}_{\R}\{\id_2,i\sigma_3\},\\
 j_{\mathrm{CI}}&=\sigma_1,\qquad j_{\mathrm{CI}}^2=+\id_2.
 \end{aligned}
 \label{eq:altremaininggenerators}
\end{equation}
These are ordinary matrix squares. Direct substitution into Eq.~\eqref{eq:altresidualcommutant} shows that the full odd space is $Fj$, and $j$ conjugates the complex structure in $F$. After reversing its square, the algebras are $\Cl_{2,0}$ for residual DIII and $\Cl_{0,2}$ for residual CI, with standard commutant labels D and C.

Together these calculations give the following map from the auxiliary residual symmetry label to the standard AZ label of the physical graded commutant:
\begin{equation}
 \boxed{\begin{aligned}
 \mathrm A&\mapsto\mathrm{AIII},&\mathrm{AIII}&\mapsto\mathrm A,\\
 \mathrm{AI}&\mapsto\mathrm{CI},&\mathrm{BDI}&\mapsto\mathrm{AI},\\
 \mathrm D&\mapsto\mathrm{BDI},&\mathrm{DIII}&\mapsto\mathrm D,\\
 \mathrm{AII}&\mapsto\mathrm{DIII},&\mathrm{CII}&\mapsto\mathrm{AII},\\
 \mathrm C&\mapsto\mathrm{CII},&\mathrm{CI}&\mapsto\mathrm C.
 \end{aligned}}
 \label{eq:altrealmap}
\end{equation}
Table~\ref{tab:altresidualmodules} records this map alongside the computed division algebras. It consists of the A/AIII exchange and an eight-cycle on the real labels; it is not an involution. It differs from $\iota$ in Eq.~\eqref{eq:AZopposite} because the residual algebra here starts from $G_0$ and uses the auxiliary grading $\eta$, whereas Eq.~\eqref{eq:AZcorrespondence} starts from $N$ and uses the inherited $c$.

For the resulting full super irrep, the represented original symmetry algebra is still $\mathscr B_V$ of Eq.~\eqref{eq:representedalgebra}, with $h$ implementing its grading. Its standard label is
\begin{equation}
 \operatorname{AZ}(\mathscr B_V)
 =\operatorname{AZ}[\mathscr D(V)^{\mathrm{sop}}]
 =\iota\!\left(\operatorname{AZ}[\mathscr D(V)]\right).
 \label{eq:altrepresentedAZ}
\end{equation}
This equals the residual symmetry label obtained by constructing the same module through $N$, as proved in Eq.~\eqref{eq:representedMorita}. It need not equal the auxiliary residual label used in this appendix: the latter grades the enlarged little-cogroup algebra by $\eta$ and does not represent the original $c$-graded symmetry algebra on the residual space.

Once Eq.~\eqref{eq:altrealmap} is established, the restriction and quotient data follow from Table~\ref{tab:realMackey}. For each auxiliary residual type in Table~\ref{tab:altresidualmodules}, use its image under Eq.~\eqref{eq:altrealmap} to select the row by the \emph{commutant AZ} column, rather than the symmetry AZ column. Equivalently, apply $\iota$ once more to that image to obtain the $N$-based residual symmetry label. The constructions through $N$ and $G_0$ organize the same irreducible $G^h$-corepresentations differently but leave their physical graded commutants and protected zero-mode content unchanged.

\clearpage
\bibliography{Chiral_Index_PRB_V0}

@article{Dyson1962,
 author={Dyson, Freeman J.}, title={The Threefold Way. Algebraic Structure of Symmetry Groups and Ensembles in Quantum Mechanics},
 journal={J. Math. Phys.}, volume={3}, pages={1199--1215}, year={1962}, doi={10.1063/1.1703863}}

@article{FreedMoore2013,
 author={Freed, Daniel S. and Moore, Gregory W.}, title={Twisted Equivariant Matter},
 journal={Ann. Henri Poincar\'e}, volume={14}, pages={1927--2023}, year={2013}, doi={10.1007/s00023-013-0236-x}, eprint={1208.5055}, archivePrefix={arXiv}, primaryClass={hep-th}}

@article{ABS1964,
 author={Atiyah, M. F. and Bott, R. and Shapiro, A.}, title={Clifford Modules},
 journal={Topology}, volume={3}, pages={3--38}, year={1964}, note={Suppl. 1}, doi={10.1016/0040-9383(64)90003-5}}

@article{AtiyahSinger1969,
 author={Atiyah, M. F. and Singer, I. M.}, title={Index theory for skew-adjoint {Fredholm} operators},
 journal={Publ. Math. Inst. Hautes \'Etudes Sci.}, volume={37}, pages={5--26}, year={1969}, doi={10.1007/BF02684885}}

@article{AltlandZirnbauer1997,
 author={Altland, Alexander and Zirnbauer, Martin R.}, title={Nonstandard symmetry classes in mesoscopic normal-superconducting hybrid structures},
 journal={Phys. Rev. B}, volume={55}, pages={1142--1161}, year={1997}, doi={10.1103/PhysRevB.55.1142}}

@article{Schnyder2008,
 author={Schnyder, Andreas P. and Ryu, Shinsei and Furusaki, Akira and Ludwig, Andreas W. W.},
 title={Classification of topological insulators and superconductors in three spatial dimensions},
 journal={Phys. Rev. B}, volume={78}, pages={195125}, year={2008}, doi={10.1103/PhysRevB.78.195125}}

@article{Ryu2010,
 author={Ryu, Shinsei and Schnyder, Andreas P. and Furusaki, Akira and Ludwig, Andreas W. W.},
 title={Topological insulators and superconductors: Tenfold way and dimensional hierarchy},
 journal={New J. Phys.}, volume={12}, pages={065010}, year={2010}, doi={10.1088/1367-2630/12/6/065010}}

@article{Kitaev2009,
 author={Kitaev, Alexei}, title={Periodic table for topological insulators and superconductors},
 journal={AIP Conf. Proc.}, volume={1134}, pages={22--30}, year={2009}, doi={10.1063/1.3149495}}

@misc{ZhaoSigma2026,
 author={Zhao, Y. X.},
 title={A {Clifford}-index approach connecting boundary {Dirac} fermions to sigma-model topological terms in the periodic table of topological insulators and superconductors},
 year={2026}, eprint={2611.01619}, archivePrefix={arXiv}, primaryClass={cond-mat.dis-nn}}

@article{Wall1964,
 author={Wall, C. T. C.}, title={Graded {Brauer} groups},
 journal={J. Reine Angew. Math.}, volume={213}, pages={187--199},
 year={1964}, doi={10.1515/crll.1964.213.187}}

@incollection{DeligneMorgan1999,
 author={Deligne, Pierre and Morgan, John W.},
 title={Notes on supersymmetry (following {Joseph Bernstein})},
 booktitle={Quantum Fields and Strings: A Course for Mathematicians},
 volume={1}, publisher={American Mathematical Society}, address={Providence, RI},
 year={1999}, pages={41--97}}

@incollection{DeligneSpinors1999,
 author={Deligne, Pierre}, title={Notes on spinors},
 booktitle={Quantum Fields and Strings: A Course for Mathematicians},
 volume={1}, publisher={American Mathematical Society}, address={Providence, RI},
 year={1999}, pages={99--135}}

@article{Mackey1958,
 author={Mackey, George W.}, title={Unitary representations of group extensions. {I}},
 journal={Acta Math.}, volume={99}, pages={265--311}, year={1958},
 doi={10.1007/BF02392428}}

@article{Sumita2019,
 author={Sumita, Shuntaro and Nomoto, Takuya and Shiozaki, Ken and Yanase, Youichi},
 title={Classification of topological crystalline superconducting nodes on high-symmetry lines: Point nodes, line nodes, and {Bogoliubov} {Fermi} surfaces},
 journal={Phys. Rev. B}, volume={99}, pages={134513}, year={2019},
 doi={10.1103/PhysRevB.99.134513}}

@article{YangLiu2024,
 author={Yang, Arist Zhenyuan and Liu, Zheng-Xin},
 title={Representation theory for massless quasiparticles in {Bogoliubov-de Gennes} systems},
 journal={Phys. Rev. B}, volume={110}, pages={054507}, year={2024},
 doi={10.1103/PhysRevB.110.054507}}

@article{GeikoMoore2021,
 author={Geiko, Roman and Moore, Gregory W.},
 title={{Dyson}'s classification and real division superalgebras},
 journal={J. High Energy Phys.}, volume={2021}, number={4}, pages={299}, year={2021},
 doi={10.1007/JHEP04(2021)299}}

@article{RumyninTaylor2021,
 author={Rumynin, Dmitriy and Taylor, James},
 title={Real representations of {$C_2$}-graded groups: The antilinear theory},
 journal={Linear Algebra Appl.}, volume={610}, pages={135--168}, year={2021},
 doi={10.1016/j.laa.2020.09.040}, eprint={2006.09765}, archivePrefix={arXiv}, primaryClass={math.RT}}

@article{Wigner1939,
 author={Wigner, E. P.},
 title={On unitary representations of the inhomogeneous {Lorentz} group},
 journal={Ann. Math.}, volume={40}, pages={149--204}, year={1939},
 doi={10.2307/1968551}}

@article{ZhaoProjectivePT2021,
 author={Zhao, Y. X. and Chen, Cong and Sheng, Xian-Lei and Yang, Shengyuan A.},
 title={Switching spinless and spinful topological phases with projective {$PT$} symmetry},
 journal={Phys. Rev. Lett.}, volume={126}, pages={196402}, year={2021},
 doi={10.1103/PhysRevLett.126.196402}}

@article{ShaoGauge2021,
 author={Shao, L. B. and Liu, Q. and Xiao, R. and Yang, Shengyuan A. and Zhao, Y. X.},
 title={Gauge-field extended {$k\cdot p$} method and novel topological phases},
 journal={Phys. Rev. Lett.}, volume={127}, pages={076401}, year={2021},
 doi={10.1103/PhysRevLett.127.076401}}

@article{ChenGauge2023,
 author={Chen, Z. Y. and Zhang, Zheng and Yang, Shengyuan A. and Zhao, Y. X.},
 title={Classification of time-reversal-invariant crystals with gauge structures},
 journal={Nat. Commun.}, volume={14}, pages={743}, year={2023},
 doi={10.1038/s41467-023-36447-7}, eprint={2302.00187}, archivePrefix={arXiv}, primaryClass={cond-mat.mes-hall}}

@article{ZhaoWang2014,
 author={Zhao, Y. X. and Wang, Z. D.},
 title={Exotic topological types of {Majorana} zero modes and their universal quantum manipulation},
 journal={Phys. Rev. B}, volume={90}, pages={115158}, year={2014},
 doi={10.1103/PhysRevB.90.115158}}

@article{Parthasarathy1969,
 author={Parthasarathy, K. R.},
 title={Projective unitary antiunitary representations of locally compact groups},
 journal={Commun. Math. Phys.}, volume={15}, pages={305--328}, year={1969},
 doi={10.1007/BF01645530}}

@article{ShawLever1974,
 author={Shaw, Ronald and Lever, Joseph},
 title={Irreducible multiplier corepresentations and generalized inducing},
 journal={Commun. Math. Phys.}, volume={38}, pages={257--277}, year={1974},
 doi={10.1007/BF01607948}}

@article{AtiyahSinger1968,
 author={Atiyah, M. F. and Singer, I. M.},
 title={The index of elliptic operators. {I}},
 journal={Ann. Math.}, volume={87}, pages={484--530}, year={1968},
 doi={10.2307/1970715}}

@article{Fujikawa1979,
 author={Fujikawa, Kazuo},
 title={Path-integral measure for gauge-invariant fermion theories},
 journal={Phys. Rev. Lett.}, volume={42}, pages={1195--1198}, year={1979},
 doi={10.1103/PhysRevLett.42.1195}}

@article{SuSchriefferHeeger1979,
 author={Su, W. P. and Schrieffer, J. R. and Heeger, A. J.},
 title={Solitons in polyacetylene},
 journal={Phys. Rev. Lett.}, volume={42}, pages={1698--1701}, year={1979},
 doi={10.1103/PhysRevLett.42.1698}}

@article{SuSchriefferHeeger1980,
 author={Su, W. P. and Schrieffer, J. R. and Heeger, A. J.},
 title={Soliton excitations in polyacetylene},
 journal={Phys. Rev. B}, volume={22}, pages={2099--2111}, year={1980},
 doi={10.1103/PhysRevB.22.2099}}

@article{RyuHatsugai2002,
 author={Ryu, Shinsei and Hatsugai, Yasuhiro},
 title={Topological origin of zero-energy edge states in particle-hole symmetric systems},
 journal={Phys. Rev. Lett.}, volume={89}, pages={077002}, year={2002},
 doi={10.1103/PhysRevLett.89.077002}, eprint={cond-mat/0112197}, archivePrefix={arXiv}}

@article{Kitaev2001,
 author={Kitaev, A. Yu.},
 title={Unpaired {Majorana} fermions in quantum wires},
 journal={Phys. Usp.}, volume={44}, number={10S}, pages={131--136}, year={2001},
 doi={10.1070/1063-7869/44/10S/S29}, eprint={cond-mat/0010440}, archivePrefix={arXiv}}

@article{ZhangKaneMele2013,
 author={Zhang, Fan and Kane, C. L. and Mele, E. J.},
 title={Time-reversal-invariant topological superconductivity and {Majorana} {Kramers} pairs},
 journal={Phys. Rev. Lett.}, volume={111}, pages={056402}, year={2013},
 doi={10.1103/PhysRevLett.111.056402}}

@article{RyuMudryObuseFurusaki2007,
 author={Ryu, Shinsei and Mudry, Christopher and Obuse, Hideaki and Furusaki, Akira},
 title={{$Z_2$} topological term, the global anomaly, and the two-dimensional symplectic symmetry class of {Anderson} localization},
 journal={Phys. Rev. Lett.}, volume={99}, pages={116601}, year={2007},
 doi={10.1103/PhysRevLett.99.116601}}

@misc{Zhang2026,
 author={Zhang, Zheng and Wang, Peiyuan and Zhao, Y. X.},
 title={Projective representation theory of spin space groups},
 year={2026}, eprint={2609.21289}, archivePrefix={arXiv}, primaryClass={cond-mat.mes-hall}}
\end{document}